\documentclass[nofootinbib,prx,twocolumn,showpacs,superscriptaddress,notitlepage,superscriptaddress,amsmath]{revtex4-2}
\usepackage[T1]{fontenc}

\usepackage{lipsum}  
\usepackage{dsfont}
\usepackage{amsmath}
\usepackage{braket}
\usepackage{amssymb}
\usepackage{amsthm}
\usepackage{algpseudocode}
\usepackage{algorithm}
\usepackage{siunitx}
\usepackage{amsfonts}
\usepackage{comment}%https://www.overleaf.com/project/603d67a5abedda342ae4c069
\usepackage[normalem]{ulem}
\usepackage{graphicx}
\usepackage{color,framed}
\usepackage{hyperref}
\usepackage{enumerate}
\usepackage{lipsum}
\usepackage{slashed}
\usepackage{url}
\usepackage{bbm}
\usepackage{chngcntr}
\counterwithout{equation}{section}
\usepackage{tikz,pgfplots}
\usepackage{pgfplotstable}
\usepackage{siunitx}
\usepackage{comment}
\usepackage{graphicx}

\makeatletter
\newcommand{\setlabel}[1]{\edef\@currentlabel{#1}\label}
\makeatother

\usepgfplotslibrary{fillbetween}

\hypersetup{
    colorlinks=true, %set true if you want colored links
    linktoc=all,     %set to all if you want both sections and subsections linked
    linkcolor=blue,  %choose some color if you want links to stand out
}

\def \beq {\begin{equation}}
\def \eeq {\end{equation}}
\def \beqa {\begin{eqnarray}}
\def \eeqa {\end{eqnarray}}
\def \bseq {\begin{subequations}}
\def \eseq {\end{subequations}}

\newcommand \up {\uparrow}
\newcommand \down {\downarrow}

\newcommand \tr {{\rm tr}\,}

\newcommand{\sign}{{\rm sgn}\,}

\newcommand{\D}[1]{\text{d}#1}

\pgfplotsset{compat=1.18}
\begin{document}

\title{AppQSim: Application-oriented benchmarks for Hamiltonian simulation on a quantum computer}
\author{Etienne Granet}\author{Henrik Dreyer}
\affiliation{Quantinuum, Leopoldstrasse 180, 80804 Munich, Germany}
\date{\today}
\begin{abstract}
    We introduce AppQSim, a benchmarking suite for quantum computers focused on applications of Hamiltonian simulation. We consider five different settings for which we define a precise task and score: condensed matter and material simulation (dynamic and static properties), nuclear magnetic resonance simulation, chemistry ground state preparation, and classical optimization. These five different benchmark tasks display different resource requirements and scalability properties. We introduce a metric to evaluate the quality of the output of a tested quantum hardware, called distinguishability cost, defined as the minimal number of gates that a perfect quantum computer would have to run to certify that the output of the benchmarked hardware is incorrect.
    
\end{abstract}
\maketitle

\section{Introduction}
\label{sec:introduction}
Quantum computing hardware has recently witnessed rapid and impressive improvements \cite{decross2024computational,bluvstein2024logical,kim2023evidence,foss-feig2023demonstration,moses2023race}. It has become clear that the difficulty of classically simulating quantum computers greatly depends on the circuits to run. While certain specific circuits are already impossible to simulate classically on the best hardware \cite{foss-feig2023demonstration,kim2023evidence,bluvstein2024logical}, many circuits that accomplish a useful, application-centered task can still be simulated as of today. For this reason, even though these difficult-to-simulate circuits give a certain measure of the overall power of a given hardware, they cannot be used to accurately evaluate their ability to solve  concrete tasks. 
% Moreover, different technologies with different architecture, qubit connectivities, native gateset and clockspeed, can display different performances for different applications. 
Now that the technology is moving from an ``abstract" quantum advantage era to a ``practical" quantum advantage era, the need for application-oriented benchmarks becomes more pressing.

The purpose of this paper is to introduce an application-oriented benchmarking suite for quantum computers focused on Hamiltonian simulation, called AppQSim. It will be partially incorporated into a more general application-oriented benchmarking suite called BenchQC \cite{benchqc}. We study different settings considered to be some of the promising applications of quantum computing, namely material simulation, quantum chemistry, Nuclear Magnetic Resonance (NMR) simulation, and classical optimization. For example, the benchmarks we define cover applications such as the simulation of neutron-scattering experiments, the computation of spectrum generated by NMR experiments, or finding the maximal cut on a graph. 

Focusing a benchmark metric on applications is somehow at odds with benchmark scalability, since benchmarking supposes to know the expected result, whereas relevant applications of quantum computing are those beyond reach of classical computers. To deal with this we proposed benchmarking settings with varied scalability properties and closeness to applications. The characteristics of the five different benchmark settings we defined are summarized in Table \ref{tab:summary}. In the following we present briefly each of these protocols.

\begin{table*}[]
    \centering
    \begin{tabular}{|c||c|c|c|c|c|}
    \hline
         & \begin{tabular}{@{}c@{}}Material simulation \\ (dynamic)\end{tabular}  & \begin{tabular}{@{}c@{}}Material simulation \\ (static)\end{tabular}  & \begin{tabular}{@{}c@{}}Nuclear Magnetic\\Resonance\end{tabular}  &  \begin{tabular}{@{}c@{}}Quantum\\chemistry\end{tabular} & \begin{tabular}{@{}c@{}}Classical\\optimization\end{tabular} \\
    \hline
    \hline
      Benchmark scalability    &   Polynomial & \begin{tabular}{@{}c@{}}Exponential,\\$N\lessapprox 30$\end{tabular} & \begin{tabular}{@{}c@{}}Exponential,\\$N\lessapprox 20$\end{tabular}  & Constant & \begin{tabular}{@{}c@{}}Exponential,\\$N\lessapprox 1000$\end{tabular} \\
    \hline
   \begin{tabular}{@{}c@{}}Minimal\\hardware requirement\end{tabular}       & \begin{tabular}{@{}c@{}}$6$ qubits,\\any error rate\end{tabular}   & \begin{tabular}{@{}c@{}}$12$ qubits,\\error rate $<10^{-3}$\end{tabular} & \begin{tabular}{@{}c@{}}$7$ qubits,\\error rate $<10^{-3}$\end{tabular} & \begin{tabular}{@{}c@{}}$4$ qubits,\\any error rate\end{tabular} & \begin{tabular}{@{}c@{}}$4$ qubits,\\any error rate\end{tabular} \\
    % \hline Minimal resources     & $12$ TQG, $10^3$ shots &  &  &  &  \\
    \hline
      Ideal connectivity  & 2D & 2D & All-to-all & All-to-all & All-to-all\\
       \hline
      Circuit geometry  & Square &Rectangle & \begin{tabular}{@{}c@{}}Small width,\\large depth\end{tabular}  & \begin{tabular}{@{}c@{}}Rectangle,\\controlled\end{tabular} & Square \\
      \hline
      Random circuit  & No &No & Yes & Yes & No\\
    \hline
      \begin{tabular}{@{}c@{}}Mid-circuit\\measurements\end{tabular}  & No &No & No & No & No\\
    \hline
    Resource requirements   & \begin{tabular}{@{}c@{}}Qubits: scalable\\Gates: scalable\\ Shots: controllable\\ (intermediate/high)\end{tabular}  & \begin{tabular}{@{}c@{}}Qubits: scalable\\Gates: high\\ Shots: low\end{tabular} &  \begin{tabular}{@{}c@{}}Qubits: low\\Gates: high\\ Shots: high\end{tabular} & \begin{tabular}{@{}c@{}}Qubits: scalable\\Gates: scalable\\ Shots: controllable\\ (intermediate/high)\end{tabular}  & \begin{tabular}{@{}c@{}}Qubits: scalable\\Gates: scalable\\ Shots: controllable\\ (low/high)\end{tabular}  \\
    % \hline
    %  \begin{tabular}{@{}c@{}}Emphasis on gate\\fidelity v.s. clockspeed\end{tabular}   & Gate fidelity &Gate fidelity & Gate fidelity & Balanced & Balanced\\
    \hline
      \begin{tabular}{@{}c@{}}Comparison possible\\beyond exact result\end{tabular}  & No &Yes & No & Yes & Yes \\
    \hline

    % \hline
    %     Application &  Benchmark scalability  &Circuit geometry &  Ideal connectivity  & Comparative \\
    % \hline
    % \hline
    %  Material simulation     & Polynomial  & Square & 2D & No\\
    % \hline
    %  Nuclear Magnetic Resonance   & Exponential, $N\lessapprox 20$  & Small width, large depth & All-to-all & No \\
    % \hline
    %  Quantum chemistry   & No limit & Square, controlled, randomized & All-to-all & Yes\\
    % \hline
    %  Classical optimization   & Exponential, $N\lessapprox 1000$ & Rectangular & All-to-all & Yes\\
    % \hline
     
    \end{tabular}
    \caption{Summary of the characteristics of the benchmarks in the AppQSim suite. \emph{Benchmark scalability} means the classical resources required to assign a score to the hardware output, as a function of system size $N$. \emph{Minimal hardware requirement} is the minimal number of qubits and two-qubit gate error rate, assuming all-to-all connectivity, to run the benchmark with non-trivial output score. \emph{Ideal connectivity} indicates the hardware connectivity that is most suited to the benchmark. \emph{Circuit geometry} indicates the aspect ratio of the circuits involved. \emph{Random circuit} indicates whether several different random circuits have to be generated to run the benchmark. \emph{Mid-circuit measurements} indicates the presence of mid-circuit measurements in the benchmark. \emph{Resource requirements} indicates the resources to run the benchmark, in terms of number of qubits, number of gates and number of shots. Scalable means the number can be varied from low to high in the benchmark. Controllable means the number is left to be set by the user, depending on the characteristics of the machine, with the range of freedom indicated in parenthesis. \emph{Comparison possible beyond exact result} indicates whether the benchmark can be used to compare different hardware, even in the regime where no exact solution can be computed classically.}
    \label{tab:summary}
\end{table*}

% We start in Section \ref{sec:background} with a brief overview of existing benchmarks and set up our motivations. Then in each of next Sections \ref{freefermionsection},\ref{sec:static},\ref{sec:nmr},\ref{sec:chemistry},\ref{sec:opt} we detail the protocol of one of the benchmarks of the AppQSim suite. 

Section \ref{freefermionsection} describes the ``flagship" benchmark of AppQSim, which is the computation of dynamic properties in conducting materials. We define a simulation setup protocol similar (but not identical) to the simulation of the Hubbard model, whose exact results can be classically computed in polynomial time. This guarantees the benchmark to be scalable. The quantities computed are those required to simulate neutron-scattering experiments, yielding an almost end-to-end application-oriented benchmark. We introduce a score called ``distinguishability cost" to measure the quality of the benchmarked hardware, that is the minimal number of gates to run on a perfect quantum computer to be able to affirm that the output of the benchmarked hardware is incorrect. Stated differently, this measures the number of computations that the benchmarked hardware can do while staying indistinguishable from a perfect hardware.  This score is a physical and meaningful number that does not require context to be interpreted, and directly informs the end user of how noisy a given hardware is for a given application.

In Section \ref{sec:static} we introduce another material-simulation benchmark focused on equilibrium state preparation. We use Hamiltonian simulation to prepare adiabatically a low-energy equilibrium state of the Heisenberg model on a Kagome lattice. This kind of adiabatic preparation of low-temperature state is known to display lower sensitivity to hardware noise \cite{schiffer2024quantum,granet2024dilution,chertkov2024robustness}, probing different capacities of the hardware. Despite the exact result being exponentially costly to compute classically, the difficulty of the preparation of the ground state in this highly quantum model ensures that the benchmark will remain relevant for years to come. Moreover, even beyond the classically simulable regime, the output of two different hardware can still be compared. 

In Section \ref{sec:nmr}, we present a benchmark of NMR experiment simulation. This is a fully end-to-end application oriented benchmark, with the score being the average precision that one can obtain on the couplings between the nuclear spins of a benzene molecule when comparing to an NMR experiment. The system sizes cannot be scaled arbitrarily, but the large circuit depth required guarantees again the benchmark to remain relevant for several years. 

In Section \ref{sec:chemistry} we then move on to the ground state preparation of molecular systems. To bypass the prohibitive cost of energy measurement in these systems to a precision that cannot be obtained with classical computers, we adopt a mirror-circuit-like approach to define a score. This allows the end user to run the benchmark for arbitrary system sizes. This comes at the cost of a more abstract score not directly related to a quantity to measure in a concrete application. The benchmark also tests the ability of the hardware to generate and run random circuits.

Finally in Section \ref{sec:opt} we present a benchmark for Hamiltonian simulation applied to classical optimization. The benchmark is not a variational algorithm (as is often implemented), but instead a deterministic heuristic protocol to solve Max-Cut that has been observed to work to at least around one hundred qubits. The score directly measures the ability of the quantum computer to find the exact optimal value, and is thus directly application-oriented. Specific classical optimization algorithms can solve the problem up to the order of one thousand qubits, which guarantees the relevance of the benchmark for a long time, at least up to the time where practical quantum advantage would be observed for that application. Even beyond the classical simulability, the output of different hardware can still be compared.

Before detailing the precise protocols in these benchmarks, we present in the following Section \ref{sec:background} a brief overview of existing benchmarks.

\section{Previous works and goals}
\label{sec:background}
% \subsection{Previous works}
% There currently exists a large variety of quantum computing hardware with very different characteristics in terms of number of qubits, native gate set, quality of gates, qubit connectivity, shot runtime. When these quantum computers are impossible or very difficult to simulate classically, developing appropriate metrics to benchmark their performance becomes crucial, both for guiding future hardware developments and for informing customers.

There exist three main approaches to evaluate the quality of quantum hardware. The first approach is a \emph{low-level benchmark}, where one directly measures the quality of basic hardware components or operations such as gate fidelity or state preparation and measurement (SPAM) errors. Well established approaches are randomized benchmarking \cite{emerson2005scalable,knill2008randomized}, gate set tomography \cite{blume2017demonstration} or cycle benchmarking \cite{erhard2019characterizing}. While these metrics provide a detailed quality assessment of the basic components of the hardware, the overall performance of an algorithm results from a complex interaction of all these error sources. These interactions can further depend on the structure of the circuit implemented and on higher-level hardware characteristics such as connectivity or
speed. There can be very significant differences in performance for different tasks with same hardware resources.

A second approach to hardware quality assessment is \emph{circuit benchmarks}, where an entire circuit is run on the hardware, instead of individual operations on isolated qubits. Well-known examples are quantum volume \cite{cross2019validating}, generation of random bit strings \cite{arute2019quantum,neill2018blueprint}, and protocols based on ``mirror circuits" \cite{proctor2022measuring}. These circuit benchmarks capture different characteristics of the hardware in a holistic way and gives a better idea of its overall capacities. However, they do not capture how much of a certain noise feature a given application can tolerate. 
% It is indeed known that even noisy states with little overlap with the exact state can carry useful information to solve a given task.

The third approach to hardware benchmarking is \emph{application-oriented benchmarks}. These benchmarks directly evaluate the ability of the hardware to solve a given real application. There already exist several application-oriented scores and benchmark suites. Benchmarks focused on simulation of physical systems include for example preparing the ground state of the 1D Fermi-Hubbard model using Variational Quantum Eigensolver (VQE) \cite{dallaire2020application,gard2022classically} or the ground state of small molecules using VQE \cite{mccaskey2019quantum}. Certain benchmarks propose implementation of Hamiltonian simulation for specific systems \cite{tomesh2022supermarq,lubinski2024quantum}. Benchmarks on classical optimization applications include solving a Max-Cut problem with Quantum Approximate Optimization Algorithm (QAOA) \cite{martiel2021benchmarking,tomesh2022supermarq}, Max-Clique problems \cite{van2024extending}, some industry-relevant problems like the robot path and vehicle optimization problems \cite{finvzgar2022quark}, as well as other benchmarking suites containing multiple problem instances \cite{mesman2021qpack,barbaresco2024bacq}, or machine-learning problems \cite{lubinski2024quantum}. Finally, some benchmarks include linear algebra routines such as Quantum Fourier Tranform (QFT), quantum matrix inversion \cite{dong2021random,cornelissen2021scalable} or linear equation solving \cite{lubinski2024quantum}. There exist works specifically proposing benchmarking libraries for Hamiltonian simulation, but without specifying a particular task \cite{sawaya2023hamlib}. 

Most of these application-oriented benchmarks rely on VQE-like algorithms. These typically involve shallow circuits with limited number of gates, but require several different circuits and sometimes a large number of measurements to optimize the VQE parameters. The actual scalability and usefulness of these variational approaches have been put in question, with serious obstacles such as the hostile optimization landscape or the effect of noise \cite{wang2021noise}. It could render these benchmarks obsolete if they become impossible to implement on near-term devices. 

In contrast, algorithms based on Hamiltonian simulation appear to be under-represented in these benchmarks. Hamiltonian simulation consists in applying a time evolution operator $e^{itH}$ on the qubit register, where $t$ is some simulation time and $H$ a Hamiltonian. It is proven to be implementable in polynomial time on a quantum computer with very simple routines like a Trotter decomposition. It appears in many algorithms, such as Quantum Phase Estimation (QPE) and adiabatic state preparation, with applications ranging from material and molecular simulation to classical optimization. Despite the high likelihood that Hamiltonian simulation will play a prominent role in the NISQ era and beyond, there seems to be no application-oriented benchmark specifically devoted to it. The purpose of the AppQSim benchmarking suite that we introduce in this paper is to fill this gap.

\section{Application: simulation of conducting materials \label{freefermionsection}}
\subsection{Context and motivation}
Electrons in material can be modeled by spinful fermions hopping from one atomic orbital to another. One of the most famous models for electrons in solids is the so-called Hubbard model. This model (or variants thereof) is  believed to be able to describe high-temperature superconductivity of the cuprates whose pairing mechanism still has not been fully understood. For this reason its solution has been the study of countless academic and industry work, and its utility has been estimated in the billions of dollars \cite{agrawal2024quantifying}. Since Hamiltonian simulation is one of the simplest tasks that a quantum computer is  likely to be able to perform exponentially faster than a classical computer, it puts the simulation of the Hubbard model at the forefront of near-term applications of quantum hardware, in the NISQ era and beyond.

Mathematically, the Hamiltonian of the Hubbard model is
\begin{equation}
    H_{\rm H}=-t\sum_{\langle i,j\rangle,\sigma} (c_{i,\sigma}^\dagger c_{j,\sigma}+c_{j,\sigma}^\dagger c_{i,\sigma})+V\sum_{i}\left(n_{i,\up}n_{i,\down}-\frac{1}{4}\right)\,,
\end{equation}
where $c_{i,\sigma}$ denotes the fermion annihilation operator at site $i=1,...,N$ and spin $\sigma\in\{\up,\down\}$, which satisfy canonical anticommutation relations, where $n_{i,\sigma}=c_{i,\sigma}^\dagger c_{i,\sigma}$ is the mode occupation number, $t,V$ some parameters, and $\langle i,j\rangle$ means that the two sites $i,j$ are neighbours on the lattice considered.

There is no known classical algorithm to simulate the time evolution of a system described by the Hubbard model, except for small system sizes (with statevector simulations) or for short times (with tensor networks or neural networks techniques). From a benchmark perspective, this is of course problematic as the result of the quantum hardware cannot be compared to the exact result beyond these cases. Even in the NISQ era, quantum computers are able to reach settings that can become challenging for classical computers \cite{kim2023evidence,nigmatullin2024experimental}, and a benchmark specifically focused on classically simulatable regimes would be too restrictive. There is however a simple way of modifying the Hamiltonian, without much modifying the circuits run on the hardware, to make the simulation classically easier. In absence of interaction between the spins, i.e. when $V=0$, the system describes free fermions, which is exactly solvable, and the computation time to classically simulate the system scales polynomially with system size and simulation time.

\subsection{The benchmark \label{sec:benchmarktrotter}}

\begin{figure}
    \centering
    \includegraphics[width=0.7\linewidth]{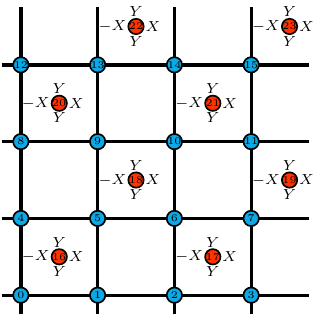}
    \caption{Square lattice after the compact fermion encoding. Blue circles indicate sites of the original lattice and red circles indicate ancillas.}
    \label{fig:freefermion}
\end{figure}
The benchmark that we propose is the implementation of the compact encoding of Ref \cite{derby2021compact} for this free fermion system.

We consider a square lattice with widths $L_x,L_y$, containing thus $L=L_xL_y$ sites, and impose periodic boundary conditions. We will restrict to only $L_x$ and $L_y$ even. We add to this lattice $L/2$ ancillas positioned in every other face of the square lattice, in a checker-board pattern as illustrated in Fig \ref{fig:freefermion}. The total system possesses thus $N=3L/2$ sites. On this system, we define the following Hamiltonian
\begin{equation}
    H=\frac{1}{2}\sum_{\langle i,j\rangle}(X_iX_j+Y_iY_j)P_a\,,
\end{equation}
where the sum runs over all the edges $\langle i,j\rangle$ of the square lattice that links neighbouring sites $i,j$, and where $a$ refers to the ancilla that is contained in the face adjacent to edge $\langle i,j\rangle$. $P_a$ is the Pauli matrix acting on the ancilla $a$, equal to $P_a=Y_a$ if $\langle i,j\rangle$ is a horizontal edge, and equal to $P_a=X_a$ (resp. $-X_a$) if $\langle i,j\rangle$ is a vertical edge on the right (resp. left) of the ancilla. These three different possibilities are sketched in Fig \ref{fig:freefermion}. 

The initial state $|\psi\rangle$ that we consider is defined as follows. We initialize the $L$ lattice sites in a product state in the $Z$ basis, with a predefined value $n_j\in\{0,1\}$ for each site $j$. We fix this function to be
\begin{equation}\label{initialstate}
    n_j=\begin{cases}
        1\qquad \text{if }j_y<L_y/2\\
        0\qquad \text{if }j_y\geq L_y/2\,,
    \end{cases}
\end{equation}
with $j_y=0,...,L_y-1$ denoting the vertical component of the site. The ancillas are initialized in a ground state of the toric code, as required by this fermionic encoding \cite{derby2021compact}. This is done as follows. For an ancilla $a$, we denote respectively $a^1,a^2,a^3$ the two ancillas on top left and top right of ancilla $a$, and the ancilla two lines on top of $a$ in the same column, applying the boundary conditions in both vertical and horizontal directions. For example in Fig \ref{fig:freefermion}, if $a=18$, then $a^1,a^2,a^3=20,21,22$. Then we define the unitary operator
\begin{equation}
    V_a=CX_{a,a^3}CX_{a,a^2}CX_{a,a^1}H_a\,,
\end{equation}
with $H_a$ denoting the Hadamard gate on ancilla $a$, and $CX_{a,b}$ the CNOT gate with control $a$ and target $b$, with the first operator applied being the rightmost. Similarly, we define
\begin{equation}
    \tilde{V}_a=CX_{a,a^2}CX_{a,a^4}CX_{a,a^5}H_a\,,
\end{equation}
with $a^4$ being the ancilla on bottom right of ancilla $a$, and $a^5$ the ancilla two columns to the right of ancilla $a$ on the same row, applying periodic boundary conditions in both directions. For example in Fig \ref{fig:freefermion}, if $a=16$, this is $a^4=22$ and $a^5=17$. We then apply the operators $V_a,\tilde{V}_a$ on some ancillas in a specific order, as prescribed and illustrated in Appendix \ref{app:toriccode}, in order to prepare the ground state of the toric code on the ancillas. In the particular case of system size $4\times 4$, applying the ordering of Appendix \ref{app:toriccode}, we would apply $V_a$ on ancillas $a=18,19$ and then $\tilde{V}_a$ on $a=16$. This operation defines the state
\begin{equation}
    |\tilde{\psi}\rangle=\prod_{a}\tilde{V}_a\prod_{a}V_a  \prod_{j=0,...,L-1}X_j^{n_j}|0\rangle\,,
\end{equation}
where the product of $V_a,\tilde{V}_a$ is as specified in Appendix \ref{app:toriccode}. Finally, we define the initial state of the quantum computer as $|\psi\rangle$ where
\begin{equation}
    |\psi\rangle=\prod_{a=1}^{L/2}Q_a |\tilde{\psi}\rangle\,,
\end{equation}
where $Q_a$ acts only on ancilla $a+L$, with $Q=HS^\dagger$ if the ancilla is on an odd row and $Q=SHS$ if the ancilla is on an even row, with $H$ denoting here the Hadamard gate and $S$ the usual $S$-gate.

We note that this state preparation protocol also holds when one of the lengths $L_x$ or $L_y$ is equal to $2$, applying strictly the periodic boundary conditions in the operator $V_a$. For example, in size $L_x=2,L_y=4$, the operator $V_9$ applies a CNOT from site $9$ to site $10$, $10$ again, and then $11$, which means a single CNOT from site $9$ to site $11$.

To entirely describe our protocol, we now define the precise Trotterization to use in the benchmark. For a given Trotter step size $\delta t$, each Trotter step operator $U$ is decomposed as
\begin{equation}
    U=U_{|,2}U_{|,1}U_{-,2}U_{-,1}\,.
\end{equation}
Here we defined
\begin{equation}
\begin{aligned}
    U_{-,1}=&\exp\left(\frac{i\delta t}{2}\sum_{\substack{\langle i,j\rangle\\\text{even row}}}Y_iY_jP_a\right)\\
    &\qquad\qquad\qquad\times\exp\left(\frac{i\delta t}{2}\sum_{\substack{\langle i,j\rangle\\\text{odd row}}}X_iX_jP_a\right)\,,
\end{aligned}
\end{equation}
as well as $U_{|,1}$ identically but with columns instead of rows, and $U_{-,2}$ identically but swapping $X_iX_j$ and $Y_iY_j$. For definiteness, we will fix the Trotter step to $\delta t=0.2$.

Finally, we measure the lattice sites in the $Z$ basis. We form the operator
\begin{equation}\label{observableO}
    \mathcal{O}=\sum_{j=1}^L f_jZ_j\,,
\end{equation}
for a given function $f_j$. We fix the following function
\begin{equation}\label{functionf}
    f_j=\begin{cases}
        -1\qquad \text{if }j_y<L_y/2\\
        1\qquad \text{if }j_y\geq L_y/2
    \end{cases}\,.
\end{equation}
With this definition, $\mathcal{O}$ measures the imbalance of fermions between the lower and upper part of the lattice.\\

\begin{figure*}
    \centering
    \includegraphics[scale=0.85]{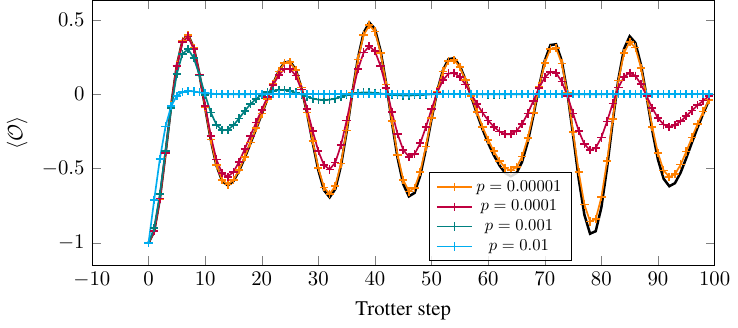}
    \includegraphics[scale=0.85]{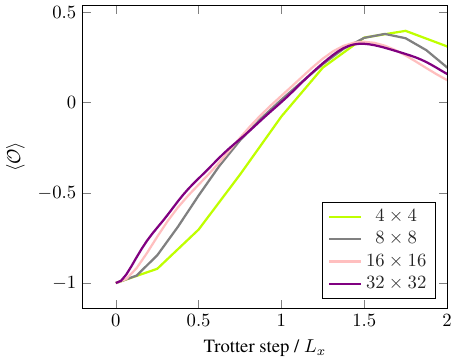}
    
    \caption{\emph{Left:} Curve $\langle \mathcal{O}\rangle$ as a function of Trotter step number, on a system of size $L=4\times 4$, for different depolarizing noise levels $p$ per two-qubit gate. The curves are averaged over $20$ different analog trajectories as described in \cite{granet2024analog}. The black continuous curve is the exact value. \emph{Right:} Exact noiseless curve $\langle \mathcal{O}\rangle$ as a function of number of Trotter steps divided by $L$, for different system sizes $L$.}
    \label{fig:free_fermion_1}
\end{figure*}

The exact outcome of the quantum circuit obtained after $n$ applications of the Trotter operator $U$ can be computed, see Appendix \ref{app:free}. The result is expressed as
\begin{equation}\label{exactformulaO}
\begin{aligned}
    &\langle \mathcal{O}(n)\rangle_{\rm exact}=\sum_{\varphi_x=0,1/2}\sum_{\varphi_y=0,1/2}\Big(\\
    &\sum_{k,q\in K_{\varphi_x,\varphi_y}}\hat{f}(k-q)(\alpha_n^*(k) \alpha_n(q)-\beta^*_n(-k) \beta_n(-q)) \hat{n}(k-q)\\
    &+\hat{f}(0)\sum_{k\in K_{\varphi_x,\varphi_y}}|\beta_n(k)|^2\Big)\,,
\end{aligned}
\end{equation}
where $K_{\varphi_x,\varphi_y}$ denotes the set of pairs
\begin{equation}
    \begin{aligned}
K_{\varphi_x,\varphi_y}=&\Big\{\left(\frac{2\pi (k_x+\varphi_x)}{L_x},\frac{2\pi (k_y+\varphi_y)}{L_y}\right)\,,\, \\
&\qquad\qquad k_{x,y}=0,...,L_{x,y}-1\Big\}\,.
    \end{aligned}
\end{equation}
The coefficients $\alpha_n(k),\beta_n(k)$ are given by
\begin{widetext}
    \begin{equation}
    \begin{aligned}
        &\alpha_n(k)=e^{-in\epsilon_k}+i\Big(-\sin(2\delta t)(\cos k_x+\cos k_y)+2\sin(2\delta t)\sin^2(\delta t)\cos k_x \cos k_y(\cos k_x+\cos k_y)+\sin\epsilon_k\Big)\frac{\sin(n\epsilon_k)}{\sin\epsilon_k}\\
        &\beta_n(k)=\Big(i\sin^2(\delta t)(\sin(2k_x)+\sin(2k_y))-2i\sin^4(\delta t)(\cos^2(k_x)\sin(2k_y)+\cos^2(k_y)\sin(2k_x))\\
        &\qquad\qquad\qquad\qquad+\sin^2(\delta t)\sin(2\delta t)(\cos(k_x)\sin(2k_y)+\cos(k_y)\sin(2k_x))\Big)\frac{\sin(n\epsilon_k)}{\sin\epsilon_k}\,,
    \end{aligned}
\end{equation}
\end{widetext}
with
\begin{equation}
\begin{aligned}
    \epsilon_k=&\sign(\cos k_x+\cos k_y)\\
    &\arccos\Big[1-2\sin^2(\delta t)(\cos k_x+\cos k_y)^2\\
    &+4\sin^4(\delta t)\cos k_x\cos k_y(1+\cos(k_x+k_y)) \Big]\,.
\end{aligned}
\end{equation}
This function $\langle \mathcal{O}(n)\rangle$ follows a non-trivial trajectory, while still being computable in a time that is polynomial in the system size. For example, we depict in the right panel of Fig \ref{fig:free_fermion_1} this observable as a function of the number of Trotter steps, for systems of different sizes up to $32\times 32=1024$ way beyond the regime that is accessible to general purpose classical methods.

\subsection{The score \label{sec:scoretrotter}}
We now would like to assign a score to a given output of a hardware to benchmark. Unlike classical computers, quantum computers can only output ``shots" over which one has to average in order to obtain an expectation value of an operator $\langle \mathcal{O}(n)\rangle$. The precision achieved on the quantum computer is thus directly related to the time spent on the computation. If because of hardware imperfections the quantum computer has a bias in the expectation value $\langle \mathcal{O}(n)\rangle$, this bias will not be \emph{detectable} if after averaging over a finite number of  shots the error bars  are larger than the bias. Hence, the noisier a hardware, the faster the imperfections can be detected as it will require averaging over fewer shots. Conversely, a given hardware with low noise will be statistically undistinguishable from noiseless, before a certain amount of resources is spent to reach the precision where the bias due to imperfections becomes visible. This suggests a physical and intuitive way of measuring the accuracy of a given quantum hardware, by answering the following question: How many gates does a perfect quantum computer have to implement (or similarly, how much time does it need), running the same circuit as the benchmarked hardware, to certify that the output of the benchmarked hardware is incorrect? We will call this quantity \emph{distinguishability cost}.

In our case, we fix the following computational task: computing the expectation values $\langle \mathcal{O}(n)\rangle$ after $n=1,...,T$ Trotter steps, with final time fixed to $T=2L_x$, in a square lattice $L_x=L_y$. Let us denote $m_n$ the estimates obtained for these expectation values on a benchmarked hardware, and consider that we run the same circuit on a perfect hardware, obtaining estimates $s_n$ with standard deviations $\sigma_n$. For the moment, we will not take into account the error bars on the estimates $m_n$ obtained from the benchmarked hardware. The output of the benchmarked hardware can be certified to be incorrect if the outcomes $m_1,...,m_T$ are statistically incompatible with the unbiased estimates $s_n$ with standard deviations $\sigma_n$. This statistical compatibility can be inferred from a chi-2 test with $T$ degrees of freedom. We will say that the output of the benchmarked hardware is certified to be incorrect if it fails the chi-2 test by $3$ sigmas, namely if
\begin{equation}
    \chi_2^{(T)}\left(\sum_{n=1}^T \frac{(s_n-m_n)^2}{\sigma_n^2}\right)>0.997\,,
\end{equation}
where $\chi_2^{(T)}$ denotes a chi-2 cumulative distribution function with $T$ degrees of freedom.

Given outputs $m_n$, it is a non-trivial problem to find the best strategy to follow on the (gedanken) perfect hardware to certify that these outputs are incorrect as quickly as possible. One would ideally run on the perfect hardware only the noisiest time point, but that time point cannot be known in advance without running other time points on the perfect hardware. 
% It is a so-called ``multi-armed bandit" problem: the most advantageous time points to run are those for which the estimate $m_n$ is far from the exact value, but this exact value is not known in advance and can only be estimated by running shots on the perfect hardware, leading to a non-trivial balance between gaining information and making use of the information gained. 
While we could implement numerically an efficient strategy for this, we prefer instead to compute the \emph{minimal} resources required to certify incorrectness of the outputs, even in the case where the user would know which points are the noisiest. This definition has the advantage of being simpler, more canonical, and not sensitive to precise details of the implementation of the strategy followed. 

Let us now determine this minimal cost. The cost of running a shot for time point $n$ is proportional to $n$, because the number of gates is proportional to $n$ (neglecting for simplicity the gates appearing in the state preparation, before applying the first Trotter step). For each shot, the expectation value of $\mathcal{O}$ is computed as an average over the $L$ different points. The variance associated to this averaging depends on the correlations between the different points. Again for simplicity and ease of the calculation of the score, we will neglect these correlations and assume that the variance $\sigma_n$ on the perfect hardware is related to the number of shots $S_n$ at time point $n$ as 
\begin{equation}
    \sigma_n^2=\frac{\sum_{j=1}^L 1-f_j^2t_{n,j}^2}{L^2S_n}\,,
\end{equation}
with $t_{n,j}$ denoting the exact expectation value of the observable $Z_j$ after $n$ Trotter steps. Hence, denoting $t_n$ the exact expectation value of $\mathcal{O}$ after $n$ Trotter steps, the cheapest way of certifying incorrectness of the outputs is to only run the time point $n_*$ that maximizes $(t_n-m_n)^2/(n\sigma_n^2)$. In that case the number of shots to run is $S_{n_*}$ that satisfies
\begin{equation}\label{equationchi2}
    \chi_2^{(T)}\left(\frac{(t_{n_*}-m_{n_*})^2}{\sum_{j=1}^L1-f_j^2t_{n,j}^2}L^2S_{n_*}\right)=0.997\,.
\end{equation}
The total number of gates is then equal to $12LS_{n_*}n_*$, namely $12Ln_*$ two-qubit gates per circuit, repeated $S_{n_*}$ times (we again did not take into account the gates used in the state preparation). Any strategy has to run at least that many gates, and even more so if one does not have access to the exact value $t_n$ beforehand. 
% Even a trained user who would be able to ``guess" beforehand which point is noisier than the others would have to run at least that many gates. 
This will thus be the definition of our score
\begin{equation}\label{scorestandard}
    \mathcal{S}(\{m_n\})=12LS_{n_*}n_*\,,
\end{equation}
where $S_{n_*}$ is the unique solution to \eqref{equationchi2}, and where $n_*$ maximizes $(t_n-m_n)^2/n$. The interpretation of $ \mathcal{S}$ is the smallest number of two-qubit gates that a perfect quantum computer would have to implement, running the same circuit as the benchmarked hardware, to certify that the output of the benchmarked hardware is incorrect. This number of two-qubit gates does not refer to the number of gates per circuit, but to the total number of gates run across different circuits and shots. We note that by ``two-qubit gate" we mean \emph{logical} two-qubit gate, namely the operation that acts on the qubits that host the quantum information (whether encoded with quantum error correction or not -- in this latter case it is the physical two-qubit gate). The gate count also should \emph{not} take into account auxiliary gates such as SWAP gates in case the hardware does not support the implementation of a two-qubit gate between arbitrary qubits. The definition of the score \eqref{scorestandard} is thus imposed to be the same for any platform, architecture or compilation scheme.

\begin{figure}
    \centering
    \includegraphics[scale=0.9]{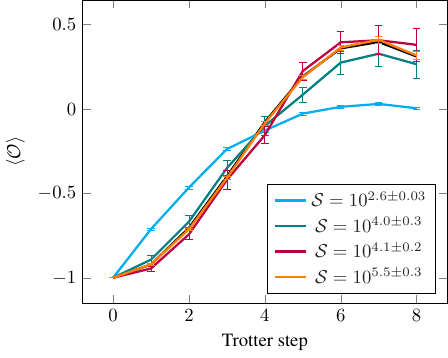}
    \includegraphics[scale=0.9]{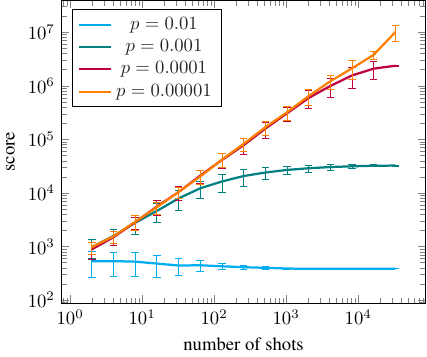}
    
    \caption{\emph{Top:} Value of the score (in the legend) obtained in size $L=4\times 4$ for different noise levels $p$ and for different number of shots $N_S$ (cyan: $p=0.01$ and $N_S=10^3$, teal: $p=0.001$ and $N_S=50$, purple: $p=0.0001$ and $N_S=50$, orange: $p=0.00001$ and $N_S=10^3$).  \emph{Bottom:} Score obtained in size $L=4\times 4$ as a function of number of shots per time point, for different noise levels. Here, the error bars indicate an estimated standard deviation of the score over different experiments (which is different from $\delta x$ in the top panel defined in \eqref{deltax}).}
    \label{fig:free_fermion_2}
\end{figure}

Let us now take into account the effect of error bars on estimates $m_n$ obtained on the benchmarked hardware. If the benchmarked hardware outputs a mean value $m_n$ with standard deviation $\tau_n$, we can approximate the output of a new run of the hardware with same number of shots as a random Gaussian variable $\xi_n$ with mean $m_n$ and standard deviation $\tau_n$. We thus define the score of the output of the hardware as
\begin{equation}
    \mathcal{S}(\{m_n;\tau_n\})=\mathbb{E}[\mathcal{S}(\{\xi_n\})]\,,
\end{equation}
where $\mathbb{E}$ denotes the statistical average with respect to the Gaussian variables $\xi_n$. To better accommodate large numbers, we present the score in an exponential form $10^x$ with $x=\log_{10}\mathcal{S}(\{m_n;\tau_n\})$. One can assign a standard deviation to the score obtained (coming from the finite number of shots performed on the benchmarked hardware) by computing the standard deviation of $\log_{10} \mathcal{S}(\{\xi_n\})$ with respect to the Gaussian random variables $\xi_n$. Namely, the standard deviation $\delta x$ assigned to the score when written $10^{x\pm \delta x}$ is defined as
\begin{equation}\label{deltax}
    \delta x=\sqrt{\mathbb{E}[\log_{10}(\mathcal{S}(\{\xi_n\}))^2]-\mathbb{E}[\log_{10}(\mathcal{S}(\{\xi_n\}))]^2}\,.
\end{equation}
In the top panel of Fig \ref{fig:free_fermion_2}, we present different curves obtained in size $L=4\times 4$ using different depolarizing noise levels and number of shots, and compute their score. In the bottom panel of Fig \ref{fig:free_fermion_2}, we show the score obtained as a function of the number of shots per time point, for different noise levels. The general behaviour of the curves is to first be proportional to the number of shots (which is expected when the number of shots is the limiting factor of the precision), and then saturate at some finite value (when the limiting factor is hardware noise). We also observe that the score is almost always an increasing function of the number of shots.

\subsection{Extensions}
\subsubsection{Neutron scattering experiments}
Neutron scattering experiments are widely used in condensed matter physics to probe the internal structure of a material. They consist in irradiating a sample material with a beam of neutrons that is then scattered by the nuclei of the material, changing their energy and momentum. The amplitude of the neutrons with momentum and energy deviation $q,\omega$ is called dynamical structure factor (DSF) $S(q,\omega)$. Mathematically, it can be computed as the Fourier transform of the dynamical correlations
\begin{equation}\label{sqomega}
    S(q,\omega)=\int \D{t}\int \D{j} e^{i(q\cdot j-\omega t)}\langle \mathcal{O}_{j}(t)\mathcal{O}_0(0) \rangle\,,
\end{equation}
where $\langle \cdot\rangle$ denotes an expectation value in some state, for example a finite-temperature equilibrium state, and where $\mathcal{O}_j(t)$ denotes  an observable, like for example particle density, at position $j$ evolved for time $t$. In a 2D material, the momentum $q=(q_x,q_y)$ is a two-dimensional vector and we defined $q\cdot j=q_xj_x+q_yj_y$. The integral (or sum if the system is finite) over $j$ is performed over all the lattice sites, and the integral over time from $-\infty$ to $\infty$. The cost in computing $\langle \mathcal{O}_j(t)\mathcal{O}_0(0) \rangle$ is, besides the preparation of the state studied, the same as computing the dynamics of the system for a time $t$ and measuring the observable $\mathcal{O}$. This is exactly what the benchmark defined in this section is testing.

In order to be able to define a benchmark that is easy to evaluate classically, we consider the same state \eqref{initialstate} as above, namely a state where all the sites of the lower half of the system are occupied, and all the sites of the upper half are empty, and set the observable of interest $\mathcal{O}=Z$. This \emph{per se} departs from a realistic description of a neutron scattering experiment, since the state is not an equilibrium state. However, it simplifies the classical computations that are necessary to benchmark the quantum computer, while still involving running very similar circuits.  Because the initial state is an eigenstate of all the $Z$ operators, we have in that case the simplification $\langle \mathcal{O}_j(t)\mathcal{O}_0(0) \rangle=-\langle Z_j(t)\rangle$. Instead of using the value \eqref{functionf} in \eqref{observableO}, we set $f_j=1$ and $f_{j'}=0$ for $j'\neq j$. Formula \eqref{exactformulaO} then holds for the exact expectation value $\langle Z_j(n)\rangle$ after $n$ Trotter steps. 

% The expression in \eqref{sqomega} involves then an integral over time, that is in principle taken between $-\infty$ to $\infty$. In practice, one has of course to truncate the integral to a finite final integration time. This truncation introduces oscillations on $S(q,\omega)$. A standard way to smooth out these oscillations is to include a damping factor $e^{-t/T_2}$ for a certain chosen $T_2>0$ in the integration. Computing $T$ Trotter steps, we thus define
% \begin{equation}
%     S(q,\omega)=-\frac{{\rm d}t}{L} \sum_{n=0}^{T} \sum_{j} e^{i(q\cdot j-\omega n {\rm d}t)}e^{-n {\rm d}t/T_2} \langle Z_j(n)\rangle\,.
% \end{equation}
% To match the benchmark at the beginning of the section, we propose to set again $T=L$, as well as $T_2=T {\rm d}t/3$.

\subsubsection{Continuous Hamiltonian simulation limit}
In the benchmark setting defined above, the Trotter step was fixed to $\delta t=0.2$. In order to recover the exact Hamiltonian dynamics, this Trotter step needs to be scaled to $0$, and the number of Trotter steps scaled as $1/\delta t$. For finite $\delta t$, an exact noiseless implementation of the circuit will display some \emph{Trotter error} compared to the continuous-time Hamiltonian simulation result. In practice, a circuit run on a hardware will thus depart from exact both because of hardware noise and Trotter error. The benchmark defined in Section \ref{sec:benchmarktrotter} only measures the amount of hardware noise in the circuit. We can generalize the benchmark to take into account as well Trotter error, the following way.

In the limit $\delta t\to 0$, the observable $\mathcal{O}$ evaluated at time $t$, i.e. after $n=t/\delta t$ Trotter steps, simplifies and is given by
\begin{equation}
\begin{aligned}
    &\langle \mathcal{O}(t)\rangle_{\rm exact}=\\
    &\sum_{k,q\in K_{0,1/2}}\hat{f}(k-q)\cos(t\varepsilon_k) \cos(t\varepsilon_q) \hat{n}(k-q)\\
    &+\sum_{k,q\in K_{1/2,0}}\hat{f}(k-q)\cos(t\varepsilon_k) \cos(t\varepsilon_q) \hat{n}(k-q)\,,
\end{aligned}
\end{equation}
with $\varepsilon_k=2(\cos(k_x)+\cos(k_y))$. We then define the benchmark as computing the value of $\langle \mathcal{O}(t)\rangle$ on the hardware for time points $t=0.2,0.4,...,0.2N$. This corresponds to the same time points (but without Trotter error) as done in the benchmark of Section \ref{sec:benchmarktrotter}. We impose that the end user chooses a Trotter step $\delta$ of the form $\delta t=0.2/k$ with $k\geq 1$ an integer, and they keep the same Trotter step for all time points. The score defined in Section \ref{sec:scoretrotter} can then be modified as follows. We now denote $t_n$ the exact expectation value without Trotter error for time point $t=0.2n$, and $m_n$ the corresponding estimate on the benchmarked hardware. We look for the time point $n_*$ that maximizes $(t_n-m_n)^2/n$ and then set $S_{n_*}$ the number of shots such that \eqref{equationchi2} holds. The total number of gates run is then $12LS_{n_*}n_* 0.2/\delta t$, with $\delta t$ the Trotter step used on the benchmarked hardware. We emphasize however that this benchmarked hardware are compared to the exact values, without Trotter error. This is the score that we assign to this exact Hamiltonian evolution benchmark.

\subsubsection{Observables with higher weight}
It is known that, under certain circumstances often met in condensed matter models, observables that are expressed in terms of long Pauli strings are more noisy than with short Pauli strings \cite{granet2024dilution}. This phenomenon, called dilution of error, has a huge impact on resource estimations, because in certain cases physical meaning can be extracted from noisy states with a very tiny overlap with the exact state. The observable $\mathcal{O}$ we considered in our free fermion benchmark has weight $1$, because it is expressed only in terms of single $Z$ Pauli matrices. However, exact formulas can also be obtained for higher weight observables, such as
\begin{equation}\label{ow}
    \mathcal{O}_{[w]}=\sum_{i_1<...<i_w}f_{i_1}...f_{i_w}Z_{i_1}...Z_{i_w}\,,
\end{equation}
for any integer $w$, and with an arbitrary given ordering on the sites. We explain how to compute the exact expectation value of these observables in Appendix \ref{app:higherweight}. Although the computation runtime increases with the weight $w$, small weights $w=1,2,3$ can still be computed in reasonable time and compared to a benchmarked hardware. This free fermion benchmark allows for comparing the noise level on observables with different weights and investigate how much dilution of error holds in the benchmarked hardware. The score obtained for observable $\mathcal{O}_{[w]}$ can thus be taken as an indication of how well observables with weight $w$ are reproduced on the hardware, in this specific benchmark model.

\section{Application: static observables at low temperature \label{sec:static}}
\subsection{Context and motivation}
Materials often display exotic properties as their temperature is lowered, with new phases requiring quantum physics in order to be described accurately, such as superconducting phases or Fermi liquids. The computation of static, equilibrium expectation values at low temperature in these many-body physics Hamiltonians can become difficult or unreliable to perform with classical computers for intermediate-size to large systems.

On a quantum computer, the adiabatic algorithm  is a generic way of preparing the ground state of a Hamiltonian. It can be formulated as follows. Given an initial Hamiltonian $H_I$ whose ground state can be prepared efficiently on a quantum computer, and a final Hamiltonian $H_F$ whose ground state is the target state, we define the time-dependent Hamiltonian
\begin{equation}\label{adiabatichamiltonian}
    H(s)=\varphi(1-s)H_I+\varphi(s)H_F\,,
\end{equation}
with $\varphi(s)$ a scheduling function that is continuous and satisfies $\varphi(0)=0$, $\varphi(1)=1$. For a given parameter $T>0$ called adiabatic time, we define then the state $|\psi_T(t)\rangle$ by the fact that $|\psi_T(t=0)\rangle$ is the ground state of $H_I$, and is evolved under the time-dependent Schr\"odinger equation
\begin{equation}
    i\partial_t |\psi_T(t)\rangle=H(t/T)|\psi_T(t)\rangle\,,
\end{equation}
for times $0\leq t\leq T$. The adiabatic theorem of quantum mechanics says that if $H(s)$ is gapped for all $0\leq s\leq 1$, then $|\psi_T(t=T)\rangle$ gets closer to the ground state of $H_F$ as $T$ grows larger, and  becomes the ground state of $H_F$ when $T\to\infty$. All the scalability aspect of the adiabatic algorithm depends on how large $T$ has to be to reach a certain precision on the ground state energy.

\subsection{The benchmark}

\begin{figure}
    \centering
    \includegraphics[width=0.8\linewidth]{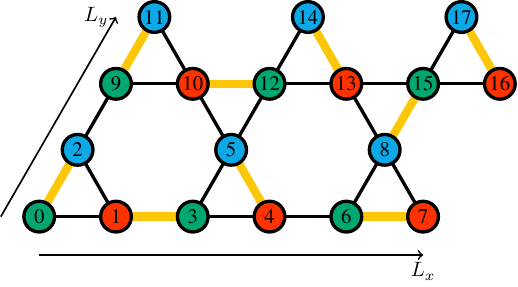}
    \includegraphics[width=\linewidth]{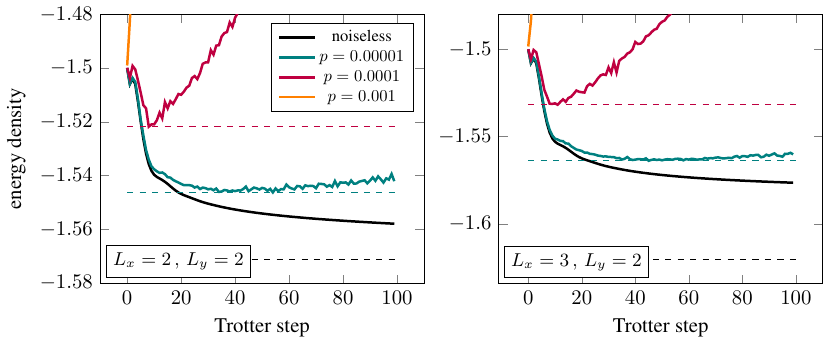}
    \caption{\emph{Top}: Kagome lattice with $L_x=3$ and $L_y=2$. \emph{Bottom}: Energy density as a function of number of Trotter steps in the benchmark setup, for different system sizes and different noise levels. Dashed lines indicate the minimum reached by the curve with the same color, and correspond to the benchmark score.}
    \label{kagome}
\end{figure}

As a benchmark, we consider the Heisenberg anti-ferromagnet model on a Kagome lattice. This model describes the material ${\rm YCu}_3 {\rm [OH(D)]}_{6.5}{\rm Br}_{2.5}$ \cite{liu2022gapless} and the precise properties of its ground state are still debated \cite{mei2017gapped,lauchli2019s}. The Hamiltonian of this system is given by
\begin{equation}
    H=-\sum_{\langle i,j\rangle}X_iX_j+Y_iY_j+Z_iZ_j\,,
\end{equation}
where $\langle i,j\rangle$ means that sites $i,j$ are neighbours on the Kagome lattice. We parametrize this lattice by two integers $L_x,L_y$ which count the number of small disjoint triangles in the vertical and horizontal directions, with $N=3L_xL_y$ sites in total, and impose open boundary conditions. We will restrict to even height $L_y$, to ensure the existence of a perfect matching on the graph. The sites are enumerated within triangles first, then along the $x$ direction, and then along the $y$ direction. An example of this Kagome lattice with site numbering and bonds between sites is represented in Fig \ref{kagome}.

To define an adiabatic path to prepare the ground state of this model, we define the initial Hamiltonian as
\begin{equation}
    H_I=-\sum_{\langle i,j\rangle'}X_iX_j+Y_iY_j+Z_iZ_j\,,
\end{equation}
where now $\langle i,j\rangle'$ means that $i,j$ are neighbours on a given \emph{perfect matching} of the Kagome lattice. We will consider the perfect matching depicted in Fig \ref{kagome} with yellow thick bonds. It contains the bonds $(0,2)$, $(1,3)$, $(4,5)$, and repeats this pattern on two neighbouring triangles in the $x$ direction over the entire lattice. If $L_x$ is odd, then for the last column of triangles in the $x$ direction, we include instead the bonds $(6,7)$, $(8,15)$, $(16,17)$ as depicted in Fig \ref{kagome}, repeated over the entire last column. The ground state of $H_I$ is given by the tensor product of singlets $\frac{1}{\sqrt{2}}(|01\rangle-|10\rangle)$ over all the $N/2$ bonds in this perfect matching. This can be prepared easily on the quantum computer. We then fix the Trotter step ${\rm dt}$ as a function of $s$ the scheduling time as
\begin{equation}
    {\rm dt}(s)=0.2\sqrt{1-s}\,.
\end{equation}
As for the scheduling function $\varphi(s)$ entering \eqref{adiabatichamiltonian}, we choose the following form
\begin{equation}
    \varphi(s)=\frac{1+\tanh (\tan(s\pi-\pi/2))}{2}\,,
\end{equation}
which interpolates smoothly between $\varphi(0)=0$ and $\varphi(1)=1$ while having all derivatives vanishing at $s=0$ and $s=1$. Finally, the ordering of the terms in the Trotter decomposition is taken to be first applying all the $XX$ terms, then all the $YY$ terms, and then all the $ZZ$ terms. 

% Which bonds are applied first within any of these three stages is irrelevant as they all commute. However, because of hardware noise, the precise ordering can have an influence on the result on actual hardware. Because this can depend on the precise platform and qubit connectivity, we leave the ordering within each of the three stages as a choice of the end user.

% The ordering of the terms in the Trotter decomposition is taken to be the following sequence of six stages: first we apply the three $XX,YY,ZZ$ terms on all the green-red bonds within the triangles, with the colors indicated in Fig \ref{kagome}, then on all the green-blue bonds within the triangles, then on all the red-blue bonds within the triangles, then on all the green-red bonds between triangles, then on all the green-blue bonds between triangles, and then on all the red-blue bonds between triangles. The energy is measured by running three times the same circuit, and measuring alternatively in the $X$, $Y$ and $Z$ basis.

\subsection{The score}

The only degree of freedom remaining is $M$ the number of Trotter steps performed. Only in the limit $M\to\infty$ is the exact adiabatic evolution implemented and the energy of the Hamiltonian $H_F=H$ minimized. On actual hardware however, noise precludes running arbitrarily deep circuits and effectively heats up the system, which competes with the cooling of the adiabatic process. At small $M$, heating due to imperfect adiabatic evolution dominates, and at large $M$, heating due to hardware noise dominates. There is thus a non-trivial optimal number of Trotter steps $M_*$ at which the energy is minimized. Given a mean energy $E_M$ obtained with $M$ Trotter steps, and with $\delta E_M$ the standard deviation, we take $E_M+2\delta E_M$ as the result energy, in order to avoid overshooting due to shot noise. We then define the benchmark score of a benchmarked hardware as
\begin{equation}
    \mathcal{S}_{\rm KH}=\underset{M\geq 1}{\min}(E_M+2\delta E_M)\,.
\end{equation}
We plot in the bottom panel of Fig \ref{kagome} the energy density obtained as a function of the number of Trotter steps, for different noise levels, together with the exact ground state. We see that for non-zero noise level $p$, the energy typically displays the expected behaviour, with an initial decrease and then an increase at large number of Trotter steps. The exact ground state energy can be obtained up to around $N\sim 30$ with classical computers, depending on the resources allocated. The comparison with the exact result is thus not scalable. However, even for system sizes beyond the classically simulable regime, the performance of the same algorithm run on different hardware can be compared, by directly comparing the energy density attained, the smaller being the best.

\section{Application: Nuclear Magnetic Resonance \label{sec:nmr}}
\subsection{Context and motivation}
Nuclear Magnetic Resonance (NMR) experiments are a key tool for material and molecular structure elucidation. They consist in polarizing all the nuclear spins of a sample material in a specific direction with a high magnetic field, and then measuring the relaxation of the magnetic field generated by the nuclear spins. The NMR spectrum of the sample material obtained by Fourier transforming the signal measured is then a signature of the bonds between the atoms supporting the nuclear spins. The classical simulation of NMR experiments can be done efficiently with dedicated softwares at high external magnetic field \cite{hogben2011spinach}. However, the simulation is more difficult in case of low external magnetic field, which is cheaper to implement experimentally. This low-field simulation of NMR experiments is one of the promising near-term applications of quantum computers \cite{sels2020quantum,seetharam2023digital,elenewski2024prospects,khedri2024impact}, although the precise settings where quantum computers would bring a practical advantage are still debated. The purpose of this present work is not to enter this debate, but instead to define a benchmark setup based on the performance of a quantum computer to infer couplings between nuclear spins in a molecule through NMR simulation.

These NMR experiments at low field are modeled as follows \cite{mr-4-87-2023}. The signal measured in an NMR experiment, called free induction decay (FID), can be written as
% \begin{equation}
%     {\rm FID}(t)=\tr[\Pi^\dagger S_z(t)\Pi e^{-\beta S_z(0)}]\,.
% \end{equation}
\begin{equation}
    {\rm FID}(t)=\tr[\Pi^\dagger S_z(t)\Pi S_z(0)]\,.
\end{equation}
% \begin{equation}
% \begin{aligned}
%     {\rm FID}(t)=&\sum_{i_1,..,i_N}\langle i_1...i_N|\Pi^\dagger S_z(t)\Pi S_z(0)|i_1...i_N\rangle\\
%    =&\sum_{i_1,..,i_N}\langle i_1...i_N|\Pi^\dagger S_z(t)\Pi|i_1...i_N\rangle m_{i_1...i_N}\,.
% \end{aligned}
% \end{equation}
Here, the total magnetization $S_z$ of the molecule is
\begin{equation}
    S_z=\sum_{j=1}^N \gamma_j Z_j\,,
\end{equation}
where each qubit corresponds to each of the $N$ nuclear spins $1/2$ contained in the molecule, and with $\gamma_j$ the gyromagnetic factor of nuclear spin $j$. The unitary operator $\Pi$ represents the initial pulse
\begin{equation}
    \Pi=e^{i\tau\sum_{j=1}^N \gamma_j X_j}\,,
\end{equation}
with $\tau$ the pulse duration. The time-evolved spin $S_z(t)$ is given by
\begin{equation}
    S_z(t)=e^{iHt}S_z e^{-iHt}\,,
\end{equation}
with $H$ the Hamiltonian describing the interactions between the $N$ spins. In absence of external magnetic field and for spin $1/2$ nuclei, this Hamiltonian can be written as
\begin{equation}
    H=\frac{1}{4}\sum_{i<j} J_{ij}(X_iX_j+Y_iY_j+Z_iZ_j)\,,
\end{equation}
with $J_{ij}$ the so-called $J$-coupling between nuclear spins $i$ and $j$, that is an effective spin-spin interaction resulting from the electron bondings in the molecule. 

From the measurement of the FID, one computes then the spectrum
\begin{equation}
    S(\omega)=\int_0^\infty e^{i\omega t} {\rm FID}(t)\D{t}\,.
\end{equation}
This amplitude $S(\omega)$ is the signal that the NMR end user is interested in. In an actual NMR experiment, the FID that is measured is the sum of all the tiny magnetic fields generated by the nuclei of all the molecules in the sample. Because of small perturbations, these slowly desynchronize with time, which results in an exponential decay in the FID. For liquid NMR, this exponential decay is very often modeled by an apodization term $e^{-t/T_2}$ multiplying ${\rm FID}(t)$, with $T_2$ a certain relaxation time.

\begin{figure}
    \centering
    \includegraphics[width=0.5\linewidth]{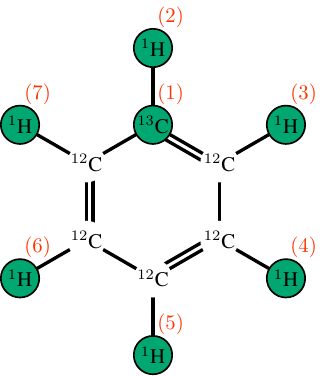}
    \caption{Depiction of the benzene molecule. The spinful atoms are indicated with a green circle. The numbering of the different spinful nuclei is given in red.}
    \label{fig:benzene}
\end{figure}

\begin{table}[]
    \centering
    \begin{tabular}{|c||c|c|c|c|c|c|}
    \hline
      $i$/$j$&  $2$ & $3$ & $4$ & $5$ & $6$ &$7$ \\
    \hline
    \hline
      $1$&  $158.354$ & $1.133$ & $7.607$ & $-1.296$ & $7.607$ &$1.133$ \\
       \hline
     $2$&   & $7.540$ & $1.380$ & $0.661$ & $1.380$ &$7.540$ \\
       \hline
     $3$&   &  & $7.543$ & $1.377$ & $0.658$ &$1.373$ \\
       \hline
     $4$&   &  &  & $7.535$ & $1.382$ &$0.658$ \\
       \hline
     $5$&   &  &  &  & $7.535$ &$1.377$ \\
       \hline
     $6$&   &  &  & &  &$7.543$ \\
       \hline
    \end{tabular}
    \caption{Coefficients $J_{ij}$ of the benzene-${}^{13}{\rm C}_1$ molecule, from \cite{wilzewski2017method}.}
    \label{nmrtable}
\end{table}

\subsection{The benchmark}
The benchmark we propose is the benzene-${}^{13}{\rm C}_1$ molecule depicted in Figure \ref{fig:benzene}. It contains $7$ nuclear spins, six hosted by the hydrogen atoms and one by the carbon-$13$ atom. The gyromagnetic factors are $\gamma_1=67.2828$ for the ${}^{13}C$ nucleus and $\gamma_j=267.522$ for $j=2,...,7$ the ${}^{1}H$ nuclei. The $J$-couplings obtained from experiments are listed in Table \ref{nmrtable}. The pulse time is taken to be $\tau=\frac{\pi}{2\gamma_1}$. The maximal simulation time is taken to be $T=50$, and we fix an arbitrary but realistic relaxation time $T_2=10$. Given the exact ${\rm FID}(t)$, we define the spectrum to which the hardware is to be compared as
\begin{equation}
    S_{\rm exact}(\omega)=\int_0^T e^{i\omega t}e^{-t/T_2}{\rm FID}(t)\D{t}\,.
\end{equation}
Since the model is defined on only $7$ qubits, this quantity can be quickly computed classically with arbitrary precision. We impose that the time evolution is implemented using a Trotter evolution, with Trotter step
\begin{equation}
    U=\prod_{i<j} e^{i\frac{\delta t}{4}X_iX_j}e^{i\frac{\delta t}{4}Y_iY_j}e^{i\frac{\delta t}{4}Z_iZ_j}\,,
\end{equation}
where $\delta t$ is a given Trotter step size. We fix the ordering of the couplings to be given by applying the gates in the following order $(1,2)$, $(3,4)$, $(5,6)$, $(1,7)$, $(2,3)$, $(4,5)$, $(6,7)$, $(1,3)$, $(4,6)$, $(2,7)$, $(3,5)$, $(1,6)$, $(2,4)$, $(5,7)$, $(1,4)$, $(1,5)$, $(2,5)$, $(2,6)$, $(3,6)$, $(3,7)$, $(4,7)$. The benchmark user is free to choose the Trotter step size $\delta t$, but is fixed to be the same for all time points.

\begin{figure*}
    \centering
    \includegraphics[scale=1.05]{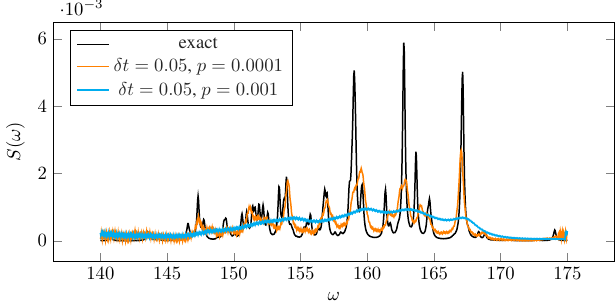}
    \includegraphics[scale=0.85]{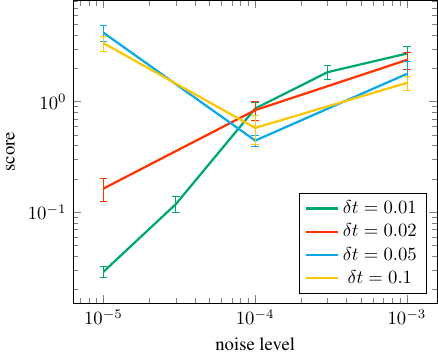}
    
    \caption{\emph{Left:} exact spectrum $S(\omega)$ obtained with $\delta t=0.01$ and $T_2=10$ (black), and noisy spectra obtained with $\delta t=0.05$ and noise levels $p=0.001$, $p=0.0001$ (cyan and orange) after optimizing $\tilde{T}_2$ and $\delta\omega$ as described in Section \ref{compatibility}. \emph{Right:} root mean square $\Delta J$ on the estimated J-couplings, as a function of the noise level, for different Trotter steps.}
    \label{fig:nmr_spectra}
\end{figure*}

\subsection{The score}
\subsubsection{Overview}
We propose to evaluate the outcomes ${\rm FID}(n\delta t)$ of the quantum computer in a most application-oriented way. NMR experiments are performed to elucidate the structure of a given molecule. In our simple use case of the benzene molecule, this would mean computing the J-couplings $J_{ij}$ between every spinful nuclei. Given a NMR spectrum obtained from experiment, we would perform simulation with some trial couplings $\tilde{J}_{ij}$, and then take as an estimate of the actual couplings $J_{ij}^{\rm est}$ the trial couplings corresponding to the spectrum that matches the experiment the most closely. A natural score is then the mean error between estimated coefficients $J_{ij}^{\rm est}$ and actual coefficients $J_{ij}$.

The computation of the score assigned to values ${\rm FID}(n\delta t)$ measured on the hardware for $n=0,...,T/\delta t$ is done in multiple stages. 

\subsubsection{Compatibility measure \label{compatibility}}
Firstly, given a candidate spectrum $S_{\rm can}(\omega)$, we would like to evaluate the compatibility with our measured time series ${\rm FID}(n\delta t)$ from the hardware. We call the output of that stage ``compatibility measure". From the time series, we compute the spectrum as
\begin{equation}
    S_{\rm hard}(\omega)=\delta t \sum_{n=0}^{T/\delta t} e^{i\omega n\delta t} e^{-n\delta t/\tilde{T}_2} \delta_n{\rm FID}(n\delta t)\,,
\end{equation}
where $\tilde{T}_2$ is a parameter, and with $\delta_n=1/2$ if $n=0$ or $n=T/\delta t$, and $\delta_n=1$ otherwise. This $\delta_n$ term removes potential baseline offset \cite{mr-4-87-2023}. The spectrum $S_{\rm hard}(\omega)$ is computed at the values $\omega$ where $S_{\rm can}(\omega)$ is available. The benchmark user is free to choose $\tilde{T}_2$ (even to take it negative) to optimize the agreement with $S_{\rm exact}(\omega)$. Hardware results are indeed going to come with noise that will already induce an exponential decay on the data: when comparing with an actual NMR experiment, such an additional exponential decay $\tilde{T}_2$ can always be incorporated to match the exponential decay observed in the NMR experiment. The benchmark user is also free to set ${\rm FID}(n\delta t)=0$ for time points that they decide not to compute. Moreover, we also allow the user to apply a shift in the frequencies, namely to redefine
\begin{equation}\label{shiftomega}
    S_{\rm hard}'(\omega)=S_{\rm hard}(\omega+\delta \omega)\,,
\end{equation}
with an arbitrary parameter $\delta \omega$ so as to optimize agrement with $S_{\rm can}(\omega)$. We indeed observed that Trotter errors coming from the finite Trotter step size tend to globally slightly shift the frequencies. While impacting significantly point-by-point agreement between $S_{\rm can}(\omega)$ and $S_{\rm hard}(\omega)$, this effect does not prevent identification of the spectrum, and so we decide to mitigate it with the above freedom to shift the frequencies. For ease of implementation and to avoid having to introduce an arbitrary scale, we impose that the shift in \eqref{shiftomega} is applied periodically on the range of $\omega$'s. The agreement between $S_{\rm hard}'(\omega)$ and $S_{\rm can}(\omega)$ is evaluated by maximizing the inner product $F(S_{\rm hard}',S_{\rm can})$ with
\begin{equation} 
    F(A,B)=\frac{\sum_{\omega}A(\omega)B(\omega)}{\sqrt{\sum_{\omega}A(\omega)^2\sum_{\omega}B(\omega)^2}}\,.
\end{equation}
Namely, the final compatibility measure between the measured time series and the candidate spectrum is the maximal value of $F(S_{\rm hard}',S_{\rm can})$ obtained when optimizing $\tilde{T}_2$ and $\delta\omega$. In the left panel of Fig \ref{fig:nmr_spectra}, we present simulated spectra $S_{\rm hard}'(\omega)$ obtained after this optimization, for different noise levels, and comparison to the exact spectrum computed with $\delta t=0.01$.

\subsubsection{Identification within a database}
We now would like to use the compatibility measure defined in the previous subsection to identify, among a database of molecular spectra, the spectrum that is the most compatible with our measured time series. We define these databases of molecular spectra as being composed of $100$ spectra of simulated benzene molecules, but with different J-couplings. One of the $100$ spectra is computed with the exact J-couplings given in Table \ref{nmrtable}. The $99$ other spectra are computed with perturbed J-couplings $\tilde{J}_{ij}$ randomly generated as follows
\begin{equation}
    \tilde{J}_{ij}=J_{ij}+0.01\cdot 2^{0.1m} \xi\,, 
\end{equation}
for $m=0,...,98$, and with $\xi$ a random Gaussian variable with mean $0$ and variance $1$ (randomly drawn for every couple $(i,j)$ and sample in the database). This ensures that there are spectra in the database that are very close to and very dissimilar from the exact spectrum of benzene. For every sample in the database, one computes the exact ${\rm FID}_{\rm can}(n\delta t)$ for $n=0,...,T/\delta t$ with noiseless numerical simulation, with $\delta t=0.01$ and $T=50$. Then one computes the spectrum associated to this $m$-th sample
\begin{equation}
    S_{\rm can}^{(m)}(\omega)=\delta t \sum_{n=0}^{T/\delta t} e^{i\omega n\delta t} e^{-n\delta t/T_2} \delta_n{\rm FID}_{\rm can}(n\delta t)\,,
\end{equation}
with $T_2=10$. We will be only interested in the frequency region $140\leq \omega \leq 175$, which contains most of the interesting features of this molecule. For definiteness, we will compute the spectrum at $1000$ equally spaced values of $\omega$ between $140$ and $175$. The compatibility measure defined in the previous subsection will thus depend only on the frequencies within this range. Once the whole database is generated, we look for the sample that has the highest compatibility measure with our time series measured on the hardware. The estimated J-couplings $J_{ij}^{\rm est}$ are then set to be the J-couplings of this most compatible sample within the database.

\subsubsection{The score}
Given estimated J-couplings $J_{ij}^{\rm est}$, we define the quality of the estimate as the mean-square error
\begin{equation}
    \Delta J=\sqrt{\frac{1}{21}\sum_{i<j} (J_{ij}-J_{ij}^{\rm est})^2}\,,
\end{equation}
where $21$ is the number of J-couplings in our particular case of benzene. Given a time series measured on the hardware, this $\Delta J$ is a random variable, since it depends on the random database generated for comparing the spectra. To define a score that is not a random variable, we then define the average
\begin{equation}
    \mathcal{S}_{\rm NMR}=\mathbb{E}[\Delta J]\,,
\end{equation}
where the statistical average $\mathbb{E}$ is over different random databases. If only a small number of databases are generated, one can provide an error bar on top of this average. This score has a very simple application-oriented meaning: it is the precision that the user can expect to obtain on the J-couplings, if the hardware was used to compare the spectrum of benzene with simulations.

In the right panel of Fig \ref{fig:nmr_spectra}, we plot the root mean-square error $\Delta J$ obtained by running the benchmark on noisy simulated circuits, for different error probability per two-qubit gate and different Trotter steps $\delta t$. At low noise level, we observe that small Trotter steps are more able to recover the true values of the J-couplings. This is expected as at low error rate, Trotter errors dominate. For these small Trotter steps, increasing the error rate blurs the NMR signal and decreases the precision. At larger error rate, larger Trotter steps perform better because in this regime, noise dominates over Trotter error, and circuits with large Trotter steps have fewer gates.

\subsection{Extensions}
Some comments on the generality of this benchmark are in order. Contrary to the previous benchmarks presented in Sections \ref{freefermionsection} and \ref{sec:static}, the size of the benchmark system we propose cannot be scaled arbitrarily. The computation of the score that we defined requires exact knowledge of the spectrum, which can be done classically with state-vector simulation only up to $\lessapprox 20$ spinful nuclei. This is justified by the fact that, firstly, there are actual potential use cases beyond classical simulability that do not require much more qubits, less than $100$ \cite{elenewski2024prospects,khedri2024impact}; and secondly, the phenomenon of dilution of error ensures that the gate fidelity required to accurately simulate NMR experiments does not scale with system size \cite{khedri2024impact,granet2024dilution}. Hence, instead, this benchmark is meant to evaluate the ability of a hardware (in the future, potentially with quantum error correction) to simulate long time-evolution with deep circuits, through a concrete application use case.

\section{Application: ground state energy of 
 molecules \label{sec:chemistry}}

\subsection{Context and motivation}
One of the main tasks of quantum chemistry is the determination of chemical reaction rates. This requires the knowledge of the ground state energy of molecules as a function of their geometry with high precision. For intermediate to large numbers of orbitals, reaching this high precision becomes a difficult or impossible task with classical computers. 

Mathematically, the Hamiltonian of a molecule decomposed onto $N$ orbitals (i.e., qubits) can be written as
\begin{equation}
    H=\sum_{i,j}h_{ij} c_i^\dagger c_j+\sum_{i,j,k,l}h_{ijkl} c_i^\dagger c^\dagger_j c_k c_l\,,
\end{equation}
with $h_{ij},h_{ijkl}$ some coefficients. This Hamiltonian is then usually expressed in terms of Pauli matrices through a Jordan-Wigner transformation. Although bearing many similarities with condensed matter systems, these chemical problems have two important specificities: the number of terms in the Hamiltonian is large and the precision required on the ground state energy is high. This necessitates using different techniques than the Trotter algorithm.

\begin{figure}
    \centering
    \includegraphics[width=0.7\linewidth]{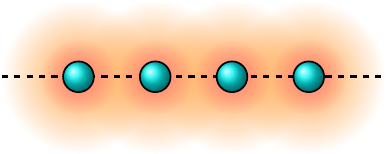}
    % \vspace{2cm}
    
    $\quad$
    \includegraphics[width=1\linewidth]{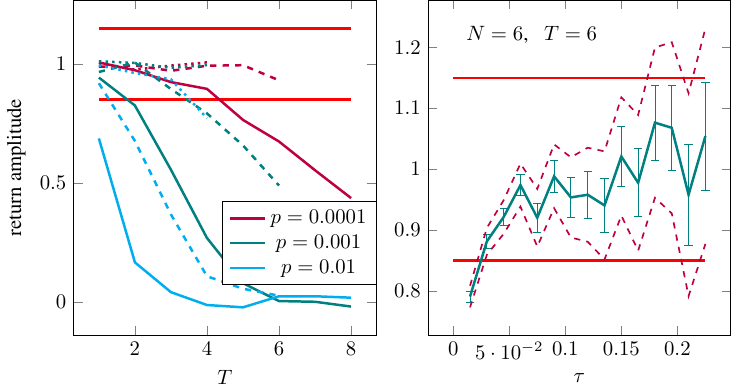}
    \caption{\emph{Top: }Depiction of the system benchmark, a hydrogen chain. \emph{Bottom left}: return amplitude measured as a function of $T$, for different error rates, at optimal gate angle $\tau$, in size $N=4$ (dotted), $N=6$ (dashed) and $N=8$ (solid). The red lines indicate the thresholds for the size to be validated at $T=N$. \emph{Bottom right}: return amplitude with error bars as a function of the gate angle $\tau$ for a fixed number of shots, with error rate $p=0.001$, size $N=6$ and time $T=N$. The purple dashed line indicates the expectation value plus or minus two error bars, which is used as the criterion for passing the test in the benchmark.}
    \label{fig:hydrogen}
\end{figure}

\subsection{The benchmark}
We define the benchmarking system to be a linear chain of $L$ hydrogen atoms, each separated by a distance $d=0.74 {\rm nm}$, decomposed in the STO-3G basis set. This system is sketched in Fig \ref{fig:hydrogen}. They are defined on $N=2L$ qubits. The fermionic basis is optimized using restricted Hartree-Fock. In case of an odd number $L$ of hydrogen atoms, we remove one electron in order to keep an even number of electrons and be able to run the restricted Hartree-Fock optimization. We decompose then the Hamiltonian into Pauli strings using a Jordan-Wigner transformation. Next, we use particle number conservation to add to the Hamiltonian the quantity $-\alpha(\sum_{j=1}^NZ_j)^2$ without changing its eigenstates. The coefficient $\alpha$ is taken to be the median of the coefficients in front of terms $Z_jZ_k$ in the Pauli string decomposition of the Hamiltonian, as this allows one to minimize the 1-norm of the Hamiltonian, i.e. the sum of the absolute values of the coefficients. In this way we obtain a decomposition of the Hamiltonian
\begin{equation}
    H_F=\sum_{n}c_n P_n\,,
\end{equation}
with $c_n$ some coefficients and $P_n$ Pauli strings on $N$ qubits. Up to changing $P_n$ into $-P_n$, we will assume $c_n>0$. These decompositions are spelled out in Appendix \ref{app:hydrogen} for $N=4,6,8$. 

We then define the time-dependent Hamiltonian
\begin{equation}\label{timedep}
    H(t)=\left(1-\frac{t}{T}\right)H_I+\frac{t}{T}H_F\,,
\end{equation}
where $0\leq t\leq T$ with $T$ a total adiabatic time, and with $H_I$ containing only the $c_n P_n$ terms of $H_F$ where the Pauli string $P_n$ is a single $Z$ term (located at any site). This time-dependent Hamiltonian has been studied in \cite{granet2024practicality}. It implements an adiabatic evolution from a diagonal Hamiltonian $H_I$, whose ground state is the Hartree-Fock state $|{\rm HF}\rangle=|1...10...0\rangle$, to the target Hamiltonian $H_F$, whose ground state energy is the sought quantity. It has been shown numerically up to $N=20$ that an adiabatic time $T=N$ is sufficient to prepare a state whose energy is within chemical accuracy of the ground state energy, i.e. such that the energy difference is smaller than $10^{-3}$. For this benchmark, we propose the implementation of this adiabatic state preparation
\begin{equation}\label{targetpsi}
    |\psi_f\rangle=U(T)|{\rm HF}\rangle\,,
\end{equation}
where $U(T)$ implements the time-dependent Hamiltonian evolution \eqref{timedep} up to time $t=T=N$. To implement this time evolution, we propose the randomized algorithm of \cite{granet2024hamiltonian}. This algorithm allows for an exact implementation of the Hamiltonian dynamics, without any Trotter error, while still displaying a finite average number of gates in each circuit. The algorithm works as follows. One chooses a gate angle $0<\tau<\pi/2$. We introduce an ancilla and initialize the total state on $N+1$ qubits in
\begin{equation}
  |\psi\rangle= \frac{1}{\sqrt{2}}\left( |0\rangle \otimes |{\rm HF}\rangle+|1\rangle \otimes |{\rm HF}\rangle\right)\,.
\end{equation}
We define the time-dependent coefficients $c_n(t)=c_n$ if $P_n$ is a single Pauli $Z$, and $c_n(t)=\frac{t}{T}c_n$ otherwise. We then evolve $|\psi\rangle$ according to the following random process. For every term in the Hamiltonian $P_n$, we apply on $|\psi\rangle$ the rotation $e^{i\tau P_n}$ conditioned on the ancilla being $1$, according to a Poisson process with time-dependent rate $c_n(t)/\sin \tau$, during a time $T$. This is described precisely in \cite{granet2024hamiltonian,granet2024practicality}. Denoting $\rho_i$ the mixed state obtained after running this random time evolution, we have
\begin{equation}
\begin{aligned}
     \rho_i&=\frac{1}{2}|0\rangle\langle 0| \otimes |{\rm HF}\rangle \langle {\rm HF}|+\frac{1}{2}|1\rangle\langle 1| \otimes \rho_{\rm unkn}\\
     &+\frac{\lambda}{2}|0\rangle \langle 1|\otimes |{\rm HF}\rangle \langle \psi_f|+\frac{\lambda}{2}|1\rangle \langle 0|\otimes  | \psi_f\rangle \langle{\rm HF}|\,,
\end{aligned}
\end{equation}
with $|\psi_f\rangle$ the target state in \eqref{targetpsi}, with $\rho_{\rm unkn}$ some unknown density matrix, and with $\lambda$ a scalar given by 
\begin{equation}\label{lambdaterm}
    \lambda=\exp\left(- \tan(\tau/2)\sum_{n}\int_0^T c_n(t)\D{t}\right)\,.
\end{equation}
Repeating the same process on this output density matrix, but conditioning the ancilla to be $0$ instead of $1$ (and of course, generating a different random Poisson process), we obtain the density matrix
\begin{equation}
\begin{aligned}
     \rho_f&=\frac{1}{2}|0\rangle\langle 0| \otimes |{\rm HF}\rangle \langle {\rm HF}|+\frac{1}{2}|1\rangle\langle 1| \otimes \rho_{\rm unkn}'\\
     &+\frac{\lambda^2}{2}|0\rangle \langle 1|\otimes |\psi_f\rangle \langle \psi_f|+\frac{\lambda^2}{2}|1\rangle \langle 0|\otimes  | \psi_f\rangle \langle\psi_f|\,,
\end{aligned}
\end{equation}
with $\rho_{\rm unkn}'$ another unknown density matrix. By taking expectation value of $X$ on the ancilla, one gets access to the exact state $|\psi_f\rangle$
\begin{equation}
    \tr[X_a \rho_f]=\lambda^2|\psi_f\rangle \langle \psi_f|\,.
\end{equation}
Namely, the expectation value of any observable $\mathcal{O}$ within $|\psi_f\rangle$ can be obtained as
\begin{equation}\label{expectationvalue}
   \langle \psi_f|\mathcal{O} |\psi_f\rangle=\lambda^{-2}\tr[(X\otimes \mathcal{O}) \rho_f]\,.
\end{equation}
By setting $\mathcal{O}=I$, the left-hand side is equal to $1$, and so $\tr[X_a\rho_f]$ must be equal to $\lambda^2$. The agreement of the benchmarked hardware with that theoretical expectation value gives a way of evaluating the quality of the computation.

In this randomized algorithm, the end user can choose the gate angle $\tau$ without influence on the result \eqref{expectationvalue}. Changing the gate angle $\tau$ however modifies the average number of gates in the circuit, and the value of $\lambda$. The number of gates in the circuit is proportional to $1/\sin \tau$, and the attenuation factor is given in \eqref{lambdaterm}. The number of shots to perform to obtain a given precision on $\langle \psi_f|\mathcal{O}|\psi_f\rangle$ in \eqref{expectationvalue} scales as $\lambda^{-4}$. Hence, changing $\tau$ allows for balancing the number of gates in the circuit (and hence the noise) and the number of shots to perform. Increasing $\tau$ decreases linearly the number of gates in the circuit, but increases exponentially the number of shots to perform. Which $\tau$ to choose depends on the hardware: fast architectures where large numbers of shots can be done prefer larger values of $\tau$; slower but more precise architectures prefer smallest values of $\tau$. There is a choice of $\tau$ that minimizes the total number of gates to implement to reach a certain precision on a noiseless perfect hardware, approximately equal to $\tau=1/(\int_0^T c_n(t)\D{t})$ \cite{granet2024hamiltonian}. However, in the presence of noise, larger values of $\tau$ might be more efficient. We therefore leave to the end user the freedom to choosing the gate angle $\tau$. This benchmark setup thus automatically balances gate fidelity and clockspeed.

\subsection{The score}
We assign the following score to the benchmark. We say that the hardware passes the test in size $N$ if the return amplitude plus or minus two error bars at time $T=N$ is contained around $1$ plus or minus a threshold value $\Theta$ that we set to $\Theta=0.15$. Namely, let us denote by $E$ the expectation value obtained for the quantity $\lambda^{-2}\tr[(X\otimes I)\rho_f]$ (i.e., $\lambda^{-2}$ times the expectation value of $X$ on the ancilla), and $\delta E$ one standard deviation on the estimate, when run at time $T=N$. Then we say that the hardware passes the test at size $N$ if
\begin{equation}
  0.85=1-\Theta \leq E\pm 2\delta E\leq 1+\Theta=1.15\,.
\end{equation}
Then, the score $\mathcal{S}_{\rm QC}$ assigned to this quantum chemistry benchmark is the largest system size $N$ for which the hardware passes the test.

In Figure \ref{fig:hydrogen}, we show a run of this benchmark for small system sizes $N=4,6,8$. In the left panel, we show the return amplitude measured as a function of $T$ for different system sizes and error rates, when choosing the gate angle $\tau$ to be the optimal value. Here, all system sizes fail the test for error rate $p=0.01$. For $p=0.001$, only $N=4$ passes the test. For $p=0.0001$, $N=4$ and $N=6$ pass the test, but not $N=8$. In the right panel, we show the effect of gate angle in the case $N=6$ and $T=6$ at error rate $p=0.001$. At optimal gate angle, the test fails. However, one sees that by increasing gate angle one can decrease the effect of noise so as to obtain a return amplitude above the threshold. But if one increases the gate angle too much, error bars grow and the test fails again. This shows that this benchmark allows the user to take advantage of a high clockspeed that allows for running a high number of shots, and so increasing the gate angle $\tau$ to mitigate the effect of hardware imperfections.

\section{Application:  classical optimization \label{sec:opt}}
\subsection{Context and motivation}
\begin{figure}
    \centering
    \includegraphics[width=0.4\linewidth]{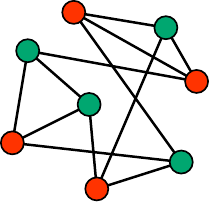}\\
    \vspace{0.5cm}
    \includegraphics[width=0.95\linewidth]{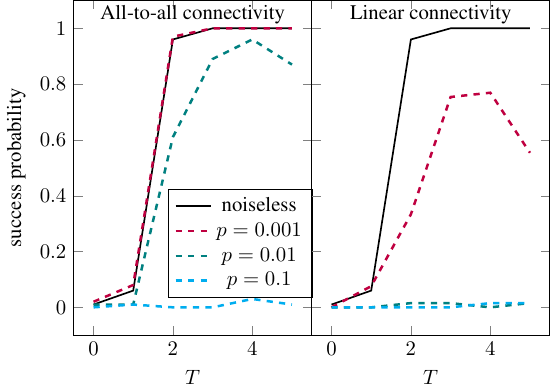}
    \caption{\emph{Top:} Example of a $3$-regular graph with an optimal coloring maximizing the number of edges between green and red nodes. \emph{Bottom:} Probability (averaged over $100$ simulations) of finding the optimal cut in a graph of size $N=20$ in $10^3$ shots, as a function of $T$ in the benchmark setting, for different noise levels $p$, comparing all-to-all connectivity and linear connectivity. The all-to-all connected simulated hardware passes the benchmark for $p=0.001$ and $p=0.01$, while the linear-connected simulated hardware passes the benchmark only for $p=0.001$.}
    \label{fig:maxcut}
\end{figure}

Classical optimization problems consist in finding the minimum of a cost function over a (usually) discrete set of configurations, such as for example the traveling salesman problem or the knapsack problem. What makes these problems attractive to quantum computing is firstly, the (quasi) guarantee that these problems cannot be solved classically in polynomial time (otherwise $P=NP$), ensuring that they will always become impossible to solve classically provided the system size is large enough; and secondly, the wide relevance of these problems to several sectors of the industry. In quantum computing, they can be formulated as finding the ground state of a \emph{classical} Hamiltonian $H$, namely that contains only $Z$ Pauli matrices. The simplest optimization problem in this formulation is the so-called Max-Cut problem, whose Hamiltonian is
\begin{equation}
    H=\sum_{\langle i,j\rangle}Z_iZ_j\,,
\end{equation}
where $\langle i,j\rangle$ means that sites $i,j$ are neighbours on a given graph. The ground state of $H$ is a product state in the $Z$ basis whose values $0,1$ partition the graph into two sub-graphs such that the number of edges connecting one sub-graph to the other is maximal. We show in Fig \ref{fig:maxcut} an example of a graph with such a maximal partition.

One way of finding the ground state of $H$ on a quantum computer is to use the adiabatic algorithm. Given a Trotter step $\delta t$ and a number of steps $T$, we implement the unitary operator
\begin{equation}
    U(T)=W_{1}(\delta t)...W_{2\delta t/T}(\delta t)W_{\delta t/T}(\delta t)\,,
\end{equation}
with
\begin{equation}
    W_s(\delta t)=\prod_{\langle j,k\rangle}e^{is\delta t Z_jZ_k} \prod_{j} e^{-i(1-s)\delta t X_j}\,.
\end{equation}
Preparing initially the quantum computer in the state $|+...+\rangle$, provided $T$ is large enough and $\delta t$ small enough, the final state obtained 
\begin{equation}\label{final}
    |\psi\rangle=U|+...+\rangle
\end{equation}
should have large overlap with the ground state of $H$. By measuring the qubits in the $Z$ basis, one obtains a list of bits that should have a non-negligible probability to provide a solution to the Max-Cut problem.

\subsection{The benchmark}
For this benchmark, we will fix the graphs to be $3$-regular graphs, i.e. graphs in which every vertex has exactly $3$ neighbours. We fix moreover the Trotter step $\delta t$ to be equal to $\delta t=0.25$. This setup has been extensively tested in \cite{granet2024benchmarking} and it has been observed that taking $T=\mathcal{O}(L)$ is enough to be able to find the ground state, for systems up to size $\approx 100$. While the ground state of these systems cannot be found classically for arbitrary system sizes, there exist classical approximate solvers that are very likely to be able to find the exact ground state in reasonable runtime up to sizes $\sim 10^3$ \cite{DunningEtAl2018,granet2024benchmarking}. This ensures that the benchmark can be implemented on hardware for probably several years to come. Even beyond the classically simulatable regime, different hardware can still be compared to each other.

\subsection{The score}
We say that a given hardware is able to solve a graph $G$ if there is experimental evidence for the existence of a value of adiabatic time $T$ and of a number of shots $N_S$, such that by measuring \eqref{final} in the $Z$ basis $N_S$ times, the optimal solution is obtained with probability larger than $1/2$. The end user is free to choose an appropriate value of adiabatic time $T$ and of number of shots $N_S$. To be able to claim that a given hardware solves the graph, we require that the user runs a minimum of $10$ groups of $N_S$ shots, and that counting $1$ for each group of shots containing the optimal solution, and $0$ otherwise, the mean value of this random variable is larger than $0.5$ by two standard deviations. 

Then, we say that a given hardware passes the benchmark in size $N$ if there exists at least one \emph{typical} (defined below) and connected $3$-regular graph on $N$ sites that the given hardware is able to solve. We define the score $\mathcal{S}_{\rm Max-Cut}$ to be the largest system size $N$ for which the hardware passes the test. A refinement of the score can be made by giving, for that value of $N$ and $T$, the average time-to-solution defined as $N_S$ times the average runtime of one shot.

We note that by increasing the number of shots $N_S$, we can always obtain a non-negligible probability of measuring the optimal solution by just random guess. This feature is not a loophole of the benchmark. While this strategy can be implemented for small system sizes, it would require scaling $N_S$ exponentially with $N$ for larger $N$ and quickly becomes impractical. From an application point of view, a hardware that is able to run a large number of shots quickly should indeed be considered more powerful than a slow hardware, all other things being equal. Imposing a number of shots $N_S$ would set an intrinsic time scale that could be detrimental to certain hardware or become obsolete in the future, if machines become faster or instead slower due to e.g. error correction. 

The constraint of typicality is defined as follows. We consider the algorithm of Steger and Wormald to generate random regular graphs \cite{steger1999generating}, that is implemented in the NetworkX Python package. For a graph $G$, we define $\lambda_1\leq \lambda_2\leq ...\leq \lambda_N$ the eigenvalues of its adjacency matrix, and $m_j=\mathbb{E}[\lambda_j]$ their mean value where $\mathbb{E}$ denotes the statistical average over random graphs of size $N$ generated with the Steger and Wormald algorithm. Then we define the variance of graph $G$ as
\begin{equation}
    v(G)=\sum_{j=1}^N (\lambda_j-m_j)^2\,,
\end{equation}
and define the mean variance over all regular graphs as $\bar{v}=\mathbb{E}[v]$. We say that the graph $G$ is typical if its variance $v(G)$ satisfies $v(G)\leq 2\bar{v}$. Numerically, we observe that only a proportion of around $\sim \frac{1}{N}$ of the graphs generated with the Steger and Wormald algorithm are not typical, so this constraint is not stringent, but we impose it only to avoid exceptional cases. 

In Fig \ref{fig:maxcut} we present some numerical noisy simulations of this benchmark, showing the probability of finding the optimal cut as a function of $T$ in a given graph of size $N=20$, for a number of shots $N_S=10^3$. We compare two simulated hardware architectures, one where all qubits are connected to each other, and another one where gates can be applied only between neighbouring qubits on a line, requiring the implementation of additional SWAP gates to connect arbitrary qubits. With all-to-all connectivity, the simulated hardware would pass the benchmark for two-qubit error rate $p=0.001$ and $p=0.01$, but would fail for $p=0.1$, because no value of $T$ leads to a success probability larger than $1/2$. With linear connectivity, only for $p=0.001$ would it pass the benchmark.

\section{Conclusion}
We have introduced an application-oriented benchmarking suite for quantum computers that is focused on Hamiltonian simulation. We have defined five different benchmark settings, that correspond to some of the most prominent potential applications of quantum computing, namely material and condensed matter physics simulation (dynamic problems and static problems), Nuclear Magnetic Resonance, quantum chemistry, and classical optimization. Specifically, we presented explicit benchmark settings for (i) computing the dynamics of electronic systems, including the simulation of neutron scattering experiments, (ii) computing the values of static observables of condensed matter physics at low temperature, (iii) computing the spectrum generated by nuclear magnetic resonance experiments, (iv) preparing the ground state of a hydrogen chain in quantum chemistry, and (v) solving the Max-Cut problem on $3$-regular graphs. 

A scalable application-oriented benchmark can be sometimes contradictory, as benchmarking supposes to know the exact result, whereas the best applications of quantum computing are those beyond reach of classical computers. We tried to slalom between these contradictions and defined different settings that, although not all scalable and not all implementing an end-to-end quantum computing application, address a variety of circuit geometries, application practicality, qubit connectivities and scalability properties that altogether should draw an accurate overview of the ability of a given quantum computing hardware to solve some real-world applications. 

Besides these benchmarks, we introduced a new metric to measure the capabilities of a quantum computing hardware at a given task that involves computing the expectation value of an observable. The metric is based on the idea that, since a certain minimal number of shots has to be performed on the quantum computer to reach a given precision on the expectation value, a systematic bias coming from noise might not be detectable before a certain number of shots have been performed. Stated differently, given a certain gate budget, a noisy quantum computing hardware can be in practice indistinguishable from a perfect quantum computer at a given task, if the effect of hardware imperfections is below the shot noise. We thus introduced the notion of \emph{distinguishability cost} to measure the quality of a quantum computing hardware at a given task, as the minimal number of gates that a perfect quantum computer has to run to certify that the output of the benchmarked hardware is incorrect. The appeal of this score is that it is universally applicable to any problem involving expectation values, and outputs a number with direct physical and practical meaning.

\section*{Acknowledgements}
We thank Yi Hsiang Chen, Daniel Mills and Kushal Seetharam for comments on the draft. The project was funded by the Bavarian Ministry of Economic Affairs, Regional Development and Energy (StMWi) under project Bench-QC (DIK0425/01).

\appendix
\section{Toric code state preparation}\label{app:toriccode}
\begin{figure*}
    \centering
    \includegraphics[width=0.27\linewidth]{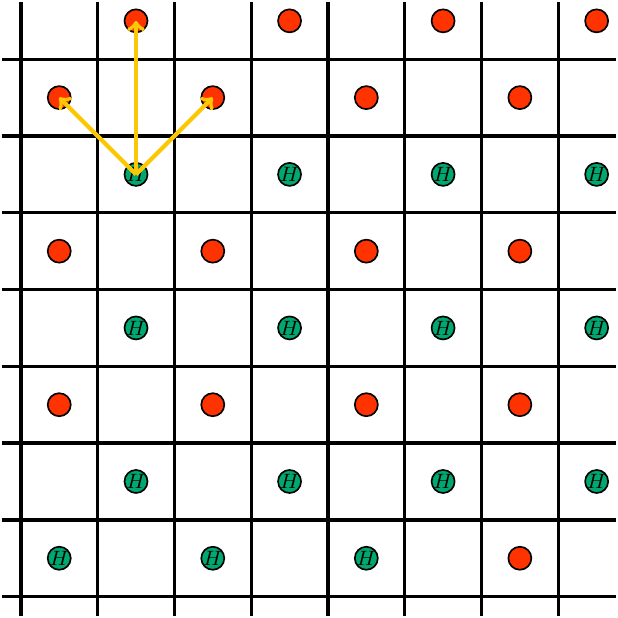}
   $\quad$
    \includegraphics[width=0.27\linewidth]{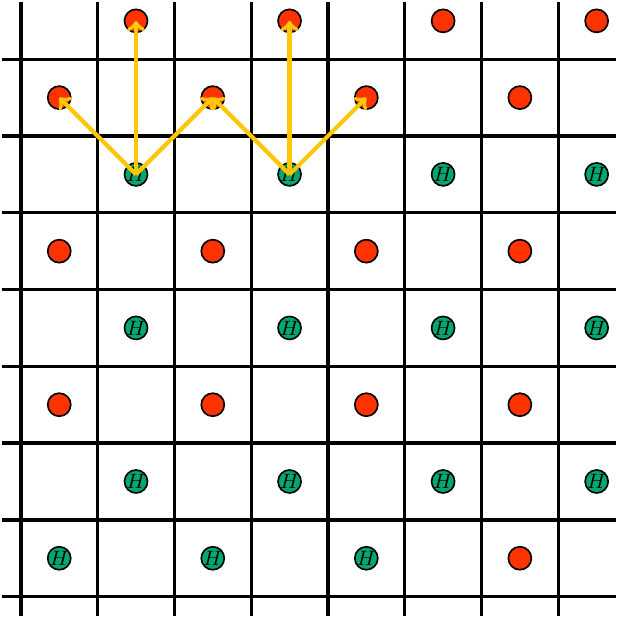}
   $\quad$
    \includegraphics[width=0.27\linewidth]{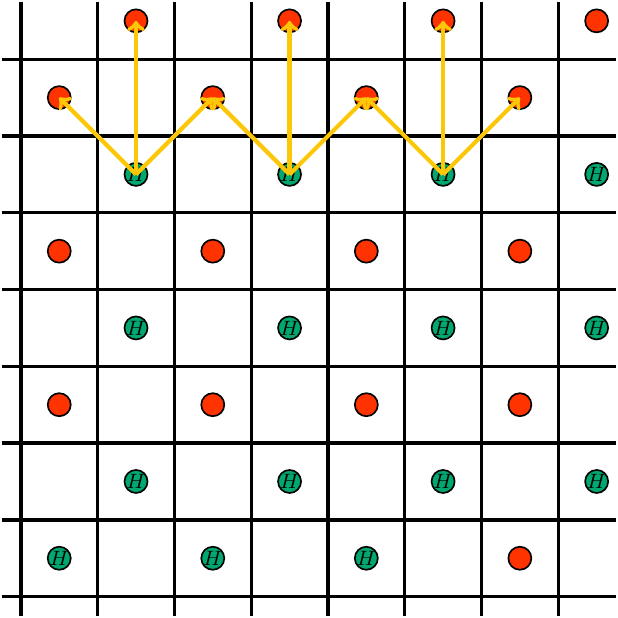}\\
    \vspace{0.5cm}
    \includegraphics[width=0.27\linewidth]{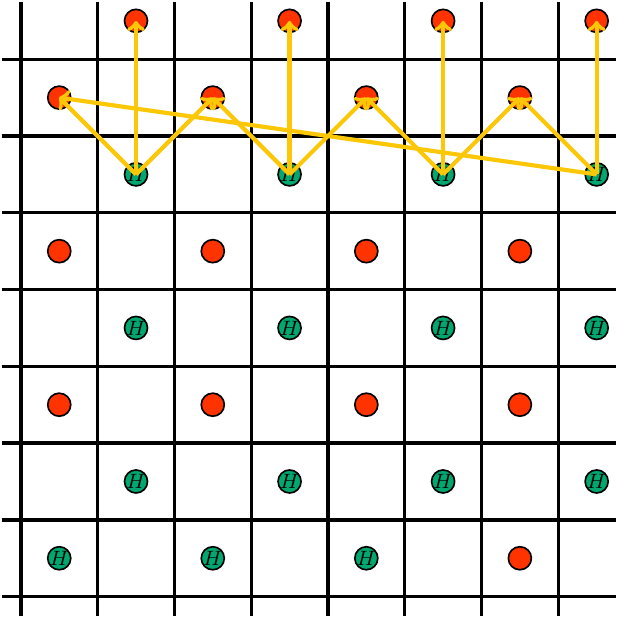}
   $\quad$
    \includegraphics[width=0.27\linewidth]{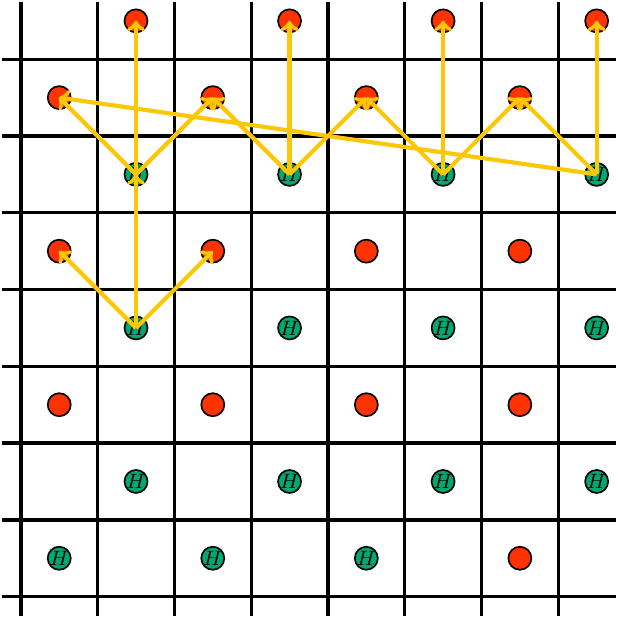}
   $\pmb{\dots}$ 
    \includegraphics[width=0.27\linewidth]{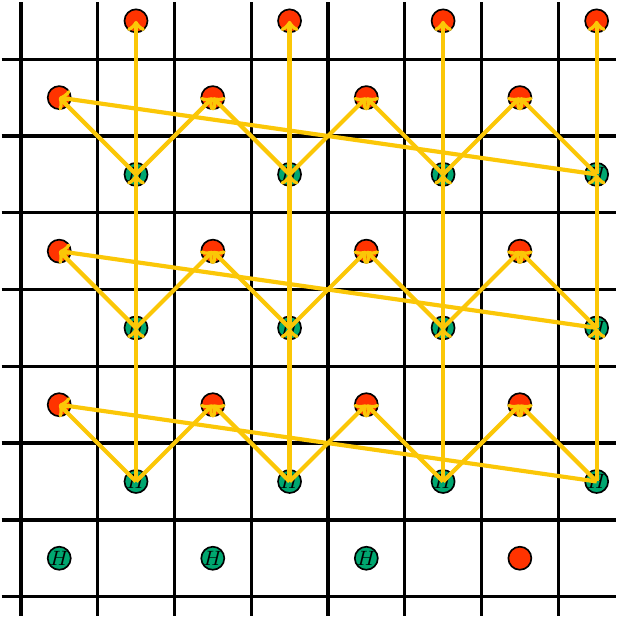}\\
    \vspace{0.5cm}
    \includegraphics[width=0.27\linewidth]{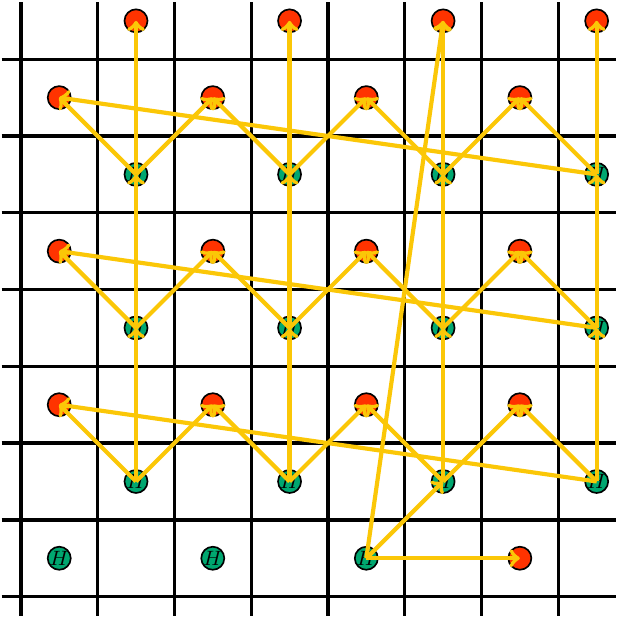}
   $\quad$
    \includegraphics[width=0.27\linewidth]{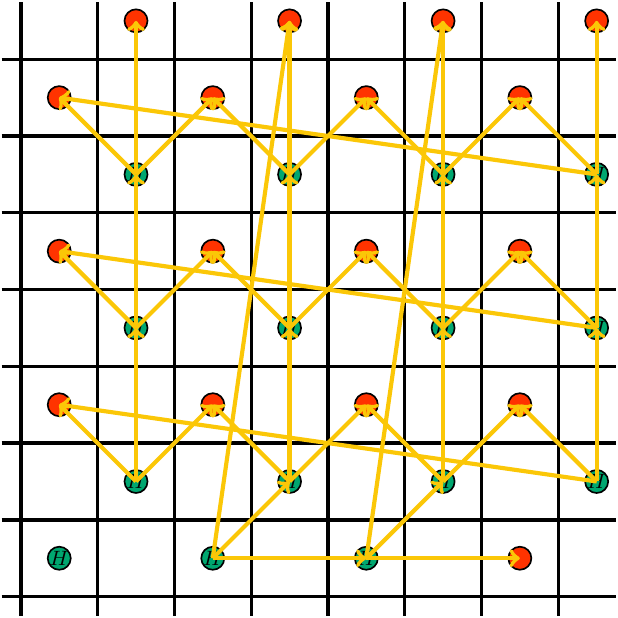}
   $\quad$
    \includegraphics[width=0.27\linewidth]{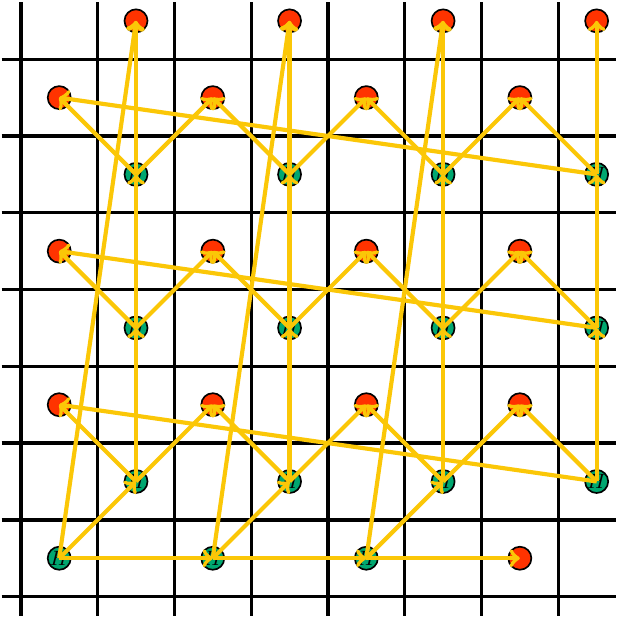}\\
    \caption{The different steps in the toric code state preparation in size $8\times 8$, ordered in reading direction. The circles represent the ancillas, with green circles representing those where a Hadamard gate is applied at the beginning. Then the yellow arrows indicate the application of CNOTs, with the arrow pointing towards the target qubit. The system qubits represented by blue circles in Fig \ref{fig:freefermion} are not shown in this picture.}
    \label{fig:toriccodestateprep}
\end{figure*}

In Fig \ref{fig:toriccodestateprep} we represent graphically the toric code state preparation used in Section \ref{sec:benchmarktrotter}, for the case $L_x=L_y=8$. The toric code state preparation for other even dimensions $L_x,L_y$ is readily deduced from Fig \ref{fig:toriccodestateprep}.

\section{Free fermion calculations}\label{app:free}
\subsection{Generalities}
We consider the free fermion Hamiltonian
\begin{equation}
    H=\sum_{\langle i,j\rangle}c_i^\dagger c_j+c_j^\dagger c_i
\end{equation}
where the sum runs over the edges of a $L_x\times L_y$ square lattice with periodic boundary conditions. After a Jordan-Wigner transformation, the term $c_i^\dagger c_j+c_j^\dagger c_i$ becomes
\begin{equation}
    c_i^\dagger c_j+c_j^\dagger c_i=\frac{1}{2}(X_iX_j+Y_iY_j)\prod_{i\ll k\ll j}Z_k\,,
\end{equation}
and we have
\begin{equation}
    c_i c_j-c_i^\dagger c_j^\dagger=\frac{1}{2}(X_iX_j-Y_iY_j)\prod_{i\ll k\ll j}Z_k\,,
\end{equation}
where $i\ll k\ll j$ means that site $k$ is comprised between sites $i$ and $j$ in a given ordering of all the sites. We decompose
\begin{equation}
    H=H_{\up,X}+H_{\up,Y}+H_{\rightarrow,X}+H_{\rightarrow,Y}\,,
\end{equation}
 with
 \begin{equation}
     H_{\up,X}=\frac{1}{2}\sum_{\substack{\langle i,j\rangle \\ {\rm vertical}}} X_iX_j\prod_{i\ll k\ll j}Z_k\,,
 \end{equation}
 and identically for $\rightarrow$ meaning that $\langle i,j\rangle$ is a horizontal bond, and with $H_{\up,Y}$ being the same but with $Y_iY_j$ instead of $X_iX_j$. In terms of fermions, we have
 \begin{equation}
     H_{\rightarrow,X/Y}=\frac{1}{2}\sum_{\substack{\langle i,j\rangle \\ {\rm horizontal}}} c_i^\dagger c_j+c_j^\dagger c_i\pm (c_ic_j-c_i^\dagger c_j^\dagger)\,.
 \end{equation}
\subsection{Time evolution of Fourier modes}
 We rewrite this Hamiltonian with the Fourier transform
 \begin{equation}
     c_j=\frac{1}{\sqrt{L}}\sum_{k\in K}c(k) e^{ijk}\,,
 \end{equation}
 where $K=\{\frac{2\pi (k_x,k_y)}{L_x},k_{x,y}=0,...,L_{x,y}-1\}$. In this expression, we see the site $j$ as a couple $j=(j_x,j_y)$ with $j_{x,y}=0,...,L_{x,y}-1$ and the scalar product defined as $jk=j_xk_x+j_yk_y$. 
 This yields
 \begin{equation}
 \begin{aligned}
H_{\rightarrow,X/Y}=\sum_{k\in K}& c^\dagger(k) c(k)\cos k_x \\
&\pm\frac{i}{2}\sin k_x(c^\dagger(k) c^\dagger(-k)-c(k)c(-k))\,.
 \end{aligned}
 \end{equation}
 We have
 \begin{equation}
     \begin{aligned}
&[H_{\rightarrow,X/Y},c(k)]=-\cos k_x c(k)\mp i\sin k_x c^\dagger(-k)\\
&[H_{\rightarrow,X/Y},c^\dagger(-k)]=\cos k_x c^\dagger(-k)\pm i\sin k_x c(k)\,.
     \end{aligned}
 \end{equation}
 Hence we have the evolution equation under $H_{\rightarrow,X/Y}$
 \begin{equation}
     \partial_t \left(\begin{matrix}
         c(k)\\c^\dagger(-k)
     \end{matrix}\right)=\left(\begin{matrix}
         -i\cos k_x & \mp \sin k_x\\
        \pm \sin k_x & i\cos k_x
     \end{matrix} \right)\left(\begin{matrix}
         c(k)\\c^\dagger(-k)
     \end{matrix}\right)\,.
 \end{equation}
 Let us perform this evolution for a time $\delta t$ for $H_{\rightarrow,X}$, and then for a time $\delta t$ for $H_{\rightarrow,Y}$. Using a symbolic software, we find that the new vector after this evolution is
 \begin{equation}
   U_{k_x}\left(\begin{matrix}
         c(k)\\c^\dagger(-k)
     \end{matrix}\right)\,,
 \end{equation}
 with
\begin{widetext}
\begin{equation}
     U_k=\left(\begin{matrix}
      1-2\sin^2(\delta t)\cos^2 k-i\sin(2\delta t)\cos k   & i\sin^2(\delta t)\sin(2k) \\
       i\sin^2(\delta t)\sin(2k)  &1-2\sin^2(\delta t)\cos^2 k+i\sin(2\delta t)\cos k   
     \end{matrix}\right)\,.
 \end{equation}
\end{widetext}
The same equations hold true for $H_{\up,X/Y}$, with $k_x$ replaced by $k_y$. It follows that after a full Trotter step, the operators $c(k),c^\dagger(-k)$ are mapped to
\begin{equation}
    \left(\begin{matrix}
        c(k)\\c^\dagger(-k)
    \end{matrix}\right)\mapsto U_{k_y}U_{k_x} \left(\begin{matrix}
        c(k)\\c^\dagger(-k)
    \end{matrix}\right)\,.
\end{equation}

%  The eigenvalues of that matrix are found to be $e^{\pm i\delta t\epsilon_k}$ with
%  \begin{equation}
%      \cos(\delta t \epsilon_k)=1-2\sin^2(\delta t)\cos^2(k_x)\,,
%  \end{equation}
% and eigenvectors
% \begin{equation}
%     \begin{aligned}
%         \gamma_x(k)&=\cos\theta_{k_x} c(k)+\sin\theta_{k_x} c^\dagger(-k)\\
%         \bar{\gamma}_x(k)&=\sin\theta_{k_x} c(k)-\cos\theta_{k_x} c^\dagger(-k)\,,
%     \end{aligned}
% \end{equation}
% with
% \begin{equation}
%     \tan\theta_{k}=\sign(k)\sqrt{\frac{\sqrt{1-\sin^2(\delta t)\cos^2(k)}-\cos (\delta t)}{\sqrt{1-\sin^2(\delta t)\cos^2(k)}+\cos (\delta t)}}\,.
% \end{equation}

% Hence, after $n$ evolutions, we can write
% \begin{equation}
% \begin{aligned}
%     c(k)(t=n\delta t) =& (\cos^2(\theta_k) e^{it\epsilon_k}+\sin^2(\theta_k)e^{-it\epsilon_k} )c(k)\\ 
% &+i\sin(2\theta_k) \sin(t\epsilon_k)c^\dagger(-k)\,.
% \end{aligned}
% \end{equation}

\subsection{Observable with two fermions}
Let us now consider an observable of the form
\begin{equation}
    \mathcal{O}=\sum_{j}f_jc_j^\dagger c_j\,,
\end{equation}
with $f_j$ some function of the site $j$. We have
\begin{equation}
    \mathcal{O}=\sum_{k,q\in K}\hat{f}(k-q)c^\dagger(k)c(q)\,,
\end{equation}
with
\begin{equation}
    \hat{f}(k)=\frac{1}{L}\sum_{j}e^{ijk}f_j\,.
\end{equation}
After application of $n$ Trotter steps, let us write the decomposition
\begin{equation}
   ( U_{k_y}U_{k_x})^n =\left(\begin{matrix}
       \alpha_n(k) & \beta_n(k)\\
       -\beta_n^*(k) & \alpha_n^*(k)
   \end{matrix}\right)\,,
\end{equation}
with $\alpha_n(k),\beta_n(k)$ coefficients. Writing explicitly the unitary matrix $U_{k_x}U_{k_y}$, one finds that its eigenvalues are $e^{\pm i\epsilon_k}$ with
\begin{equation}
\begin{aligned}
    \epsilon_k=&\sign(\cos k) \arccos\Big[1-2\sin^2(\delta t)(\cos k+\cos q)^2\\
    &+4\sin^4(\delta t)\cos k\cos q(1+\cos(k+q)) \Big]\,.
\end{aligned}
\end{equation}
One then knows that the coefficients $\alpha_n(k),\beta_n(k)$ are a linear combination of $e^{in\epsilon_k}$ and $e^{-in\epsilon_k}$. From the cases $n=0,n=1$ one finds then
\begin{widetext}
    \begin{equation}
    \begin{aligned}
        &\alpha_n(k)=e^{-in\epsilon_k}+i\Big(-\sin(2\delta t)(\cos k_x+\cos k_y)+2\sin(2\delta t)\sin^2(\delta t)\cos k_x \cos k_y(\cos k_x+\cos k_y)+\sin\epsilon_k\Big)\frac{\sin(n\epsilon_k)}{\sin\epsilon_k}\\
        &\beta_n(k)=\Big(i\sin^2(\delta t)(\sin(2k_x)+\sin(2k_y))-2i\sin^4(\delta t)(\cos^2(k_x)\sin(2k_y)+\cos^2(k_y)\sin(2k_x))\\
        &\qquad\qquad\qquad\qquad+\sin^2(\delta t)\sin(2\delta t)(\cos(k_x)\sin(2k_y)+\cos(k_y)\sin(2k_x))\Big)\frac{\sin(n\epsilon_k)}{\sin\epsilon_k}\,.
    \end{aligned}
\end{equation}
\end{widetext}

We thus have after $n$ Trotter steps
\begin{equation}
\begin{aligned}\label{ondt}
    \mathcal{O}&(n\delta t)=\sum_{k,q\in K}\hat{f}(k-q)\Big[\alpha^*_n(k) \alpha_n(q) c^\dagger(k)c(q)\\
    &+\beta^*_n(k) \alpha_n(q) c(-k)c(q)+\alpha^*_n(k)\beta_n(q) c^\dagger(k)c^\dagger(-q)\\
    &+\beta_n^*(k)\beta_n(q)c(-k)c^\dagger(-q)\Big]\,.
\end{aligned}
\end{equation}
Let us evaluate it in a product state with mode occupation $n_j$ on site $j$. Introducing
\begin{equation}
    \hat{n}(k)=\frac{1}{L}\sum_j e^{ijk}n_j\,,
\end{equation}
we get that $\langle c^\dagger(k) c(q)\rangle=\hat{n}(k-q)$,  $\langle c(-k) c(q)\rangle=0$, $\langle c^\dagger(k) c^\dagger(-q)\rangle=0$ and $\langle c(-k) c^\dagger(-q)\rangle=\delta_{k,q}-\hat{n}(k-q)$. Hence
\begin{equation}
\begin{aligned}
    &\langle \mathcal{O}(n\delta t)\rangle=\\
    &\sum_{k,q\in K}\hat{f}(k-q)(\alpha_n^*(k) \alpha_n(q)-\beta^*_n(-k) \beta_n(-q)) \hat{n}(k-q)\\
    &+\hat{f}(0)\sum_{k\in K}|\beta_n(k)|^2\,.
\end{aligned}
\end{equation}

% \subsection{Observable with four fermions}
% Let us now consider the square of the observable $\mathcal{O}^2$. We would like to evaluate the expectation value of $ \mathcal{O}(n\delta t)^2$ after $n$ Trotter steps. This involves computing expectation values of observables with four fermions. Using Wick's theorem, we have
% \begin{equation}
%     \begin{aligned}
%         \langle c^\dagger(k)c(q)c^\dagger(k')c(q') \rangle&=\hat{n}(k-q)\hat{n}(k'-q')\\
%         \langle c^\dagger(k)c(q)c(-k')c^\dagger(-q') \rangle&=\hat{n}(k-q)\hat{n}(k'-q')\\
%     \end{aligned}
% \end{equation}

\subsection{Higher-weight observables \label{app:higherweight}}
We now consider the higher-weight observables $\mathcal{O}_{[w]}$ defined in \eqref{ow}. The observable is exactly the coefficient in front of $\epsilon^w$ in the Taylor expansion of
\begin{equation}
    F(\epsilon)=(1+\epsilon f_1Z_1)(1+\epsilon f_2 Z_2)...(1+\epsilon f_LZ_L)\,.
\end{equation}
Let us write
\begin{equation}
    \begin{aligned}
        F(\epsilon)&=\exp\left(\sum_{j=1}^L \log(1+\epsilon f_jZ_j)\right)\\
&=\exp\left(\frac{1}{2}\sum_{j=1}^L \log(1-\epsilon^2 f_j^2)+\log \frac{1+\epsilon f_j}{1-\epsilon f_j}Z_j\right)\,.
    \end{aligned}
\end{equation}
Introducing
\begin{equation}
    S_{2w}=\sum_{j=1}^L f_j^{2w}\,, \qquad S_{2w+1}=\sum_{j=1}^L f_j^{2w+1}Z_j\,,
\end{equation}
we have
\begin{equation}
    F(\epsilon)=\exp\left(- \sum_{w\geq 1}S_w\frac{(-\epsilon)^{w}}{w}\right)\,.
\end{equation}
From this expression, the observable $\mathcal{O}_{[w]}$ \eqref{ow} for all $w$ can be expressed in terms of the $S_{w}$'s. For example, the first few terms are
\begin{equation}
    \begin{aligned}
        \mathcal{O}_{[1]}&=S_1\\
        \mathcal{O}_{[2]}&=\frac{S_1^2-S_2}{2}\\
        \mathcal{O}_{[3]}&=\frac{S_1^3-3S_1S_2+2S_3}{6}\\
        \mathcal{O}_{[4]}&=\frac{S_1^4-6S_1^2S_2+3S_2^2+8S_1S_3-6S_4}{24}\,.
    \end{aligned}
\end{equation}
The problem of computing $\mathcal{O}_{[w]}$ is thus reduced to that of computing the powers $S_w^p$. When $w$ is even, this is only a scalar. When $w$ is odd, in terms of the fermions, this can written as
\begin{equation}
   S_{2w+1}^p=\left(\sum_{j=1}^L (1-2c^\dagger_jc_j)f_j^{2w+1}\right)^p\,.
\end{equation}
We expand it as
\begin{equation}
    S_{2w+1}^p=\sum_{q=0}^p {p\choose q} (-2)^q \tilde{S}^q_{2w+1} \Sigma_{2w+1}^{p-q}\,,
\end{equation}
with
\begin{equation}
  \tilde{S}_{2w+1}=\sum_{j=1}^L f_j^{2w+1} c_j^\dagger c_j\,,\qquad \Sigma_{2w+1}=\sum_{j=1}^Lf_j^{2w+1}\,.
\end{equation}
After $n$ Trotter steps, using \eqref{ondt} with $\hat{f}$ replaced by the Fourier transform of $f^{2w+1}$, the powers $\tilde{S}_{2w+1}^p$ are expressed as sums of terms of the type
\begin{equation}
c^\dagger(k_{i_1})c(k_{j_1})...c^\dagger(k_{i_p})c(k_{j_p})\,,
\end{equation}
as well as with any $c$ replaced by $c^\dagger$ and conversely. The expectation value of these expressions are computed using Wick's theorem. Namely we have the recursive formula for any $m$
\begin{equation}
\begin{aligned}
    \langle\bar{c}(k_1)...\bar{c}(k_{2m})\rangle&=\sum_{i=2}^{2m}(-1)^{i-1}\langle \bar{c}(k_1)\bar{c}(k_i)\rangle \times\\
    &\langle \bar{c}(k_2)...\bar{c}(k_{i-1})\bar{c}(k_{i+1})...\bar{c}(k_{2m})\rangle\,,
\end{aligned}
\end{equation}
where by $\bar{c}$ we mean any of $c$ or $c^\dagger$. 

% The computational cost of computing $\langle \mathcal{O}_{[w]}\rangle$ is exponential in the weight $w$ of the operator, but polynomial in the number of sites $L$.

\section{Hydrogen chain Hamiltonians \label{app:hydrogen}}
In this Appendix we provide the Pauli string decomposition of the hydrogen chains implemented in the benchmark.

\subsection{$N=4$}
\begin{center}
     \begin{tabular}{c|c}
ZIII &0.1714128264477691\\
IZII& 0.17141282644776906\\
IIZI& -0.2234315369081344\\
IIIZ& -0.2234315369081344\\
ZZII& 0.0027611313659086645\\
ZIZI& -0.04530261550379926\\
IZIZ& -0.04530261550379926\\
IIZZ& 0.008485025784912364\\
XXYY& -0.04530261550379926\\
XYYX& 0.04530261550379926\\
YXXY& 0.04530261550379926\\
YYXX& -0.04530261550379926
\end{tabular}
\end{center}

\subsection{$N=6$}
\begin{center}
     \begin{tabular}{c|c}
ZIIIII& 0.21618381471527334\\
IZIIII& 0.21618381471527331\\
IIZIII& -0.008325684680054873\\
IIIZII& -0.008325684680054887\\
IIIIZI& -0.4600463793181071\\
IIIIIZ& -0.4600463793181071\\
ZZIIII& 0.021554650579316437\\
ZIZIII& -0.03633262001772418\\
XZZZXI& 0.02373733926074964\\
YZZZYI& 0.02373733926074964\\
ZIIIZI& -0.008397304018527951\\
ZIIIIZ& 0.02371353325802969\\
\end{tabular}
\end{center}
\begin{center}
     \begin{tabular}{c|c}
IZIZII& -0.03633262001772418\\
IZIIZI& 0.02371353325802969\\
IXZZZX& 0.023737339260749637\\
IYZZZY& 0.023737339260749637\\
IZIIIZ& -0.008397304018527951\\
IIZZII& 0.009130923176207006\\
IIZIZI& -0.02937805273886611\\
IIZIIZ& 0.008648868840876262\\
IIIZZI& 0.008648868840876262\\
IIIZIZ& -0.02937805273886611\\
IIIIZZ& 0.037696007424849604\\
ZXZZZX& -0.02695362135916983\\
\end{tabular}
\end{center}
\begin{center}
     \begin{tabular}{c|c}
ZYZZZY& -0.02695362135916983\\
XIZZXI& -0.02695362135916983\\
YIZZYI& -0.02695362135916983\\
XXYYII& -0.03633262001772418\\
XYYXII& 0.03633262001772418\\
YXXYII& 0.03633262001772418\\
YYXXII& -0.03633262001772418\\
XXIIYY& -0.03211083727655765\\
XYIIYX& 0.03211083727655765\\
YXIIXY& 0.03211083727655765\\
YYIIXX& -0.03211083727655765\\
XZIZXI& -0.02975613461866213\\
\end{tabular}
\end{center}
\begin{center}
     \begin{tabular}{c|c}
YZIZYI& -0.02975613461866213\\
XZXXZX &-0.03330235266776721\\
XZXYZY& -0.03330235266776721\\
YZYXZX& -0.03330235266776721\\
YZYYZY& -0.03330235266776721\\
XZZIXI& 0.003546218049105083\\
YZZIYI& 0.003546218049105083\\
XZZZXZ& -0.024481114159709084\\
YZZZYZ& -0.024481114159709084\\
IXIZZX& 0.003546218049105083\\
IYIZZY& 0.003546218049105083\\
IXXYYI& 0.03330235266776721\\
\end{tabular}
\end{center}
\begin{center}
     \begin{tabular}{c|c}
IXYYXI& -0.03330235266776721\\
IYXXYI& -0.03330235266776721\\
IYYXXI& 0.03330235266776721\\
IXZIZX& -0.02975613461866213\\
IYZIZY& -0.02975613461866213\\
IXZZIX& -0.024481114159709084\\
IYZZIY& -0.024481114159709084\\
IIXXYY& -0.03802692157974238\\
IIXYYX& 0.03802692157974238\\
IIYXXY& 0.03802692157974238\\
IIYYXX& -0.03802692157974238\\
\end{tabular}
\end{center}

\subsection{$N=8$}

\begin{center}
     \begin{tabular}{c|c}
ZIIIIIII& 0.23402690958875838\\
IZIIIIII& 0.23402690958875838\\
IIZIIIII& 0.0878497543264086\\
IIIZIIII& 0.0878497543264086\\
IIIIZIII& -0.17401158373028885\\
IIIIIZII& -0.1740115837302888\\
IIIIIIZI& -0.641779436923978\\
IIIIIIIZ& -0.641779436923978\\
ZZIIIIII& 0.013411568108160798\\
ZIZIIIII& -0.04308333056495532\\
ZIIZIIII& -0.004360265661550927\\
XZZZXIII& 0.0022482895325020465\\
\end{tabular}
\end{center}
\begin{center}
     \begin{tabular}{c|c}
YZZZYIII& 0.0022482895325020465\\
ZIIIZIII& -0.0263239441653568\\
ZIIIIZII& 0.0004336370049061733\\
ZIIIIIIZ& 0.023256526056827265\\
IZZIIIII& -0.004360265661550927\\
IZIZIIII& -0.04308333056495532\\
IZIIZIII& 0.0004336370049061733\\
IXZZZXII& 0.0022482895325020395\\
IYZZZYII& 0.0022482895325020395\\
IZIIIZII& -0.0263239441653568\\
IZIIIIZI& 0.023256526056827265\\
IIZZIIII& 0.00016350215060983997\\
\end{tabular}
\end{center}
\begin{center}
     \begin{tabular}{c|c}
IIZIZIII& -0.0361290527748267\\
IIZIIZII& -0.0013524195353956658\\
IIXZZZXI& -0.02396872391879997\\
IIYZZZYI& -0.02396872391879997\\
IIZIIIZI& -0.019313058428047147\\
IIZIIIIZ& 0.0058975739553504825\\
IIIZZIII& -0.0013524195353956658\\
IIIZIZII& -0.0361290527748267\\
IIIZIIZI& 0.0058975739553504825\\
IIIXZZZX& -0.02396872391879997\\
IIIYZZZY& -0.02396872391879997\\
IIIZIIIZ& -0.019313058428047147\\
\end{tabular}
\end{center}
\begin{center}
     \begin{tabular}{c|c}
IIIIZZII& 0.005666441733850391\\
IIIIZIZI& -0.02622140176373078\\
IIIIZIIZ& 0.013052719406888458\\
IIIIIZZI& 0.013052719406888458\\
IIIIIZIZ& -0.02622140176373078\\
IIIIIIZZ& 0.04615408116019942\\
ZXZZZXII& 0.02352725017982058\\
ZYZZZYII& 0.02352725017982058\\
XIZZXIII& 0.02352725017982058\\
YIZZYIII& 0.02352725017982058\\
XXYYIIII& -0.03872306490340441\\
XYYXIIII& 0.03872306490340441\\
\end{tabular}
\end{center}
\begin{center}
     \begin{tabular}{c|c}
YXXYIIII& 0.03872306490340441\\
YYXXIIII& -0.03872306490340441\\
XXYZZZZY& 0.012136462148764519\\
XYYZZZZX& -0.012136462148764519\\
YXXZZZZY& -0.012136462148764519\\
YYXZZZZX& 0.012136462148764519\\
XXIXZZXI& 0.012136462148764519\\
XYIYZZXI& 0.012136462148764519\\
YXIXZZYI& 0.012136462148764519\\
YYIYZZYI& 0.012136462148764519\\
XXIIYYII& -0.026757581170262966\\
XYIIYXII& 0.026757581170262966\\
\end{tabular}
\end{center}
\begin{center}
     \begin{tabular}{c|c}
YXIIXYII& 0.026757581170262966\\
YYIIXXII& -0.026757581170262966\\
XXIIIIYY& -0.02325652605682727\\
XYIIIIYX& 0.02325652605682727\\
YXIIIIXY& 0.02325652605682727\\
YYIIIIXX& -0.02325652605682727\\
ZIXZZZXI& 0.012327633512530762\\
ZIYZZZYI& 0.012327633512530762\\
XZIZXIII& 0.02594262738030117\\
YZIZYIII& 0.02594262738030117\\
XZXIXZXI& 0.02650309648399575\\
XZXIYZYI& 0.012727232827558596\\
\end{tabular}
\end{center}
\begin{center}
     \begin{tabular}{c|c}
XZYIYZXI& 0.013775863656437154\\
YZXIXZYI& 0.013775863656437154\\
YZYIXZXI& 0.012727232827558596\\
YZYIYZYI& 0.02650309648399575\\
ZIIXZZZX& 0.024464095661295277\\
ZIIYZZZY& 0.024464095661295277\\
XZXXZXII& 0.0264506721239868\\
XZXYZYII& 0.0264506721239868\\
YZYXZXII& 0.0264506721239868\\
YZYYZYII& 0.0264506721239868\\
XZZIXIII& -0.0005080447436856309\\
YZZIYIII& -0.0005080447436856309\\
\end{tabular}
\end{center}
\begin{center}
     \begin{tabular}{c|c}
XZZXYZZY& -0.02320745867088144\\
XZZYYZZX &0.02320745867088144\\
YZZXXZZY &0.02320745867088144\\
YZZYXZZX &-0.02320745867088144\\
XZZXIXXI &-0.009431595014444284\\
XZZYIYXI &-0.009431595014444284\\
YZZXIXYI &-0.009431595014444284\\
YZZYIYYI &-0.009431595014444284\\
XZXIIXZX &0.03593469149844004\\
XZXIIYZY &0.03593469149844004\\
YZYIIXZX &0.03593469149844004\\
YZYIIYZY &0.03593469149844004\\
\end{tabular}
\end{center}
\begin{center}
     \begin{tabular}{c|c}
XZZZXZII& 0.006475016715462559\\
YZZZYZII &0.006475016715462559\\
XZZZZXYY &0.011706349170573015\\
XZZZZYYX &-0.011706349170573015\\
YZZZZXXY &-0.011706349170573015\\
YZZZZYXX &0.011706349170573015\\
XZZZXIZI &0.01426609323822967\\
YZZZYIZI &0.01426609323822967\\
XZZZXIIZ &0.025972442408802685\\
YZZZYIIZ &0.025972442408802685\\
IZXZZZXI &0.024464095661295277\\
IZYZZZYI &0.024464095661295277\\
\end{tabular}
\end{center}
\begin{center}
     \begin{tabular}{c|c}
IXIZZXII& -0.0005080447436856309\\
IYIZZYII &-0.0005080447436856309\\
IXXYYIII &-0.0264506721239868\\
IXYYXIII &0.0264506721239868\\
IYXXYIII &0.0264506721239868\\
IYYXXIII &-0.0264506721239868\\
IXXIXZZX &-0.009431595014444284\\
IXYIYZZX &-0.009431595014444284\\
IYXIXZZY &-0.009431595014444284\\
IYYIYZZY &-0.009431595014444284\\
IXXIIYYI &-0.02320745867088144\\
IXYIIYXI &0.02320745867088144\\
\end{tabular}
\end{center}
\begin{center}
     \begin{tabular}{c|c}
IYXIIXYI& 0.02320745867088144\\
IYYIIXXI& -0.02320745867088144\\
IZIXZZZX& 0.012327633512530762\\
IZIYZZZY& 0.012327633512530762\\
IXZIZXII& 0.02594262738030117\\
IYZIZYII& 0.02594262738030117\\
IXZXIXZX& 0.02650309648399575\\
IXZXIYZY& 0.012727232827558596\\
IXZYIYZX& 0.013775863656437154\\
IYZXIXZY& 0.013775863656437154\\
IYZYIXZX& 0.012727232827558596\\
IYZYIYZY& 0.02650309648399575\\
\end{tabular}
\end{center}
\begin{center}
     \begin{tabular}{c|c}
IXZXXZXI& 0.03593469149844004\\
IXZXYZYI& 0.03593469149844004\\
IYZYXZXI& 0.03593469149844004\\
IYZYYZYI& 0.03593469149844004\\
IXZZIXII& 0.006475016715462559\\
IYZZIYII& 0.006475016715462559\\
IXZZXIXX& 0.011706349170573015\\
IXZZYIYX& 0.011706349170573015\\
IYZZXIXY& 0.011706349170573015\\
IYZZYIYY& 0.011706349170573015\\
IXZZZXZI& 0.025972442408802685\\
IYZZZYZI& 0.025972442408802685\\
\end{tabular}
\end{center}
\begin{center}
     \begin{tabular}{c|c}
IXZZZXIZ& 0.01426609323822967\\
IYZZZYIZ& 0.01426609323822967\\
IIZXZZZX& 0.004454413742177123\\
IIZYZZZY& 0.004454413742177123\\
IIXIZZXI& 0.004454413742177123\\
IIYIZZYI& 0.004454413742177123\\
IIXXYYII& -0.03477663323943102\\
IIXYYXII& 0.03477663323943102\\
IIYXXYII& 0.03477663323943102\\
IIYYXXII& -0.03477663323943102\\
IIXXIIYY& -0.025210632383397637\\
IIXYIIYX& 0.025210632383397637\\
\end{tabular}
\end{center}
\begin{center}
     \begin{tabular}{c|c}
IIYXIIXY& 0.025210632383397637\\
IIYYIIXX& -0.025210632383397637\\
IIXZIZXI& 0.031157773938421854\\
IIYZIZYI& 0.031157773938421854\\
IIXZXXZX& 0.025863272971678446\\
IIXZXYZY& 0.025863272971678446\\
IIYZYXZX& 0.025863272971678446\\
IIYZYYZY& 0.025863272971678446\\
IIXZZIXI& 0.00529450096674341\\
IIYZZIYI& 0.00529450096674341\\
IIXZZZXZ& 0.028762584646736263\\
IIYZZZYZ& 0.028762584646736263\\
\end{tabular}
\end{center}
\begin{center}
     \begin{tabular}{c|c}
IIIXIZZX& 0.00529450096674341\\
IIIYIZZY& 0.00529450096674341\\
IIIXXYYI& -0.025863272971678446\\
IIIXYYXI& 0.025863272971678446\\
IIIYXXYI& 0.025863272971678446\\
IIIYYXXI& -0.025863272971678446\\
IIIXZIZX& 0.031157773938421854\\
IIIYZIZY& 0.031157773938421854\\
IIIXZZIX& 0.028762584646736263\\
IIIYZZIY& 0.028762584646736263\\
IIIIXXYY& -0.03927412117061923\\
IIIIXYYX& 0.03927412117061923\\
IIIIYXXY& 0.03927412117061923\\
IIIIYYXX& -0.03927412117061923
\end{tabular}
\end{center}


\begin{thebibliography}{50}%
\makeatletter
\providecommand \@ifxundefined [1]{%
 \@ifx{#1\undefined}
}%
\providecommand \@ifnum [1]{%
 \ifnum #1\expandafter \@firstoftwo
 \else \expandafter \@secondoftwo
 \fi
}%
\providecommand \@ifx [1]{%
 \ifx #1\expandafter \@firstoftwo
 \else \expandafter \@secondoftwo
 \fi
}%
\providecommand \natexlab [1]{#1}%
\providecommand \enquote  [1]{``#1''}%
\providecommand \bibnamefont  [1]{#1}%
\providecommand \bibfnamefont [1]{#1}%
\providecommand \citenamefont [1]{#1}%
\providecommand \href@noop [0]{\@secondoftwo}%
\providecommand \href [0]{\begingroup \@sanitize@url \@href}%
\providecommand \@href[1]{\@@startlink{#1}\@@href}%
\providecommand \@@href[1]{\endgroup#1\@@endlink}%
\providecommand \@sanitize@url [0]{\catcode `\\12\catcode `\$12\catcode
  `\&12\catcode `\#12\catcode `\^12\catcode `\_12\catcode `\%12\relax}%
\providecommand \@@startlink[1]{}%
\providecommand \@@endlink[0]{}%
\providecommand \url  [0]{\begingroup\@sanitize@url \@url }%
\providecommand \@url [1]{\endgroup\@href {#1}{\urlprefix }}%
\providecommand \urlprefix  [0]{URL }%
\providecommand \Eprint [0]{\href }%
\providecommand \doibase [0]{http://dx.doi.org/}%
\providecommand \selectlanguage [0]{\@gobble}%
\providecommand \bibinfo  [0]{\@secondoftwo}%
\providecommand \bibfield  [0]{\@secondoftwo}%
\providecommand \translation [1]{[#1]}%
\providecommand \BibitemOpen [0]{}%
\providecommand \bibitemStop [0]{}%
\providecommand \bibitemNoStop [0]{.\EOS\space}%
\providecommand \EOS [0]{\spacefactor3000\relax}%
\providecommand \BibitemShut  [1]{\csname bibitem#1\endcsname}%
\let\auto@bib@innerbib\@empty
%</preamble>
\bibitem [{\citenamefont {DeCross}\ \emph {et~al.}(2024)\citenamefont
  {DeCross}, \citenamefont {Haghshenas}, \citenamefont {Liu}, \citenamefont
  {Rinaldi}, \citenamefont {Gray}, \citenamefont {Alexeev}, \citenamefont
  {Baldwin}, \citenamefont {Bartolotta}, \citenamefont {Bohn}, \citenamefont
  {Chertkov} \emph {et~al.}}]{decross2024computational}%
  \BibitemOpen
  \bibfield  {author} {\bibinfo {author} {\bibfnamefont {M.}~\bibnamefont
  {DeCross}}, \bibinfo {author} {\bibfnamefont {R.}~\bibnamefont {Haghshenas}},
  \bibinfo {author} {\bibfnamefont {M.}~\bibnamefont {Liu}}, \bibinfo {author}
  {\bibfnamefont {E.}~\bibnamefont {Rinaldi}}, \bibinfo {author} {\bibfnamefont
  {J.}~\bibnamefont {Gray}}, \bibinfo {author} {\bibfnamefont {Y.}~\bibnamefont
  {Alexeev}}, \bibinfo {author} {\bibfnamefont {C.~H.}\ \bibnamefont
  {Baldwin}}, \bibinfo {author} {\bibfnamefont {J.~P.}\ \bibnamefont
  {Bartolotta}}, \bibinfo {author} {\bibfnamefont {M.}~\bibnamefont {Bohn}},
  \bibinfo {author} {\bibfnamefont {E.}~\bibnamefont {Chertkov}},  \emph
  {et~al.},\ }\href {\doibase 10.48550/arXiv.2406.02501} {\bibfield  {journal}
  {\bibinfo  {journal} {arXiv preprint arXiv:2406.02501}\ } (\bibinfo {year}
  {2024}),\ 10.48550/arXiv.2406.02501}\BibitemShut {NoStop}%
\bibitem [{\citenamefont {Bluvstein}\ \emph {et~al.}(2024)\citenamefont
  {Bluvstein}, \citenamefont {Evered}, \citenamefont {Geim}, \citenamefont
  {Li}, \citenamefont {Zhou}, \citenamefont {Manovitz}, \citenamefont {Ebadi},
  \citenamefont {Cain}, \citenamefont {Kalinowski}, \citenamefont {Hangleiter}
  \emph {et~al.}}]{bluvstein2024logical}%
  \BibitemOpen
  \bibfield  {author} {\bibinfo {author} {\bibfnamefont {D.}~\bibnamefont
  {Bluvstein}}, \bibinfo {author} {\bibfnamefont {S.~J.}\ \bibnamefont
  {Evered}}, \bibinfo {author} {\bibfnamefont {A.~A.}\ \bibnamefont {Geim}},
  \bibinfo {author} {\bibfnamefont {S.~H.}\ \bibnamefont {Li}}, \bibinfo
  {author} {\bibfnamefont {H.}~\bibnamefont {Zhou}}, \bibinfo {author}
  {\bibfnamefont {T.}~\bibnamefont {Manovitz}}, \bibinfo {author}
  {\bibfnamefont {S.}~\bibnamefont {Ebadi}}, \bibinfo {author} {\bibfnamefont
  {M.}~\bibnamefont {Cain}}, \bibinfo {author} {\bibfnamefont {M.}~\bibnamefont
  {Kalinowski}}, \bibinfo {author} {\bibfnamefont {D.}~\bibnamefont
  {Hangleiter}},  \emph {et~al.},\ }\href {\doibase 10.1038/s41586-023-06927-3}
  {\bibfield  {journal} {\bibinfo  {journal} {Nature}\ }\textbf {\bibinfo
  {volume} {626}},\ \bibinfo {pages} {58} (\bibinfo {year} {2024})}\BibitemShut
  {NoStop}%
\bibitem [{\citenamefont {Kim}\ \emph {et~al.}(2023)\citenamefont {Kim},
  \citenamefont {Eddins}, \citenamefont {Anand}, \citenamefont {Wei},
  \citenamefont {Van Den~Berg}, \citenamefont {Rosenblatt}, \citenamefont
  {Nayfeh}, \citenamefont {Wu}, \citenamefont {Zaletel}, \citenamefont {Temme}
  \emph {et~al.}}]{kim2023evidence}%
  \BibitemOpen
  \bibfield  {author} {\bibinfo {author} {\bibfnamefont {Y.}~\bibnamefont
  {Kim}}, \bibinfo {author} {\bibfnamefont {A.}~\bibnamefont {Eddins}},
  \bibinfo {author} {\bibfnamefont {S.}~\bibnamefont {Anand}}, \bibinfo
  {author} {\bibfnamefont {K.~X.}\ \bibnamefont {Wei}}, \bibinfo {author}
  {\bibfnamefont {E.}~\bibnamefont {Van Den~Berg}}, \bibinfo {author}
  {\bibfnamefont {S.}~\bibnamefont {Rosenblatt}}, \bibinfo {author}
  {\bibfnamefont {H.}~\bibnamefont {Nayfeh}}, \bibinfo {author} {\bibfnamefont
  {Y.}~\bibnamefont {Wu}}, \bibinfo {author} {\bibfnamefont {M.}~\bibnamefont
  {Zaletel}}, \bibinfo {author} {\bibfnamefont {K.}~\bibnamefont {Temme}},
  \emph {et~al.},\ }\href {\doibase 10.1038/s41586-023-06096-3} {\bibfield
  {journal} {\bibinfo  {journal} {Nature}\ }\textbf {\bibinfo {volume} {618}},\
  \bibinfo {pages} {500} (\bibinfo {year} {2023})}\BibitemShut {NoStop}%
\bibitem [{\citenamefont {Foss-Feig}\ \emph {et~al.}(2023)\citenamefont
  {Foss-Feig}, \citenamefont {Tikku}, \citenamefont {Lu}, \citenamefont
  {Mayer}, \citenamefont {Iqbal}, \citenamefont {Gatterman}, \citenamefont
  {Gresh}, \citenamefont {Hankin}, \citenamefont {Hewitt}, \citenamefont
  {Horst} \emph {et~al.}}]{foss-feig2023demonstration}%
  \BibitemOpen
  \bibfield  {author} {\bibinfo {author} {\bibfnamefont {M.}~\bibnamefont
  {Foss-Feig}}, \bibinfo {author} {\bibfnamefont {A.}~\bibnamefont {Tikku}},
  \bibinfo {author} {\bibfnamefont {T.-C.}\ \bibnamefont {Lu}}, \bibinfo
  {author} {\bibfnamefont {K.}~\bibnamefont {Mayer}}, \bibinfo {author}
  {\bibfnamefont {M.}~\bibnamefont {Iqbal}}, \bibinfo {author} {\bibfnamefont
  {T.~M.}\ \bibnamefont {Gatterman}}, \bibinfo {author} {\bibfnamefont
  {D.}~\bibnamefont {Gresh}}, \bibinfo {author} {\bibfnamefont
  {A.}~\bibnamefont {Hankin}}, \bibinfo {author} {\bibfnamefont
  {N.}~\bibnamefont {Hewitt}}, \bibinfo {author} {\bibfnamefont {C.~V.}\
  \bibnamefont {Horst}},  \emph {et~al.},\ }\href {\doibase 10.48550/arXiv.2302.03029} {\bibfield  {journal} {\bibinfo  {journal} {arXiv
  preprint arXiv:2302.03029}\ } (\bibinfo {year} {2023}),\
  10.48550/arXiv.2302.03029}\BibitemShut {NoStop}%
\bibitem [{\citenamefont {Moses}\ \emph {et~al.}(2023)\citenamefont {Moses},
  \citenamefont {Baldwin}, \citenamefont {Allman}, \citenamefont {Ancona},
  \citenamefont {Ascarrunz}, \citenamefont {Barnes}, \citenamefont
  {Bartolotta}, \citenamefont {Bjork}, \citenamefont {Blanchard}, \citenamefont
  {Bohn} \emph {et~al.}}]{moses2023race}%
  \BibitemOpen
  \bibfield  {author} {\bibinfo {author} {\bibfnamefont {S.~A.}\ \bibnamefont
  {Moses}}, \bibinfo {author} {\bibfnamefont {C.~H.}\ \bibnamefont {Baldwin}},
  \bibinfo {author} {\bibfnamefont {M.~S.}\ \bibnamefont {Allman}}, \bibinfo
  {author} {\bibfnamefont {R.}~\bibnamefont {Ancona}}, \bibinfo {author}
  {\bibfnamefont {L.}~\bibnamefont {Ascarrunz}}, \bibinfo {author}
  {\bibfnamefont {C.}~\bibnamefont {Barnes}}, \bibinfo {author} {\bibfnamefont
  {J.}~\bibnamefont {Bartolotta}}, \bibinfo {author} {\bibfnamefont
  {B.}~\bibnamefont {Bjork}}, \bibinfo {author} {\bibfnamefont
  {P.}~\bibnamefont {Blanchard}}, \bibinfo {author} {\bibfnamefont
  {M.}~\bibnamefont {Bohn}},  \emph {et~al.},\ }\href {\doibase 10.1103/PhysRevX.13.041052} {\bibfield  {journal} {\bibinfo  {journal}
  {Physical Review X}\ }\textbf {\bibinfo {volume} {13}},\ \bibinfo {pages}
  {041052} (\bibinfo {year} {2023})}\BibitemShut {NoStop}%
\bibitem [{ben(pear)}]{benchqc}%
  \BibitemOpen
  \href@noop {} {\enquote {\bibinfo {title} {Bench{QC} - scalable and modular
  benchmarking of modern quantum computing applications},}\ } (\bibinfo {year}
  {to appear})\BibitemShut {NoStop}%
\bibitem [{\citenamefont {Schiffer}\ \emph {et~al.}(2024)\citenamefont
  {Schiffer}, \citenamefont {Rubio}, \citenamefont {Trivedi},\ and\
  \citenamefont {Cirac}}]{schiffer2024quantum}%
  \BibitemOpen
  \bibfield  {author} {\bibinfo {author} {\bibfnamefont {B.~F.}\ \bibnamefont
  {Schiffer}}, \bibinfo {author} {\bibfnamefont {A.~F.}\ \bibnamefont {Rubio}},
  \bibinfo {author} {\bibfnamefont {R.}~\bibnamefont {Trivedi}}, \ and\
  \bibinfo {author} {\bibfnamefont {J.~I.}\ \bibnamefont {Cirac}},\ }\href
  {\doibase 10.48550/arXiv.2404.15397} {\bibfield  {journal} {\bibinfo
  {journal} {arXiv preprint arXiv:2404.15397}\ } (\bibinfo {year} {2024}),\
  10.48550/arXiv.2404.15397}\BibitemShut {NoStop}%
\bibitem [{\citenamefont {Granet}\ and\ \citenamefont
  {Dreyer}(2025)}]{granet2024dilution}%
  \BibitemOpen
  \bibfield  {author} {\bibinfo {author} {\bibfnamefont {E.}~\bibnamefont
  {Granet}}\ and\ \bibinfo {author} {\bibfnamefont {H.}~\bibnamefont
  {Dreyer}},\ }\href {\doibase 10.1103/PRXQuantum.6.010333} {\bibfield
  {journal} {\bibinfo  {journal} {PRX Quantum}\ }\textbf {\bibinfo {volume}
  {6}},\ \bibinfo {pages} {010333} (\bibinfo {year} {2025})}\BibitemShut
  {NoStop}%
\bibitem [{\citenamefont {Chertkov}\ \emph {et~al.}(2024)\citenamefont
  {Chertkov}, \citenamefont {Chen}, \citenamefont {Lubasch}, \citenamefont
  {Hayes},\ and\ \citenamefont {Foss-Feig}}]{chertkov2024robustness}%
  \BibitemOpen
  \bibfield  {author} {\bibinfo {author} {\bibfnamefont {E.}~\bibnamefont
  {Chertkov}}, \bibinfo {author} {\bibfnamefont {Y.-H.}\ \bibnamefont {Chen}},
  \bibinfo {author} {\bibfnamefont {M.}~\bibnamefont {Lubasch}}, \bibinfo
  {author} {\bibfnamefont {D.}~\bibnamefont {Hayes}}, \ and\ \bibinfo {author}
  {\bibfnamefont {M.}~\bibnamefont {Foss-Feig}},\ }\href {\doibase 10.48550/arXiv.2410.10794} {\bibfield  {journal} {\bibinfo  {journal} {arXiv
  preprint arXiv:2410.10794}\ } (\bibinfo {year} {2024}),\
  10.48550/arXiv.2410.10794}\BibitemShut {NoStop}%
\bibitem [{\citenamefont {Emerson}\ \emph {et~al.}(2005)\citenamefont
  {Emerson}, \citenamefont {Alicki},\ and\ \citenamefont
  {{\.Z}yczkowski}}]{emerson2005scalable}%
  \BibitemOpen
  \bibfield  {author} {\bibinfo {author} {\bibfnamefont {J.}~\bibnamefont
  {Emerson}}, \bibinfo {author} {\bibfnamefont {R.}~\bibnamefont {Alicki}}, \
  and\ \bibinfo {author} {\bibfnamefont {K.}~\bibnamefont {{\.Z}yczkowski}},\
  }\href {\doibase 10.1088/1464-4266/7/10/021} {\bibfield  {journal} {\bibinfo
  {journal} {Journal of Optics B: Quantum and Semiclassical Optics}\ }\textbf
  {\bibinfo {volume} {7}},\ \bibinfo {pages} {S347} (\bibinfo {year}
  {2005})}\BibitemShut {NoStop}%
\bibitem [{\citenamefont {Knill}\ \emph {et~al.}(2008)\citenamefont {Knill},
  \citenamefont {Leibfried}, \citenamefont {Reichle}, \citenamefont {Britton},
  \citenamefont {Blakestad}, \citenamefont {Jost}, \citenamefont {Langer},
  \citenamefont {Ozeri}, \citenamefont {Seidelin},\ and\ \citenamefont
  {Wineland}}]{knill2008randomized}%
  \BibitemOpen
  \bibfield  {author} {\bibinfo {author} {\bibfnamefont {E.}~\bibnamefont
  {Knill}}, \bibinfo {author} {\bibfnamefont {D.}~\bibnamefont {Leibfried}},
  \bibinfo {author} {\bibfnamefont {R.}~\bibnamefont {Reichle}}, \bibinfo
  {author} {\bibfnamefont {J.}~\bibnamefont {Britton}}, \bibinfo {author}
  {\bibfnamefont {R.~B.}\ \bibnamefont {Blakestad}}, \bibinfo {author}
  {\bibfnamefont {J.~D.}\ \bibnamefont {Jost}}, \bibinfo {author}
  {\bibfnamefont {C.}~\bibnamefont {Langer}}, \bibinfo {author} {\bibfnamefont
  {R.}~\bibnamefont {Ozeri}}, \bibinfo {author} {\bibfnamefont
  {S.}~\bibnamefont {Seidelin}}, \ and\ \bibinfo {author} {\bibfnamefont
  {D.~J.}\ \bibnamefont {Wineland}},\ }\href {\doibase PhysRevA.77.012307}
  {\bibfield  {journal} {\bibinfo  {journal} {Physical Review A}\ }\textbf
  {\bibinfo {volume} {77}},\ \bibinfo {pages} {012307} (\bibinfo {year}
  {2008})}\BibitemShut {NoStop}%
\bibitem [{\citenamefont {Blume-Kohout}\ \emph {et~al.}(2017)\citenamefont
  {Blume-Kohout}, \citenamefont {Gamble}, \citenamefont {Nielsen},
  \citenamefont {Rudinger}, \citenamefont {Mizrahi}, \citenamefont {Fortier},\
  and\ \citenamefont {Maunz}}]{blume2017demonstration}%
  \BibitemOpen
  \bibfield  {author} {\bibinfo {author} {\bibfnamefont {R.}~\bibnamefont
  {Blume-Kohout}}, \bibinfo {author} {\bibfnamefont {J.~K.}\ \bibnamefont
  {Gamble}}, \bibinfo {author} {\bibfnamefont {E.}~\bibnamefont {Nielsen}},
  \bibinfo {author} {\bibfnamefont {K.}~\bibnamefont {Rudinger}}, \bibinfo
  {author} {\bibfnamefont {J.}~\bibnamefont {Mizrahi}}, \bibinfo {author}
  {\bibfnamefont {K.}~\bibnamefont {Fortier}}, \ and\ \bibinfo {author}
  {\bibfnamefont {P.}~\bibnamefont {Maunz}},\ }\href {\doibase 10.1038/ncomms14485} {\bibfield  {journal} {\bibinfo  {journal} {Nature
  communications}\ }\textbf {\bibinfo {volume} {8}},\ \bibinfo {pages} {14485}
  (\bibinfo {year} {2017})}\BibitemShut {NoStop}%
\bibitem [{\citenamefont {Erhard}\ \emph {et~al.}(2019)\citenamefont {Erhard},
  \citenamefont {Wallman}, \citenamefont {Postler}, \citenamefont {Meth},
  \citenamefont {Stricker}, \citenamefont {Martinez}, \citenamefont
  {Schindler}, \citenamefont {Monz}, \citenamefont {Emerson},\ and\
  \citenamefont {Blatt}}]{erhard2019characterizing}%
  \BibitemOpen
  \bibfield  {author} {\bibinfo {author} {\bibfnamefont {A.}~\bibnamefont
  {Erhard}}, \bibinfo {author} {\bibfnamefont {J.~J.}\ \bibnamefont {Wallman}},
  \bibinfo {author} {\bibfnamefont {L.}~\bibnamefont {Postler}}, \bibinfo
  {author} {\bibfnamefont {M.}~\bibnamefont {Meth}}, \bibinfo {author}
  {\bibfnamefont {R.}~\bibnamefont {Stricker}}, \bibinfo {author}
  {\bibfnamefont {E.~A.}\ \bibnamefont {Martinez}}, \bibinfo {author}
  {\bibfnamefont {P.}~\bibnamefont {Schindler}}, \bibinfo {author}
  {\bibfnamefont {T.}~\bibnamefont {Monz}}, \bibinfo {author} {\bibfnamefont
  {J.}~\bibnamefont {Emerson}}, \ and\ \bibinfo {author} {\bibfnamefont
  {R.}~\bibnamefont {Blatt}},\ }\href {\doibase 10.1038/s41467-019-13068-7}
  {\bibfield  {journal} {\bibinfo  {journal} {Nature communications}\ }\textbf
  {\bibinfo {volume} {10}},\ \bibinfo {pages} {5347} (\bibinfo {year}
  {2019})}\BibitemShut {NoStop}%
\bibitem [{\citenamefont {Cross}\ \emph {et~al.}(2019)\citenamefont {Cross},
  \citenamefont {Bishop}, \citenamefont {Sheldon}, \citenamefont {Nation},\
  and\ \citenamefont {Gambetta}}]{cross2019validating}%
  \BibitemOpen
  \bibfield  {author} {\bibinfo {author} {\bibfnamefont {A.~W.}\ \bibnamefont
  {Cross}}, \bibinfo {author} {\bibfnamefont {L.~S.}\ \bibnamefont {Bishop}},
  \bibinfo {author} {\bibfnamefont {S.}~\bibnamefont {Sheldon}}, \bibinfo
  {author} {\bibfnamefont {P.~D.}\ \bibnamefont {Nation}}, \ and\ \bibinfo
  {author} {\bibfnamefont {J.~M.}\ \bibnamefont {Gambetta}},\ }\href {\doibase 10.1103/PhysRevA.100.032328} {\bibfield  {journal} {\bibinfo  {journal}
  {Physical Review A}\ }\textbf {\bibinfo {volume} {100}},\ \bibinfo {pages}
  {032328} (\bibinfo {year} {2019})}\BibitemShut {NoStop}%
\bibitem [{\citenamefont {Arute}\ \emph {et~al.}(2019)\citenamefont {Arute},
  \citenamefont {Arya}, \citenamefont {Babbush}, \citenamefont {Bacon},
  \citenamefont {Bardin}, \citenamefont {Barends}, \citenamefont {Biswas},
  \citenamefont {Boixo}, \citenamefont {Brandao}, \citenamefont {Buell} \emph
  {et~al.}}]{arute2019quantum}%
  \BibitemOpen
  \bibfield  {author} {\bibinfo {author} {\bibfnamefont {F.}~\bibnamefont
  {Arute}}, \bibinfo {author} {\bibfnamefont {K.}~\bibnamefont {Arya}},
  \bibinfo {author} {\bibfnamefont {R.}~\bibnamefont {Babbush}}, \bibinfo
  {author} {\bibfnamefont {D.}~\bibnamefont {Bacon}}, \bibinfo {author}
  {\bibfnamefont {J.~C.}\ \bibnamefont {Bardin}}, \bibinfo {author}
  {\bibfnamefont {R.}~\bibnamefont {Barends}}, \bibinfo {author} {\bibfnamefont
  {R.}~\bibnamefont {Biswas}}, \bibinfo {author} {\bibfnamefont
  {S.}~\bibnamefont {Boixo}}, \bibinfo {author} {\bibfnamefont {F.~G.}\
  \bibnamefont {Brandao}}, \bibinfo {author} {\bibfnamefont {D.~A.}\
  \bibnamefont {Buell}},  \emph {et~al.},\ }\href {\doibase 10.1038/s41586-019-1666-5} {\bibfield  {journal} {\bibinfo  {journal}
  {Nature}\ }\textbf {\bibinfo {volume} {574}},\ \bibinfo {pages} {505}
  (\bibinfo {year} {2019})}\BibitemShut {NoStop}%
\bibitem [{\citenamefont {Neill}\ \emph {et~al.}(2018)\citenamefont {Neill},
  \citenamefont {Roushan}, \citenamefont {Kechedzhi}, \citenamefont {Boixo},
  \citenamefont {Isakov}, \citenamefont {Smelyanskiy}, \citenamefont {Megrant},
  \citenamefont {Chiaro}, \citenamefont {Dunsworth}, \citenamefont {Arya} \emph
  {et~al.}}]{neill2018blueprint}%
  \BibitemOpen
  \bibfield  {author} {\bibinfo {author} {\bibfnamefont {C.}~\bibnamefont
  {Neill}}, \bibinfo {author} {\bibfnamefont {P.}~\bibnamefont {Roushan}},
  \bibinfo {author} {\bibfnamefont {K.}~\bibnamefont {Kechedzhi}}, \bibinfo
  {author} {\bibfnamefont {S.}~\bibnamefont {Boixo}}, \bibinfo {author}
  {\bibfnamefont {S.~V.}\ \bibnamefont {Isakov}}, \bibinfo {author}
  {\bibfnamefont {V.}~\bibnamefont {Smelyanskiy}}, \bibinfo {author}
  {\bibfnamefont {A.}~\bibnamefont {Megrant}}, \bibinfo {author} {\bibfnamefont
  {B.}~\bibnamefont {Chiaro}}, \bibinfo {author} {\bibfnamefont
  {A.}~\bibnamefont {Dunsworth}}, \bibinfo {author} {\bibfnamefont
  {K.}~\bibnamefont {Arya}},  \emph {et~al.},\ }\href {\doibase 10.1126/science.aao4309} {\bibfield  {journal} {\bibinfo  {journal}
  {Science}\ }\textbf {\bibinfo {volume} {360}},\ \bibinfo {pages} {195}
  (\bibinfo {year} {2018})}\BibitemShut {NoStop}%
\bibitem [{\citenamefont {Proctor}\ \emph {et~al.}(2022)\citenamefont
  {Proctor}, \citenamefont {Rudinger}, \citenamefont {Young}, \citenamefont
  {Nielsen},\ and\ \citenamefont {Blume-Kohout}}]{proctor2022measuring}%
  \BibitemOpen
  \bibfield  {author} {\bibinfo {author} {\bibfnamefont {T.}~\bibnamefont
  {Proctor}}, \bibinfo {author} {\bibfnamefont {K.}~\bibnamefont {Rudinger}},
  \bibinfo {author} {\bibfnamefont {K.}~\bibnamefont {Young}}, \bibinfo
  {author} {\bibfnamefont {E.}~\bibnamefont {Nielsen}}, \ and\ \bibinfo
  {author} {\bibfnamefont {R.}~\bibnamefont {Blume-Kohout}},\ }\href {\doibase 10.1038/s41567-021-01409-7} {\bibfield  {journal} {\bibinfo  {journal}
  {Nature Physics}\ }\textbf {\bibinfo {volume} {18}},\ \bibinfo {pages} {75}
  (\bibinfo {year} {2022})}\BibitemShut {NoStop}%
\bibitem [{\citenamefont {Dallaire-Demers}\ \emph {et~al.}(2020)\citenamefont
  {Dallaire-Demers}, \citenamefont {St{\k{e}}ch{\l}y}, \citenamefont
  {Gonthier}, \citenamefont {Bashige}, \citenamefont {Romero},\ and\
  \citenamefont {Cao}}]{dallaire2020application}%
  \BibitemOpen
  \bibfield  {author} {\bibinfo {author} {\bibfnamefont {P.-L.}\ \bibnamefont
  {Dallaire-Demers}}, \bibinfo {author} {\bibfnamefont {M.}~\bibnamefont
  {St{\k{e}}ch{\l}y}}, \bibinfo {author} {\bibfnamefont {J.~F.}\ \bibnamefont
  {Gonthier}}, \bibinfo {author} {\bibfnamefont {N.~T.}\ \bibnamefont
  {Bashige}}, \bibinfo {author} {\bibfnamefont {J.}~\bibnamefont {Romero}}, \
  and\ \bibinfo {author} {\bibfnamefont {Y.}~\bibnamefont {Cao}},\ }\href
  {\doibase 10.48550/arXiv.2003.01862} {\bibfield  {journal} {\bibinfo
  {journal} {arXiv preprint arXiv:2003.01862}\ } (\bibinfo {year} {2020}),\
  10.48550/arXiv.2003.01862}\BibitemShut {NoStop}%
\bibitem [{\citenamefont {Gard}\ and\ \citenamefont
  {Meier}(2022)}]{gard2022classically}%
  \BibitemOpen
  \bibfield  {author} {\bibinfo {author} {\bibfnamefont {B.~T.}\ \bibnamefont
  {Gard}}\ and\ \bibinfo {author} {\bibfnamefont {A.~M.}\ \bibnamefont
  {Meier}},\ }\href {\doibase 10.1103/PhysRevA.105.042602} {\bibfield
  {journal} {\bibinfo  {journal} {Physical Review A}\ }\textbf {\bibinfo
  {volume} {105}},\ \bibinfo {pages} {042602} (\bibinfo {year}
  {2022})}\BibitemShut {NoStop}%
\bibitem [{\citenamefont {McCaskey}\ \emph {et~al.}(2019)\citenamefont
  {McCaskey}, \citenamefont {Parks}, \citenamefont {Jakowski}, \citenamefont
  {Moore}, \citenamefont {Morris}, \citenamefont {Humble},\ and\ \citenamefont
  {Pooser}}]{mccaskey2019quantum}%
  \BibitemOpen
  \bibfield  {author} {\bibinfo {author} {\bibfnamefont {A.~J.}\ \bibnamefont
  {McCaskey}}, \bibinfo {author} {\bibfnamefont {Z.~P.}\ \bibnamefont {Parks}},
  \bibinfo {author} {\bibfnamefont {J.}~\bibnamefont {Jakowski}}, \bibinfo
  {author} {\bibfnamefont {S.~V.}\ \bibnamefont {Moore}}, \bibinfo {author}
  {\bibfnamefont {T.~D.}\ \bibnamefont {Morris}}, \bibinfo {author}
  {\bibfnamefont {T.~S.}\ \bibnamefont {Humble}}, \ and\ \bibinfo {author}
  {\bibfnamefont {R.~C.}\ \bibnamefont {Pooser}},\ }\href {\doibase 10.48550/arXiv.1905.01534} {\bibfield  {journal} {\bibinfo  {journal} {npj
  Quantum Information}\ }\textbf {\bibinfo {volume} {5}},\ \bibinfo {pages}
  {99} (\bibinfo {year} {2019})}\BibitemShut {NoStop}%
\bibitem [{\citenamefont {Tomesh}\ \emph {et~al.}(2022)\citenamefont {Tomesh},
  \citenamefont {Gokhale}, \citenamefont {Omole}, \citenamefont {Ravi},
  \citenamefont {Smith}, \citenamefont {Viszlai}, \citenamefont {Wu},
  \citenamefont {Hardavellas}, \citenamefont {Martonosi},\ and\ \citenamefont
  {Chong}}]{tomesh2022supermarq}%
  \BibitemOpen
  \bibfield  {author} {\bibinfo {author} {\bibfnamefont {T.}~\bibnamefont
  {Tomesh}}, \bibinfo {author} {\bibfnamefont {P.}~\bibnamefont {Gokhale}},
  \bibinfo {author} {\bibfnamefont {V.}~\bibnamefont {Omole}}, \bibinfo
  {author} {\bibfnamefont {G.~S.}\ \bibnamefont {Ravi}}, \bibinfo {author}
  {\bibfnamefont {K.~N.}\ \bibnamefont {Smith}}, \bibinfo {author}
  {\bibfnamefont {J.}~\bibnamefont {Viszlai}}, \bibinfo {author} {\bibfnamefont
  {X.-C.}\ \bibnamefont {Wu}}, \bibinfo {author} {\bibfnamefont
  {N.}~\bibnamefont {Hardavellas}}, \bibinfo {author} {\bibfnamefont {M.~R.}\
  \bibnamefont {Martonosi}}, \ and\ \bibinfo {author} {\bibfnamefont {F.~T.}\
  \bibnamefont {Chong}},\ }in\ \href {\doibase 10.48550/arXiv.2202.11045}
  {\emph {\bibinfo {booktitle} {2022 IEEE International Symposium on
  High-Performance Computer Architecture (HPCA)}}}\ (\bibinfo {organization}
  {IEEE},\ \bibinfo {year} {2022})\ pp.\ \bibinfo {pages}
  {587--603}\BibitemShut {NoStop}%
\bibitem [{\citenamefont {Lubinski}\ \emph {et~al.}(2024)\citenamefont
  {Lubinski}, \citenamefont {Goings}, \citenamefont {Mayer}, \citenamefont
  {Johri}, \citenamefont {Reddy}, \citenamefont {Mehta}, \citenamefont
  {Bhatia}, \citenamefont {Rappaport}, \citenamefont {Mills}, \citenamefont
  {Baldwin} \emph {et~al.}}]{lubinski2024quantum}%
  \BibitemOpen
  \bibfield  {author} {\bibinfo {author} {\bibfnamefont {T.}~\bibnamefont
  {Lubinski}}, \bibinfo {author} {\bibfnamefont {J.~J.}\ \bibnamefont
  {Goings}}, \bibinfo {author} {\bibfnamefont {K.}~\bibnamefont {Mayer}},
  \bibinfo {author} {\bibfnamefont {S.}~\bibnamefont {Johri}}, \bibinfo
  {author} {\bibfnamefont {N.}~\bibnamefont {Reddy}}, \bibinfo {author}
  {\bibfnamefont {A.}~\bibnamefont {Mehta}}, \bibinfo {author} {\bibfnamefont
  {N.}~\bibnamefont {Bhatia}}, \bibinfo {author} {\bibfnamefont
  {S.}~\bibnamefont {Rappaport}}, \bibinfo {author} {\bibfnamefont
  {D.}~\bibnamefont {Mills}}, \bibinfo {author} {\bibfnamefont {C.~H.}\
  \bibnamefont {Baldwin}},  \emph {et~al.},\ }\href {\doibase 10.48550/arXiv.2402.08985} {\bibfield  {journal} {\bibinfo  {journal} {arXiv
  preprint arXiv:2402.08985}\ } (\bibinfo {year} {2024}),\
  10.48550/arXiv.2402.08985}\BibitemShut {NoStop}%
\bibitem [{\citenamefont {Martiel}\ \emph {et~al.}(2021)\citenamefont
  {Martiel}, \citenamefont {Ayral},\ and\ \citenamefont
  {Allouche}}]{martiel2021benchmarking}%
  \BibitemOpen
  \bibfield  {author} {\bibinfo {author} {\bibfnamefont {S.}~\bibnamefont
  {Martiel}}, \bibinfo {author} {\bibfnamefont {T.}~\bibnamefont {Ayral}}, \
  and\ \bibinfo {author} {\bibfnamefont {C.}~\bibnamefont {Allouche}},\ }\href
  {\doibase 10.1109/TQE.2021.3090207} {\bibfield  {journal} {\bibinfo
  {journal} {IEEE Transactions on Quantum Engineering}\ }\textbf {\bibinfo
  {volume} {2}},\ \bibinfo {pages} {1} (\bibinfo {year} {2021})}\BibitemShut
  {NoStop}%
\bibitem [{\citenamefont {van~der Schoot}\ \emph {et~al.}(2024)\citenamefont
  {van~der Schoot}, \citenamefont {Wezeman}, \citenamefont {Neumann},
  \citenamefont {Phillipson},\ and\ \citenamefont {Kooij}}]{van2024extending}%
  \BibitemOpen
  \bibfield  {author} {\bibinfo {author} {\bibfnamefont {W.}~\bibnamefont
  {van~der Schoot}}, \bibinfo {author} {\bibfnamefont {R.}~\bibnamefont
  {Wezeman}}, \bibinfo {author} {\bibfnamefont {N.}~\bibnamefont {Neumann}},
  \bibinfo {author} {\bibfnamefont {F.}~\bibnamefont {Phillipson}}, \ and\
  \bibinfo {author} {\bibfnamefont {R.}~\bibnamefont {Kooij}},\ }in\ \href
  {\doibase 10.48550/arXiv.2302.00639} {\emph {\bibinfo {booktitle} {2024 IEEE
  International Conference on Quantum Computing and Engineering (QCE)}}},\
  Vol.~\bibinfo {volume} {1}\ (\bibinfo {organization} {IEEE},\ \bibinfo {year}
  {2024})\ pp.\ \bibinfo {pages} {941--951}\BibitemShut {NoStop}%
\bibitem [{\citenamefont {Fin{\v{z}}gar}\ \emph {et~al.}(2022)\citenamefont
  {Fin{\v{z}}gar}, \citenamefont {Ross}, \citenamefont {H{\"o}lscher},
  \citenamefont {Klepsch},\ and\ \citenamefont {Luckow}}]{finvzgar2022quark}%
  \BibitemOpen
  \bibfield  {author} {\bibinfo {author} {\bibfnamefont {J.~R.}\ \bibnamefont
  {Fin{\v{z}}gar}}, \bibinfo {author} {\bibfnamefont {P.}~\bibnamefont {Ross}},
  \bibinfo {author} {\bibfnamefont {L.}~\bibnamefont {H{\"o}lscher}}, \bibinfo
  {author} {\bibfnamefont {J.}~\bibnamefont {Klepsch}}, \ and\ \bibinfo
  {author} {\bibfnamefont {A.}~\bibnamefont {Luckow}},\ }in\ \href {\doibase 10.1109/QCE53715.2022.00042} {\emph {\bibinfo {booktitle} {2022 IEEE
  international conference on quantum computing and engineering (QCE)}}}\
  (\bibinfo {organization} {IEEE},\ \bibinfo {year} {2022})\ pp.\ \bibinfo
  {pages} {226--237}\BibitemShut {NoStop}%
\bibitem [{\citenamefont {Mesman}\ \emph {et~al.}(2021)\citenamefont {Mesman},
  \citenamefont {Al-Ars},\ and\ \citenamefont {M{\"o}ller}}]{mesman2021qpack}%
  \BibitemOpen
  \bibfield  {author} {\bibinfo {author} {\bibfnamefont {K.}~\bibnamefont
  {Mesman}}, \bibinfo {author} {\bibfnamefont {Z.}~\bibnamefont {Al-Ars}}, \
  and\ \bibinfo {author} {\bibfnamefont {M.}~\bibnamefont {M{\"o}ller}},\
  }\href {\doibase 10.48550/arXiv.2103.17193} {\bibfield  {journal} {\bibinfo
  {journal} {arXiv preprint arXiv:2103.17193}\ } (\bibinfo {year} {2021}),\
  10.48550/arXiv.2103.17193}\BibitemShut {NoStop}%
\bibitem [{\citenamefont {Barbaresco}\ \emph {et~al.}(2024)\citenamefont
  {Barbaresco}, \citenamefont {Rioux}, \citenamefont {Labreuche}, \citenamefont
  {Nowak}, \citenamefont {Olivier}, \citenamefont {Nicolazic}, \citenamefont
  {Hess}, \citenamefont {Guilmin}, \citenamefont {Wang}, \citenamefont
  {Sassolas} \emph {et~al.}}]{barbaresco2024bacq}%
  \BibitemOpen
  \bibfield  {author} {\bibinfo {author} {\bibfnamefont {F.}~\bibnamefont
  {Barbaresco}}, \bibinfo {author} {\bibfnamefont {L.}~\bibnamefont {Rioux}},
  \bibinfo {author} {\bibfnamefont {C.}~\bibnamefont {Labreuche}}, \bibinfo
  {author} {\bibfnamefont {M.}~\bibnamefont {Nowak}}, \bibinfo {author}
  {\bibfnamefont {N.}~\bibnamefont {Olivier}}, \bibinfo {author} {\bibfnamefont
  {D.}~\bibnamefont {Nicolazic}}, \bibinfo {author} {\bibfnamefont
  {O.}~\bibnamefont {Hess}}, \bibinfo {author} {\bibfnamefont {A.-L.}\
  \bibnamefont {Guilmin}}, \bibinfo {author} {\bibfnamefont {R.}~\bibnamefont
  {Wang}}, \bibinfo {author} {\bibfnamefont {T.}~\bibnamefont {Sassolas}},
  \emph {et~al.},\ }\href {\doibase 10.48550/arXiv.2403.12205} {\bibfield
  {journal} {\bibinfo  {journal} {arXiv preprint arXiv:2403.12205}\ } (\bibinfo
  {year} {2024}),\ 10.48550/arXiv.2403.12205}\BibitemShut {NoStop}%
\bibitem [{\citenamefont {Dong}\ and\ \citenamefont
  {Lin}(2021)}]{dong2021random}%
  \BibitemOpen
  \bibfield  {author} {\bibinfo {author} {\bibfnamefont {Y.}~\bibnamefont
  {Dong}}\ and\ \bibinfo {author} {\bibfnamefont {L.}~\bibnamefont {Lin}},\
  }\href {\doibase 10.1103/PhysRevA.103.062412} {\bibfield  {journal} {\bibinfo
   {journal} {Physical Review A}\ }\textbf {\bibinfo {volume} {103}},\ \bibinfo
  {pages} {062412} (\bibinfo {year} {2021})}\BibitemShut {NoStop}%
\bibitem [{\citenamefont {Cornelissen}\ \emph {et~al.}(2021)\citenamefont
  {Cornelissen}, \citenamefont {Bausch},\ and\ \citenamefont
  {Gily{\'e}n}}]{cornelissen2021scalable}%
  \BibitemOpen
  \bibfield  {author} {\bibinfo {author} {\bibfnamefont {A.}~\bibnamefont
  {Cornelissen}}, \bibinfo {author} {\bibfnamefont {J.}~\bibnamefont {Bausch}},
  \ and\ \bibinfo {author} {\bibfnamefont {A.}~\bibnamefont {Gily{\'e}n}},\
  }\href {\doibase 10.48550/arXiv.2104.10698} {\bibfield  {journal} {\bibinfo
  {journal} {arXiv preprint arXiv:2104.10698}\ } (\bibinfo {year} {2021}),\
  10.48550/arXiv.2104.10698}\BibitemShut {NoStop}%
\bibitem [{\citenamefont {Sawaya}\ \emph {et~al.}(2024)\citenamefont {Sawaya},
  \citenamefont {Marti-Dafcik}, \citenamefont {Ho}, \citenamefont {Tabor},
  \citenamefont {Neira}, \citenamefont {Magann}, \citenamefont {Premaratne},
  \citenamefont {Dubey}, \citenamefont {Matsuura}, \citenamefont {Bishop} \emph
  {et~al.}}]{sawaya2023hamlib}%
  \BibitemOpen
  \bibfield  {author} {\bibinfo {author} {\bibfnamefont {N.~P.}\ \bibnamefont
  {Sawaya}}, \bibinfo {author} {\bibfnamefont {D.}~\bibnamefont
  {Marti-Dafcik}}, \bibinfo {author} {\bibfnamefont {Y.}~\bibnamefont {Ho}},
  \bibinfo {author} {\bibfnamefont {D.~P.}\ \bibnamefont {Tabor}}, \bibinfo
  {author} {\bibfnamefont {D.~E.~B.}\ \bibnamefont {Neira}}, \bibinfo {author}
  {\bibfnamefont {A.~B.}\ \bibnamefont {Magann}}, \bibinfo {author}
  {\bibfnamefont {S.}~\bibnamefont {Premaratne}}, \bibinfo {author}
  {\bibfnamefont {P.}~\bibnamefont {Dubey}}, \bibinfo {author} {\bibfnamefont
  {A.}~\bibnamefont {Matsuura}}, \bibinfo {author} {\bibfnamefont
  {N.}~\bibnamefont {Bishop}},  \emph {et~al.},\ }\href {\doibase 10.22331/q-2024-12-11-1559} {\bibfield  {journal} {\bibinfo  {journal}
  {Quantum}\ }\textbf {\bibinfo {volume} {8}},\ \bibinfo {pages} {1559}
  (\bibinfo {year} {2024})}\BibitemShut {NoStop}%
\bibitem [{\citenamefont {Wang}\ \emph {et~al.}(2021)\citenamefont {Wang},
  \citenamefont {Fontana}, \citenamefont {Cerezo}, \citenamefont {Sharma},
  \citenamefont {Sone}, \citenamefont {Cincio},\ and\ \citenamefont
  {Coles}}]{wang2021noise}%
  \BibitemOpen
  \bibfield  {author} {\bibinfo {author} {\bibfnamefont {S.}~\bibnamefont
  {Wang}}, \bibinfo {author} {\bibfnamefont {E.}~\bibnamefont {Fontana}},
  \bibinfo {author} {\bibfnamefont {M.}~\bibnamefont {Cerezo}}, \bibinfo
  {author} {\bibfnamefont {K.}~\bibnamefont {Sharma}}, \bibinfo {author}
  {\bibfnamefont {A.}~\bibnamefont {Sone}}, \bibinfo {author} {\bibfnamefont
  {L.}~\bibnamefont {Cincio}}, \ and\ \bibinfo {author} {\bibfnamefont {P.~J.}\
  \bibnamefont {Coles}},\ }\href {\doibase 10.1038/s41467-021-27045-6}
  {\bibfield  {journal} {\bibinfo  {journal} {Nature communications}\ }\textbf
  {\bibinfo {volume} {12}},\ \bibinfo {pages} {6961} (\bibinfo {year}
  {2021})}\BibitemShut {NoStop}%
\bibitem [{\citenamefont {Agrawal}\ \emph {et~al.}(2024)\citenamefont
  {Agrawal}, \citenamefont {Job}, \citenamefont {Wilson}, \citenamefont
  {Saadatmand}, \citenamefont {Hodson}, \citenamefont {Mutus}, \citenamefont
  {Caesura}, \citenamefont {Johnson}, \citenamefont {Elenewski}, \citenamefont
  {Morrell} \emph {et~al.}}]{agrawal2024quantifying}%
  \BibitemOpen
  \bibfield  {author} {\bibinfo {author} {\bibfnamefont {A.~A.}\ \bibnamefont
  {Agrawal}}, \bibinfo {author} {\bibfnamefont {J.}~\bibnamefont {Job}},
  \bibinfo {author} {\bibfnamefont {T.~L.}\ \bibnamefont {Wilson}}, \bibinfo
  {author} {\bibfnamefont {S.}~\bibnamefont {Saadatmand}}, \bibinfo {author}
  {\bibfnamefont {M.~J.}\ \bibnamefont {Hodson}}, \bibinfo {author}
  {\bibfnamefont {J.~Y.}\ \bibnamefont {Mutus}}, \bibinfo {author}
  {\bibfnamefont {A.}~\bibnamefont {Caesura}}, \bibinfo {author} {\bibfnamefont
  {P.~D.}\ \bibnamefont {Johnson}}, \bibinfo {author} {\bibfnamefont {J.~E.}\
  \bibnamefont {Elenewski}}, \bibinfo {author} {\bibfnamefont {K.~J.}\
  \bibnamefont {Morrell}},  \emph {et~al.},\ }\href {\doibase 10.48550/arXiv.2406.06511} {\bibfield  {journal} {\bibinfo  {journal} {arXiv
  preprint arXiv:2406.06511}\ } (\bibinfo {year} {2024}),\
  10.48550/arXiv.2406.06511}\BibitemShut {NoStop}%
\bibitem [{\citenamefont {Nigmatullin}\ \emph {et~al.}(2024)\citenamefont
  {Nigmatullin}, \citenamefont {Hemery}, \citenamefont {Ghanem}, \citenamefont
  {Moses}, \citenamefont {Gresh}, \citenamefont {Siegfried}, \citenamefont
  {Mills}, \citenamefont {Gatterman}, \citenamefont {Hewitt}, \citenamefont
  {Granet} \emph {et~al.}}]{nigmatullin2024experimental}%
  \BibitemOpen
  \bibfield  {author} {\bibinfo {author} {\bibfnamefont {R.}~\bibnamefont
  {Nigmatullin}}, \bibinfo {author} {\bibfnamefont {K.}~\bibnamefont {Hemery}},
  \bibinfo {author} {\bibfnamefont {K.}~\bibnamefont {Ghanem}}, \bibinfo
  {author} {\bibfnamefont {S.}~\bibnamefont {Moses}}, \bibinfo {author}
  {\bibfnamefont {D.}~\bibnamefont {Gresh}}, \bibinfo {author} {\bibfnamefont
  {P.}~\bibnamefont {Siegfried}}, \bibinfo {author} {\bibfnamefont
  {M.}~\bibnamefont {Mills}}, \bibinfo {author} {\bibfnamefont
  {T.}~\bibnamefont {Gatterman}}, \bibinfo {author} {\bibfnamefont
  {N.}~\bibnamefont {Hewitt}}, \bibinfo {author} {\bibfnamefont
  {E.}~\bibnamefont {Granet}},  \emph {et~al.},\ }\href {\doibase 10.48550/arXiv.2409.06789} {\bibfield  {journal} {\bibinfo  {journal} {arXiv
  preprint arXiv:2409.06789}\ } (\bibinfo {year} {2024}),\
  10.48550/arXiv.2409.06789}\BibitemShut {NoStop}%
\bibitem [{\citenamefont {Derby}\ \emph {et~al.}(2021)\citenamefont {Derby},
  \citenamefont {Klassen}, \citenamefont {Bausch},\ and\ \citenamefont
  {Cubitt}}]{derby2021compact}%
  \BibitemOpen
  \bibfield  {author} {\bibinfo {author} {\bibfnamefont {C.}~\bibnamefont
  {Derby}}, \bibinfo {author} {\bibfnamefont {J.}~\bibnamefont {Klassen}},
  \bibinfo {author} {\bibfnamefont {J.}~\bibnamefont {Bausch}}, \ and\ \bibinfo
  {author} {\bibfnamefont {T.}~\bibnamefont {Cubitt}},\ }\href {\doibase 10.1103/PhysRevB.104.035118} {\bibfield  {journal} {\bibinfo  {journal}
  {Physical Review B}\ }\textbf {\bibinfo {volume} {104}},\ \bibinfo {pages}
  {035118} (\bibinfo {year} {2021})}\BibitemShut {NoStop}%
\bibitem [{\citenamefont {Granet}\ \emph
  {et~al.}(2025{\natexlab{a}})\citenamefont {Granet}, \citenamefont
  {H{\'e}mery},\ and\ \citenamefont {Dreyer}}]{granet2024analog}%
  \BibitemOpen
  \bibfield  {author} {\bibinfo {author} {\bibfnamefont {E.}~\bibnamefont
  {Granet}}, \bibinfo {author} {\bibfnamefont {K.}~\bibnamefont {H{\'e}mery}},
  \ and\ \bibinfo {author} {\bibfnamefont {H.}~\bibnamefont {Dreyer}},\ }\href
  {\doibase 10.1103/PhysRevResearch.7.013213} {\bibfield  {journal} {\bibinfo
  {journal} {Physical Review Research}\ }\textbf {\bibinfo {volume} {7}},\
  \bibinfo {pages} {013213} (\bibinfo {year} {2025}{\natexlab{a}})}\BibitemShut
  {NoStop}%
\bibitem [{\citenamefont {Liu}\ \emph {et~al.}(2022)\citenamefont {Liu},
  \citenamefont {Gong}, \citenamefont {Li}, \citenamefont {Poilblanc},
  \citenamefont {Chen},\ and\ \citenamefont {Gu}}]{liu2022gapless}%
  \BibitemOpen
  \bibfield  {author} {\bibinfo {author} {\bibfnamefont {W.-Y.}\ \bibnamefont
  {Liu}}, \bibinfo {author} {\bibfnamefont {S.-S.}\ \bibnamefont {Gong}},
  \bibinfo {author} {\bibfnamefont {Y.-B.}\ \bibnamefont {Li}}, \bibinfo
  {author} {\bibfnamefont {D.}~\bibnamefont {Poilblanc}}, \bibinfo {author}
  {\bibfnamefont {W.-Q.}\ \bibnamefont {Chen}}, \ and\ \bibinfo {author}
  {\bibfnamefont {Z.-C.}\ \bibnamefont {Gu}},\ }\href {\doibase 10.1016/j.scib.2022.03.010} {\bibfield  {journal} {\bibinfo  {journal}
  {Science bulletin}\ }\textbf {\bibinfo {volume} {67}},\ \bibinfo {pages}
  {1034} (\bibinfo {year} {2022})}\BibitemShut {NoStop}%
\bibitem [{\citenamefont {Mei}\ \emph {et~al.}(2017)\citenamefont {Mei},
  \citenamefont {Chen}, \citenamefont {He},\ and\ \citenamefont
  {Wen}}]{mei2017gapped}%
  \BibitemOpen
  \bibfield  {author} {\bibinfo {author} {\bibfnamefont {J.-W.}\ \bibnamefont
  {Mei}}, \bibinfo {author} {\bibfnamefont {J.-Y.}\ \bibnamefont {Chen}},
  \bibinfo {author} {\bibfnamefont {H.}~\bibnamefont {He}}, \ and\ \bibinfo
  {author} {\bibfnamefont {X.-G.}\ \bibnamefont {Wen}},\ }\href {\doibase 10.1103/PhysRevB.95.235107} {\bibfield  {journal} {\bibinfo  {journal}
  {Physical Review B}\ }\textbf {\bibinfo {volume} {95}},\ \bibinfo {pages}
  {235107} (\bibinfo {year} {2017})}\BibitemShut {NoStop}%
\bibitem [{\citenamefont {L{\"a}uchli}\ \emph {et~al.}(2019)\citenamefont
  {L{\"a}uchli}, \citenamefont {Sudan},\ and\ \citenamefont
  {Moessner}}]{lauchli2019s}%
  \BibitemOpen
  \bibfield  {author} {\bibinfo {author} {\bibfnamefont {A.~M.}\ \bibnamefont
  {L{\"a}uchli}}, \bibinfo {author} {\bibfnamefont {J.}~\bibnamefont {Sudan}},
  \ and\ \bibinfo {author} {\bibfnamefont {R.}~\bibnamefont {Moessner}},\
  }\href {\doibase 10.1103/PhysRevB.100.155142} {\bibfield  {journal} {\bibinfo
   {journal} {Physical Review B}\ }\textbf {\bibinfo {volume} {100}},\ \bibinfo
  {pages} {155142} (\bibinfo {year} {2019})}\BibitemShut {NoStop}%
\bibitem [{\citenamefont {Hogben}\ \emph {et~al.}(2011)\citenamefont {Hogben},
  \citenamefont {Krzystyniak}, \citenamefont {Charnock}, \citenamefont {Hore},\
  and\ \citenamefont {Kuprov}}]{hogben2011spinach}%
  \BibitemOpen
  \bibfield  {author} {\bibinfo {author} {\bibfnamefont {H.~J.}\ \bibnamefont
  {Hogben}}, \bibinfo {author} {\bibfnamefont {M.}~\bibnamefont {Krzystyniak}},
  \bibinfo {author} {\bibfnamefont {G.~T.}\ \bibnamefont {Charnock}}, \bibinfo
  {author} {\bibfnamefont {P.~J.}\ \bibnamefont {Hore}}, \ and\ \bibinfo
  {author} {\bibfnamefont {I.}~\bibnamefont {Kuprov}},\ }\href {\doibase 10.1016/j.jmr.2010.11.008} {\bibfield  {journal} {\bibinfo  {journal}
  {Journal of magnetic resonance}\ }\textbf {\bibinfo {volume} {208}},\
  \bibinfo {pages} {179} (\bibinfo {year} {2011})}\BibitemShut {NoStop}%
\bibitem [{\citenamefont {Sels}\ \emph {et~al.}(2020)\citenamefont {Sels},
  \citenamefont {Dashti}, \citenamefont {Mora}, \citenamefont {Demler},\ and\
  \citenamefont {Demler}}]{sels2020quantum}%
  \BibitemOpen
  \bibfield  {author} {\bibinfo {author} {\bibfnamefont {D.}~\bibnamefont
  {Sels}}, \bibinfo {author} {\bibfnamefont {H.}~\bibnamefont {Dashti}},
  \bibinfo {author} {\bibfnamefont {S.}~\bibnamefont {Mora}}, \bibinfo {author}
  {\bibfnamefont {O.}~\bibnamefont {Demler}}, \ and\ \bibinfo {author}
  {\bibfnamefont {E.}~\bibnamefont {Demler}},\ }\href {\doibase 10.48550/arXiv.1910.14221} {\bibfield  {journal} {\bibinfo  {journal} {Nature
  machine intelligence}\ }\textbf {\bibinfo {volume} {2}},\ \bibinfo {pages}
  {396} (\bibinfo {year} {2020})}\BibitemShut {NoStop}%
\bibitem [{\citenamefont {Seetharam}\ \emph {et~al.}(2023)\citenamefont
  {Seetharam}, \citenamefont {Biswas}, \citenamefont {Noel}, \citenamefont
  {Risinger}, \citenamefont {Zhu}, \citenamefont {Katz}, \citenamefont
  {Chattopadhyay}, \citenamefont {Cetina}, \citenamefont {Monroe},
  \citenamefont {Demler} \emph {et~al.}}]{seetharam2023digital}%
  \BibitemOpen
  \bibfield  {author} {\bibinfo {author} {\bibfnamefont {K.}~\bibnamefont
  {Seetharam}}, \bibinfo {author} {\bibfnamefont {D.}~\bibnamefont {Biswas}},
  \bibinfo {author} {\bibfnamefont {C.}~\bibnamefont {Noel}}, \bibinfo {author}
  {\bibfnamefont {A.}~\bibnamefont {Risinger}}, \bibinfo {author}
  {\bibfnamefont {D.}~\bibnamefont {Zhu}}, \bibinfo {author} {\bibfnamefont
  {O.}~\bibnamefont {Katz}}, \bibinfo {author} {\bibfnamefont {S.}~\bibnamefont
  {Chattopadhyay}}, \bibinfo {author} {\bibfnamefont {M.}~\bibnamefont
  {Cetina}}, \bibinfo {author} {\bibfnamefont {C.}~\bibnamefont {Monroe}},
  \bibinfo {author} {\bibfnamefont {E.}~\bibnamefont {Demler}},  \emph
  {et~al.},\ }\href {\doibase 10.1126/sciadv.adh2594} {\bibfield  {journal}
  {\bibinfo  {journal} {Science Advances}\ }\textbf {\bibinfo {volume} {9}},\
  \bibinfo {pages} {eadh2594} (\bibinfo {year} {2023})}\BibitemShut {NoStop}%
\bibitem [{\citenamefont {Elenewski}\ \emph {et~al.}(2024)\citenamefont
  {Elenewski}, \citenamefont {Camara},\ and\ \citenamefont
  {Kalev}}]{elenewski2024prospects}%
  \BibitemOpen
  \bibfield  {author} {\bibinfo {author} {\bibfnamefont {J.~E.}\ \bibnamefont
  {Elenewski}}, \bibinfo {author} {\bibfnamefont {C.~M.}\ \bibnamefont
  {Camara}}, \ and\ \bibinfo {author} {\bibfnamefont {A.}~\bibnamefont
  {Kalev}},\ }\href {\doibase 10.48550/arXiv.2406.09340} {\bibfield  {journal}
  {\bibinfo  {journal} {arXiv preprint arXiv:2406.09340}\ } (\bibinfo {year}
  {2024}),\ 10.48550/arXiv.2406.09340}\BibitemShut {NoStop}%
\bibitem [{\citenamefont {Khedri}\ \emph {et~al.}(2024)\citenamefont {Khedri},
  \citenamefont {Stadler}, \citenamefont {Bark}, \citenamefont {Lodi},
  \citenamefont {Reiner}, \citenamefont {Vogt}, \citenamefont {Marthaler},\
  and\ \citenamefont {Lepp{\"a}kangas}}]{khedri2024impact}%
  \BibitemOpen
  \bibfield  {author} {\bibinfo {author} {\bibfnamefont {A.}~\bibnamefont
  {Khedri}}, \bibinfo {author} {\bibfnamefont {P.}~\bibnamefont {Stadler}},
  \bibinfo {author} {\bibfnamefont {K.}~\bibnamefont {Bark}}, \bibinfo {author}
  {\bibfnamefont {M.}~\bibnamefont {Lodi}}, \bibinfo {author} {\bibfnamefont
  {R.}~\bibnamefont {Reiner}}, \bibinfo {author} {\bibfnamefont
  {N.}~\bibnamefont {Vogt}}, \bibinfo {author} {\bibfnamefont {M.}~\bibnamefont
  {Marthaler}}, \ and\ \bibinfo {author} {\bibfnamefont {J.}~\bibnamefont
  {Lepp{\"a}kangas}},\ }\href {\doibase 10.48550/arXiv.2404.18903} {\bibfield
  {journal} {\bibinfo  {journal} {arXiv preprint arXiv:2404.18903}\ } (\bibinfo
  {year} {2024}),\ 10.48550/arXiv.2404.18903}\BibitemShut {NoStop}%
\bibitem [{\citenamefont {Stern}\ and\ \citenamefont
  {Sheberstov}(2023)}]{mr-4-87-2023}%
  \BibitemOpen
  \bibfield  {author} {\bibinfo {author} {\bibfnamefont {Q.}~\bibnamefont
  {Stern}}\ and\ \bibinfo {author} {\bibfnamefont {K.}~\bibnamefont
  {Sheberstov}},\ }\href {\doibase 10.5194/mr-4-87-2023} {\bibfield  {journal}
  {\bibinfo  {journal} {Magnetic Resonance}\ }\textbf {\bibinfo {volume} {4}},\
  \bibinfo {pages} {87} (\bibinfo {year} {2023})}\BibitemShut {NoStop}%
\bibitem [{\citenamefont {Wilzewski}\ \emph {et~al.}(2017)\citenamefont
  {Wilzewski}, \citenamefont {Afach}, \citenamefont {Blanchard},\ and\
  \citenamefont {Budker}}]{wilzewski2017method}%
  \BibitemOpen
  \bibfield  {author} {\bibinfo {author} {\bibfnamefont {A.}~\bibnamefont
  {Wilzewski}}, \bibinfo {author} {\bibfnamefont {S.}~\bibnamefont {Afach}},
  \bibinfo {author} {\bibfnamefont {J.~W.}\ \bibnamefont {Blanchard}}, \ and\
  \bibinfo {author} {\bibfnamefont {D.}~\bibnamefont {Budker}},\ }\href
  {\doibase 10.1016/j.jmr.2017.08.016} {\bibfield  {journal} {\bibinfo
  {journal} {Journal of Magnetic Resonance}\ }\textbf {\bibinfo {volume}
  {284}},\ \bibinfo {pages} {66} (\bibinfo {year} {2017})}\BibitemShut
  {NoStop}%
\bibitem [{\citenamefont {Granet}\ \emph
  {et~al.}(2025{\natexlab{b}})\citenamefont {Granet}, \citenamefont {Ghanem},\
  and\ \citenamefont {Dreyer}}]{granet2024practicality}%
  \BibitemOpen
  \bibfield  {author} {\bibinfo {author} {\bibfnamefont {E.}~\bibnamefont
  {Granet}}, \bibinfo {author} {\bibfnamefont {K.}~\bibnamefont {Ghanem}}, \
  and\ \bibinfo {author} {\bibfnamefont {H.}~\bibnamefont {Dreyer}},\ }\href
  {\doibase 10.1103/PhysRevA.111.022428} {\bibfield  {journal} {\bibinfo
  {journal} {Phys. Rev. A}\ }\textbf {\bibinfo {volume} {111}},\ \bibinfo
  {pages} {022428} (\bibinfo {year} {2025}{\natexlab{b}})}\BibitemShut
  {NoStop}%
\bibitem [{\citenamefont {Granet}\ and\ \citenamefont
  {Dreyer}(2024{\natexlab{a}})}]{granet2024hamiltonian}%
  \BibitemOpen
  \bibfield  {author} {\bibinfo {author} {\bibfnamefont {E.}~\bibnamefont
  {Granet}}\ and\ \bibinfo {author} {\bibfnamefont {H.}~\bibnamefont
  {Dreyer}},\ }\href {\doibase 10.1038/s41534-024-00877-y} {\bibfield
  {journal} {\bibinfo  {journal} {npj Quantum Information}\ }\textbf {\bibinfo
  {volume} {10}},\ \bibinfo {pages} {82} (\bibinfo {year}
  {2024}{\natexlab{a}})}\BibitemShut {NoStop}%
\bibitem [{\citenamefont {Granet}\ and\ \citenamefont
  {Dreyer}(2024{\natexlab{b}})}]{granet2024benchmarking}%
  \BibitemOpen
  \bibfield  {author} {\bibinfo {author} {\bibfnamefont {E.}~\bibnamefont
  {Granet}}\ and\ \bibinfo {author} {\bibfnamefont {H.}~\bibnamefont
  {Dreyer}},\ }\href {\doibase 10.48550/arXiv.2404.16001} {\bibfield  {journal}
  {\bibinfo  {journal} {arXiv preprint arXiv:2404.16001}\ } (\bibinfo {year}
  {2024}{\natexlab{b}}),\ 10.48550/arXiv.2404.16001}\BibitemShut {NoStop}%
\bibitem [{\citenamefont {Dunning}\ \emph {et~al.}(2018)\citenamefont
  {Dunning}, \citenamefont {Gupta},\ and\ \citenamefont
  {Silberholz}}]{DunningEtAl2018}%
  \BibitemOpen
  \bibfield  {author} {\bibinfo {author} {\bibfnamefont {I.}~\bibnamefont
  {Dunning}}, \bibinfo {author} {\bibfnamefont {S.}~\bibnamefont {Gupta}}, \
  and\ \bibinfo {author} {\bibfnamefont {J.}~\bibnamefont {Silberholz}},\
  }\href {\doibase 10.1287/ijoc.2017.0798} {\bibfield  {journal} {\bibinfo
  {journal} {{INFORMS} Journal on Computing}\ }\textbf {\bibinfo {volume} {30}}
  (\bibinfo {year} {2018}),\ 10.1287/ijoc.2017.0798}\BibitemShut {NoStop}%
\bibitem [{\citenamefont {Steger}\ and\ \citenamefont
  {Wormald}(1999)}]{steger1999generating}%
  \BibitemOpen
  \bibfield  {author} {\bibinfo {author} {\bibfnamefont {A.}~\bibnamefont
  {Steger}}\ and\ \bibinfo {author} {\bibfnamefont {N.~C.}\ \bibnamefont
  {Wormald}},\ }\href {\doibase 10.1017/S0963548399003867} {\bibfield
  {journal} {\bibinfo  {journal} {Combinatorics, Probability and Computing}\
  }\textbf {\bibinfo {volume} {8}},\ \bibinfo {pages} {377} (\bibinfo {year}
  {1999})}\BibitemShut {NoStop}%
\end{thebibliography}
\end{document}